\documentclass[12pt]{article}
\usepackage{graphicx}
\usepackage{amssymb}
\usepackage{amsmath}
\usepackage{amsfonts}
\usepackage{bm}
\usepackage[square,numbers,sort&compress]{natbib}

\newcommand\vecl[1]{\overrightarrow{#1}}
\newcommand\cev[1]{\overleftarrow{#1}}

\usepackage{longtable}

\newcommand{\be}{\begin{eqnarray}}
\newcommand{\ee}{\end{eqnarray}}

\newcommand{\ket}{\rangle}
\newcommand{\bra}{\langle}

\begin{document}

\title{Exotic Hadrons with Heavy Flavors -- $X$, $Y$, $Z$ and Related States --}

\maketitle

\begin{center}
\author{\large {Atsushi~Hosaka}$^{1,2}$, Toru~Iijima$^{3,4}$, 
Kenkichi~Miyabayashi$^{5}$, 
Yoshihide~Sakai$^{6,7}$, 
Shigehiro~Yasui$^{8}$}
\end{center}
\vspace*{5mm}

\begin{center}
{ \it 
$^{1}${Research Center for Nuclear Physics (RCNP), Osaka University, Ibaraki, Osaka, 567-0047, Japan}\\ 
$^{2}${J-PARC Branch, KEK Theory Center, Institute of Particle and Nuclear Studies, KEK, Tokai, Ibaraki, 319-1106, Japan}\\ 
$^{3}${Graduate School of Science, Nagoya University, Nagoya, 464-8602, Japan}\\ 
$^{4}${Kobayashi-Maskawa Institute, Nagoya University, Nagoya, 464-8602, Japan}\\ 
$^{5}${Nara Women's University, Nara, 630-8506, Japan}\\ 
$^{6}${High Energy Accelerator Research Organization (KEK), Tsukuba, 305-0801, Japan}\\ 
$^{7}${SOKENDAI (The Graduate University for Advanced Studies), Hayama, 240-0193, Japan}\\ 
$^{8}${Department of Physics, Tokyo Institute of Technology, Tokyo, 152-8551, Japan} 
}

\end{center}

\begin{center}
\begin{abstract}%
In the past decade, exotic hadrons with charm and bottom flavors have been extensively studied 
both in experiments and in theories. In this review, we provide topical discussions by selecting $X,Y,Z$ particles, to which Belle has made important contributions. 
These are $X(3872)$, $Y(4260)$, $Z_c(4430)^+$, $Z_c(3900)^+$, $Z_{b}(10610)^+$ and  $Z_{b}(10650)^+$. Based on the current experimental observations, we discuss those states with emphasis on hadronic molecule whose dynamics is governed by chiral symmetry and heavy-quark symmetry of QCD. We also mention briefly various interpretations and some theoretical predictions for the yet undiscovered exotic hadrons.
\end{abstract}
\end{center}


\newpage

\tableofcontents

\section{Introduction} 
\label{sec:Introduction_all}

The nature of the strong interaction does not allow quarks be ``bare" and ``alone", 
thus they are confined into hadrons. 
Many hadron states are categorized into
mesons and baryons containing constituent quark-antiquark ($q \bar q$) and 
three quarks ($qqq$), respectively~\cite{GellMann:1964nj,Zweig:1964jf}. 
On the other hand, there is no proof in the fundamental theory of the 
strong interaction, Quantum Chromo Dynamics (QCD),
to exclude the hadrons having other 
structure than the ordinary mesons and baryons.
In fact, various candidates of exotic hadrons were proposed already from 1970's or earlier; 
such as  tetraquarks having two quarks and two antiquarks as 
constituents,  hybrids containing 
gluonic excitations as additional degrees of freedom,  
mesonic molecules which are bound or resonant  states of two mesons 
and so on~\cite{Jaffe:1976ig,Jaffe:1976ih,Isgur:1984bm,Isgur:1985vy,Dalitz:1959dn,DeRujula:1976qd}.  
However, due to the lack of clear experimental evidences, existence of 
the exotic hadrons have been a ``smoking gun'' issue for a long time. 

The situation has largely changed by the high statistics $e^+e^-$ collision
data accumulated by $B$-factory experiments~\cite{Bevan:2014iga}. 
Originally their data were collected to perform comprehensive study of 
the $CP$ violation in $B$ meson decays. 
At the same time, the high statistics data with abundant charm and bottom 
productions brought an ideal playground to perform heavy hadron spectroscopy.
Especially the states called ``$X$, $Y$, $Z$'' are thought to be 
candidates of exotic hadrons, which have been  attracting a lot of attentions 
to reveal unvisited areas of QCD.

Such activities were initiated around 2003 which is the year when 
the first observation of $X(3872)$ was reported~\cite{Choi:2003ue}.  
Interestingly, an observation of another candidate of exotic state, 
the pentaquark $\Theta^+$ baryon, was also reported in the same period, which together 
with $X(3872)$ triggered diverse activities of both theoretical and 
experimental studies~\cite{Nakano:2003qx,Nakano:2008ee}.  
In this review, we describe the ``$X$, $Y$, $Z$'' states as best established 
states discovered by Belle as well as other relevant experiments and discuss  
their theoretical interpretations.

Historically,  exotic states, in fact, multiquark states have  already been 
pointed out  by Gell-Mann in his original paper of the quark model
in 1964~\cite{GellMann:1964nj}.  
Before the discovery of the charm (heavy) quark, the situation for the light 
quark sector of $u, d, s$ quarks was somewhat complicated, due to their light 
masses which are comparable to the QCD scale of several hundred MeV.  
Contrary, the masses of $c$ and $b$ quarks are large, approximately 
1.5 GeV/$c^2$ and 5 GeV/$c^2$, respectively. 
Because of their large and well-separated masses from the QCD scale, 
the description based on the heavy constituent quark 
wave functions are well established. 
The success of the quark model with a static potential for heavy quarkonium systems 
may be explained by the non-relativistic QCD (NRQCD)~\cite{Bodwin:1994jh} 
or potential non-relativistic QCD (pNRQCD)~\cite{Brambilla:1999xf}.  
In the pNRQCD, the hierarchy of $m \gg mv \gg mv^2$ for small velocity $v$  
justifies the use of a potential for non-relativistically slowly moving heavy quarks.  

\begin{figure}[t]
 \begin{center}
 \includegraphics[width=0.6 \linewidth]{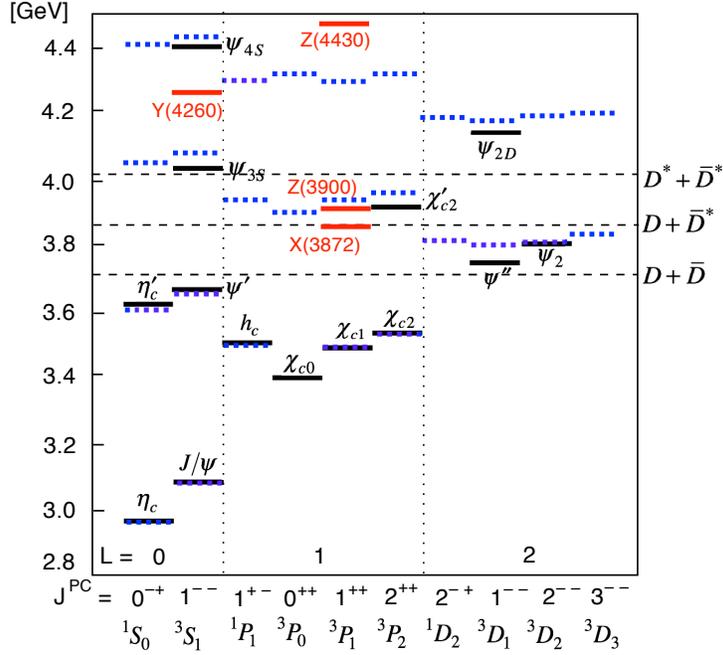}
 \end{center}
 \vspace{-5mm}
\caption{Charmonium spectrum (black solid bars) with some $X$, $Y$, $Z$ states (red solid bars) 
in comparison with a conventional 
quark model results for $c \bar c$ states (blue dashed bars)~\cite{Godfrey:1985xj}.  }
\label{fig_ccbar}
\end{figure}

In Fig.~\ref{fig_ccbar}, charmonium spectrum is shown with some $X$, $Y$, $Z$ 
particles. Experimentally observed spectrum is shown by solid bars which is 
compared with the predictions of the conventional $\bar cc$ quark 
model~\cite{Godfrey:1985xj},
and naive assignment is also shown.  
Below the open charm threshold of $D\bar D, D \bar D^*$ and $D^* \bar D^*$ 
the agreement between the experiment and theory is remarkable, 
as anticipated by the pNRQCD.  
Contrary, near and above the threshold the situation changes, 
where the clear hierarchy of small parameters may no longer hold well.  
Moreover, the coupling to the open charm  (decaying)  channels strongly affects 
the bare $\bar cc$ states.  
Due to confinement, open charm necessarily requires a creation of a light 
$q \bar q$ pair to form charmed mesons, $D$ or $D^*$.  
Thus the presence of extra quark degrees of freedom 
associated with the energy deposit above the threshold 
makes the charmonium spectrum much richer~\cite{DeRujula:1976qd}, 
where realistic theoretical description requires coupled channel treatment 
of the quarkonium like configuration to the open charm mesons.  

Among various candidates of exotic hadrons, perhaps $X(3872)$ is the most 
established one experimentally, and is well studied also by 
theories~\cite{Voloshin:2007dx}.     
The most accepted picture for $X(3872)$ by now is a molecular state of 
the $D$ and $\bar D^*$ mesons~\footnote{
To be precise, both $D\bar D^*$ and $\bar DD^*$ channels can couple.   
In what follows, whenever one of these configurations is indicated, it is 
understood that the other one is also included unless specially noticed.}.  
Because the observed $X(3872)$ is very close to the $D \bar D^*$ threshold, 
the difference between the masses of 
charged $D^{\pm} D^{* \mp}$  and the neutral $D^{0} \bar D^{* 0}$ channels 
makes a large influence on the breaking of isospin symmetry.
The fact that the neutral channel is much closer to the 
threshold than the charged one leads to the dominance of the neutral one.  
This in turn implies a large isospin mixing of $I =0$ and $I=1$ channels, 
reasonably explaining the decay branching ratio of 
$X(3872) \to J/\psi \ \pi^+ \pi^-$ and $J/\psi \ \pi^+ \pi^- \pi^0$.  
We will discuss more on $X(3872)$ in Sections~\ref{sec:exp_x3872} 
and \ref{sec:theory_X3872}.

Theoretically there are other candidates 
for exotic hadrons such as tetraquarks, hybrids or even pure glue-like states.
The tetraquark is a multiquark configuration where colored diquarks are bound by a strong color force between them, thus forming 
a rather tightly bound compact object.  
So far, we still do not have good evidences for such tetraquarks.  
The hybrid states contain the gluonic excitations, which is a consequence of 
colored dynamics of QCD, and therefore 
their excitation energies are of order of at least several hundred MeV.  
Moreover, because the gluons are flavor blind, such excitations couple to 
flavor singlet channels; in terms of hadrons, for instance, two pions.  
There are discussions that 
$Y(4260)$ and similar states
could be a candidate of such hybrid 
states~\cite{Close:2005iz,Kou:2005gt,Zhu:2005hp}, 
though we need more systematic studies not only in the 
charm but also in the bottom region.  

In general, the above different configurations may mix for a given hadronic 
state as long as its quantum numbers allow. 
This is typically what we expect in quantum many-body systems with strong 
interaction near the threshold.  
As anticipated this requires 
a treatment for coupled channel problem.  
In hadron physics, these occur with non-trivial dynamics of QCD, such as 
color confinement and spontaneous breaking of chiral symmetry.  
Perhaps,  one of aims of hadron physics is to clarify how and where different  
configurations show up, 
from which we hope to explain the  observed data with clear physical mechanism.
We then expect to approach more fundamental questions of QCD
to fill the missing link between quarks and hadrons.  

Thus in hadron physics, experimental information is extremely important, 
indicating the necessity of data to explore the strong interaction dynamics.   
In addition to $X(3872)$ and $Y(4260)$, 
more states have been well-established; 
two charged charmonium like states of $Z_c(4430)^+$ and 
$Z_c(3900)^+$~\footnote{Hereafter, inclusion of the charge conjugate 
state is implied unless otherwise stated.}
as well as the twin charged bottomonia, 
$Z_b(10610)^+$ and $Z_b(10650)^+$~\cite{Belle:2011aa}.
Due to their non-zero charge, the $Z_{c, b}^+$ states 
are truly four quark exotic hadrons.  
Experimentally, the confirmation of $Z_c(4430)^+$ by LHCb is also important 
where they performed the Argan plot analysis, verifying its resonance 
nature~\cite{Aaij:2014jqa}.  
This is 
the first evidence that shows the particle nature among exotic hadrons.
Theoretical interpretation, however, does not seem easy and its nature is 
still open.  
Concerning $Z_b(10610)^+$ and $Z_b(10650)^+$, 
KEK B-factory is the only facility which has the $\Upsilon(5S)$\footnote{In this paper we employ the notation $\Upsilon(5S)$, which is denoted as $\Upsilon(10860)$ 
in the particle data \cite{Agashe:2014kda}.} data 
to make the measurements possible.
The current observations such as the peaks in several different final 
states~\cite{Belle:2011aa}, 
finding the neutral partner~\cite{Krokovny:2013mgx}, 
analysis of decay branching ratios~\cite{Adachi:2012cx} are all well 
consistent with what are expected by the $B^{(*)} \bar B^*$ molecule picture.  

This paper is organized as follows.  
In section 2, we review the experimental status focusing on the results brought by Belle from 
the early stage activities to the most recent achievements till today.  
Anomalous behavior in the $e^+ e^-$  reaction rate accompanied by 
two pion emission, so called $Y$ states, 
is emphasized both in the charm and bottom sectors, 
which became the driving force of the new findings.  
A brief but sufficient contents for the status of the most important 
states $X(3872)$, $Y(4260)$, $Z_c(3900)^+$, $Z_c(4430)^+$, 
$Z_b(10610)^+$ and $Z_b(10650)^+$ are summarized.  
Section \ref{sec:theory} is for the theory discussions.  
As basic issues of the theory ideas, we briefly describe 
heavy quark symmetry and chiral symmetry with its spontaneous breakdown, 
as well as important theoretical methods of
QCD sum rule and lattice simulations.   
For interpretations of the exotic states, our discussions do not cover diverse  
approaches, but mostly 
concentrate on rather well accepted interpretation,
the molecular picture,
where the basic effective degrees of freedom are supplied by the scattering hadrons instead of quarks.
This is explained 
for the examples of $X(3872)$ and $Z_b$ resonances.  
We also give a brief review on other possible interpretations
for $Y(4260)$, $Z_c(4430)^+$ and $Z_c(3900)^+$ and on recent discussion based on the dynamical treatment of the scattering hadrons.
Some future prospects are also discussed.
We will then summarize our discussions in Section \ref{sec:summary_all}.

%
\section{Experiment} 
\label{sec:experiments}
%
\subsection{Belle experiment} 
\label{sec:belle_detector}

The Belle detector is the 
4$\pi$ general purpose
spectrometer with 1.5 T solenoidal magnetic field \cite{Abashian:2000cg} 
(see Ref.~ \cite{Brodzicka:2012jm} for summary of achievement).
It is designed to measure time-dependent $CP$ violation in $B$ meson 
system by equipping (1) the silicon vertex detector (SVD) consisting of 
double-sided silicon strip detectors (DSSDs) to get $B$ meson decay 
vertices with superb resolution of 61 $\mu$m in $J/\psi \to \mu^+\mu^-$
mode, (2) the central drift chamber (CDC) filled with He and C$_2$H$_6$ 
half-and-half mixture gas at atmospheric pressure to give 
a high momentum resolution of 0.36\% for 1 GeV/$c$ transverse momentum tracks,
(3) charged particle 
identification system comprised by the plastic scintillator array based
time-of-flight (TOF) counters and aerogel Cerenkov counters (ACC) to
identify charged hadrons to identify charged kaons with typically 
90\% efficiency with 10\% pion misidentification, (4) high resolution 
electromagnetic calorimeter based on the CsI(Tl)
scintillating crystals  
to realize 5 MeV/$c^2$ $\pi^0$ mass resolution and (5) the iron flux 
return instrumented by the resistive planer chambers to identify 
muons and $K^0_L$ (KLM). Electron (muon) identification is done with 
90\% (also 90\%) efficiency with a small fake rate of 0.3\% (2\%).

The KEKB collider \cite{Kurokawa:2001nw,Abe:2013kxa} stores 
8 GeV $e^-$ and 3.5 $e^+$ beams in the proper rings to produce 
$\Upsilon(4S)$ resonance to have high statistics $B$ meson data. 
The accumulated integrated luminosity in that condition is 
711 fb$^{-1}$ corresponding to 772 million $B\bar{B}$ pairs.
The accelerator operation has been done also at $\Upsilon(1S)$, 
$\Upsilon(2S)$ and $\Upsilon(5S)$ up to the integrated luminosities 
6 fb$^{-1}$, 25 fb$^{-1}$ and 121 fb$^{-1}$, respectively.
Runs at off-resonance energy as well as 
the energy scan to go above $\Upsilon(5S)$ 
were also performed.
Thanks to the KEKB accelerator world record luminosity performance, 
each subset is corresponding to the two (or one) order of magnitude
higher statistics than the older experiments.
These high statistics data resulted in so rich and interesting outcome 
for heavy flavored hadron spectroscopy.

\subsection{Selections of exotic events} 
\label{sec:exp_selections}

Heavy quarkonium (or simply quarkonium) which is the general term 
to mention charmonium ($c\bar{c}$) and bottomonium ($b\bar{b}$) can be 
regarded as a reference to discuss exotic hadrons.
Among charmonia, $J/\psi$ and $\psi(2S)$ (or  $\psi^\prime$ in 
Fig.~\ref{fig_ccbar}) decays to $e^+e^-$ or $\mu^+\mu^-$ pair.
Summing up these two decay modes, the branching fraction of interest 
for $J/\psi$ and $\psi(2S)$ amount 12 \%  and 1.6 \%, respectively.
This enables us to reconstruct them with low background.
$\psi(2S)$ can also be reconstructed by $J/\psi \pi^+\pi^-$ mode.
$J/\psi \gamma$ mode is useful to reconstruct $\chi_{c1}$ and 
$\chi_{c2}$ states. 
Similarly, we can have clear signature by $\mu^+\mu^-$ modes for 
$\Upsilon(1S)$, $\Upsilon(2S)$ and $\Upsilon(3S)$ in bottomonia cases.
The exotic hadron candidate states mentioned in this report are mostly 
discovered in the data analyses involving those quarkonia and decays.

For charmonia above the $D$ meson pair (open charm) 
threshold (3.74 GeV/$c^2$) and 
bottomonia above $B$ meson pair (open bottom) threshold (10.56 GeV/$c^2$), 
the strong decays to those heavy meson pairs become dominant and 
consequently decay width would be broad unless special suppression 
mechanism works. 
Therefore we can identify an exotic hadron candidate as 
the quarkonium-like state above the corresponding meson pair threshold 
satisfying one or more criteria out of the following conditions; 
(1)
large branching fraction to the mode other than the heavy meson pair,
(2)
extraordinary narrow decay width, 
(3)
special decay mode which can not be explained by two constituent quarks, 
(4)
the observed mass does not match with any of the predicted ones,
and
(5) 
having an electric charge.

\begin{figure}[t]
 \begin{center}
 \includegraphics[width=0.6 \linewidth]{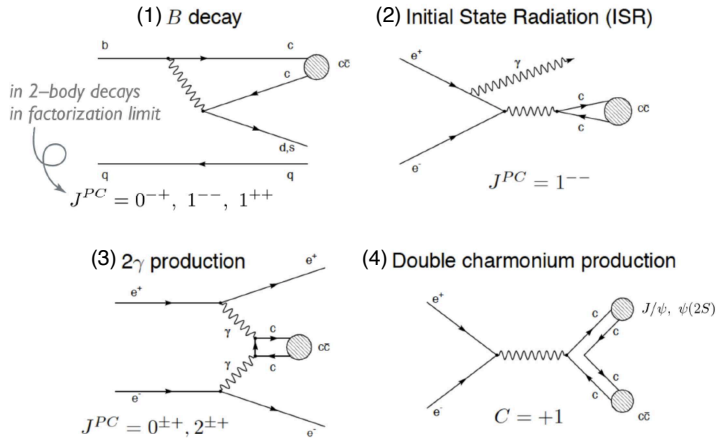}
 \end{center}
 \vspace{-5mm}
\caption{Four processes of productions in $e^{+}e^{-}$ colliders. 
See the text for the explanation. 
}
\label{fig:NHSchool2013_iijima_charmonium_four_peocess}
\end{figure}

Various different reaction processes which can take place at $e^+e^-$ 
collider would 
also help a lot to find and identify new quarkonium(-like) states. 
Depending on production processes, allowed or favored quantum numbers 
such as spin ($J$), parity ($P$) and charge-conjugation ($C$) are 
different. 
(1) 
The charmonium(-like) states of quantum numbers 
$J^{PC}=0^{-+}, 1^{--}$ and $1^{++}$ are favored to be 
produced 
in the two-body $B$ meson decays, 
because the factorization hypothesis works well, 
i.e.  with small final state interaction due to the relatively large $B$ mass.
(2) 
Initial state radiation produces $J^{PC}=1^{--}$ states via a virtual 
photon exchange with varying effective center-mass energy.
(3) Two photon collisions produce $J^{PC}=0^{\pm +}$ and $2^{\pm +}$ states. 
(4) Only $C=+1$ states can be produced in double charmonia production in 
$e^+e^-$ annihilation 
since $J/\psi$ or $\psi(2S)$ is used to tag a candidate event of this reaction
(see Fig.~\ref{fig:NHSchool2013_iijima_charmonium_four_peocess}). 
Such selection rules of quantum numbers are beneficial to discuss  possible interpretation of 
the observed states.

\subsection{$X(3872)$} 
\label{sec:exp_x3872}
\begin{figure}[t]
\centering\includegraphics[width=2.5in]{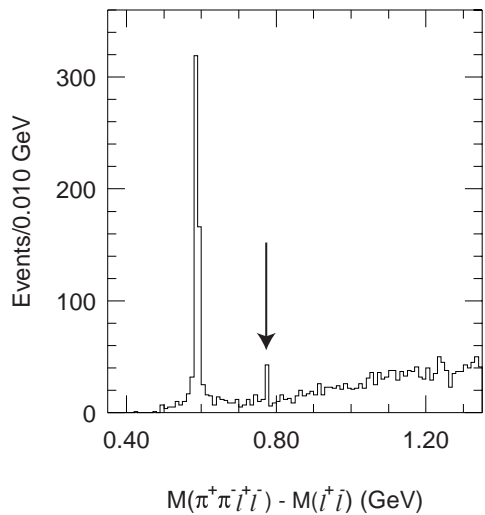}
\caption{The distribution of mass difference between $J/\psi \pi^+\pi^-$
and $J/\psi$ in $B^{\pm} \to J/\psi \pi^+\pi^- K^{\pm}$ decays.
The peak at 0.59 GeV/$c^2$ is due to the conventional charmonium, 
$\psi(2S)$. The peak corresponding to the $X(3872)$ is indicated by a 
vertical arrow \cite{Choi:2003ue}. 
}
\label{fig_x3872_dis}
\end{figure}

In 2003, Belle found very narrow peak in $J/\psi \pi^+\pi^-$ mass spectrum
at 3872 MeV/$c^2$ in $B^{\pm} \to J/\psi \pi^+\pi^- K^{\pm}$ decays 
as shown in Fig. \ref{fig_x3872_dis} using the data 
corresponding to an integrated luminosity of 140 fb$^{-1}$ which contains
152 million $B\bar{B}$ pairs \cite{Choi:2003ue}.
The discovered state is named $X(3872)$.
The letter ``$X$'' was chosen because of its extraordinary properties;
in spite of its mass well above $D\bar{D}$ threshold, 
its decay width is surprisingly narrow, the observed decay mode is 
$J/\psi \pi^+\pi^-$ and there was no obvious assignment to 
a known charmonium.

\begin{figure}[t]
\centering\includegraphics[width=5in]{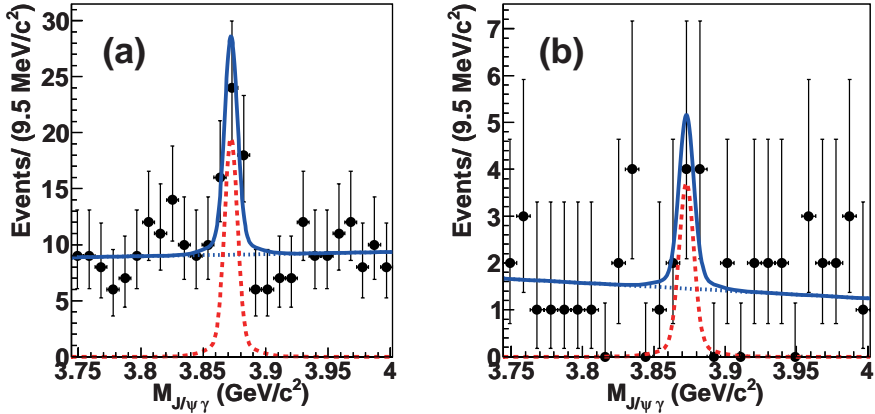}
\caption{The $J/\psi \gamma$ invariant mass distributions for
(left) $B^{\pm} \to J/\psi \gamma K^{\pm}$ and 
(right) $B^0 \to J/\psi \gamma K^0_S$ decays \cite{Bhardwaj:2011dj}. 
}
\label{fig_xtojpsigam}
\end{figure}

Immediately after this narrow state is reported, a lot of discussions 
have been exchanged attempting to give a proper interpretation.
What experimentalists should do to reveal $X(3872)$'s nature would be 
to determine its quantum number $J^{PC}$. 
The $X(3872) \to J/\psi \gamma$ is established by both 
Belle \cite{Bhardwaj:2011dj} and BaBar \cite{Aubert:2008ae}
measurements. 
The Belle result is shown in Fig. \ref{fig_xtojpsigam}, 
thus it is confirmed that the charge conjugation of $X(3872)$ is $C=+1$. 

It is also possible to determine the spin and parity by the angular 
distribution of decay products of $X(3872)$. 
The studies for $J/\psi \pi^+\pi^-$ mode by CDF \cite{Abulencia:2006ma}
and Belle \cite{Choi:2011fc} using three decay angular variables 
as well as the 3$\pi$ invariant mass spectrum in $J/\psi \pi^+\pi^- \pi^0$ 
mode by BaBar \cite{delAmoSanchez:2010jr} give 
constraint on $J^{PC}$ to be either $1^{++}$ and $2^{-+}$ but do not reach 
a definitive determination. 
A full five-dimensional amplitude analysis is performed by 
LHCb collaboration \cite{Aaij:2013zoa} for 
$X(3872) \to J/\psi \pi^+\pi^-$ mode and $J^{PC}=1^{++}$ has 
unambiguously assigned.

In a tetraquark hypothesis for $X(3872)$, a mass splitting due to
the mixing between $c\bar{c}u\bar{u}$ and $c\bar{c}d\bar{d}$ was 
theoretically predicted \cite{Maiani:2004vq}, 
$M(X_h) - M(X_l) = (7 \pm 2)/\cos(2\theta)$ MeV, 
where $M(X_{h,l})$ are the masses of the higher and lower states 
after the mixing of the two states.  
The difference is expected to 
appear as the difference in the $X$ masses separately measured in 
$B^{\pm} \to X K^{\pm}$
and $B^0 \to X K^0$. The Belle result for this difference in 
$J/\psi \pi^+\pi^-$ mode is found to be 
($-0.71 \pm 0.96 \mbox{(stat)} \pm 0.19 \mbox{(syst)}$) MeV/$c^2$ 
\cite{Choi:2011fc} thus it strongly disfavors the tetraquark interpretation. 
The BaBar measurement for this quantity is reported to be 
($2.7 \pm 1.6 \mbox{(stat)} \pm 0.4 \mbox{(syst)}$) MeV/$c^2$
\cite{Aubert:2008gu}.
Belle also finds no signature for the charged partner state in 
$J/\psi \pi^{\pm} \pi^0$ mode. 

\begin{figure}[t]
\centering\includegraphics[width=5in]{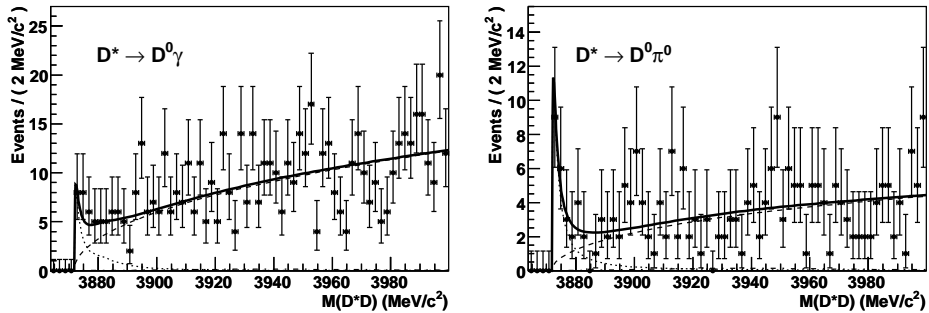}
\caption{The $D^0 \bar{D}{}^{*0}$ invariant mass distributions for
(left) $\bar{D}{}^{*0} \to \bar{D}{}^0 \gamma$ and 
(right) $\bar{D}{}^{*0} \to \bar{D}{}^0 \pi^0$ modes in 
$B \to D^0\bar{D}{}^{*0} K$ decays \cite{Adachi:2008sua}. 
}
\label{fig_xtoddstar}
\end{figure}

On the other hand, a $D^0 \bar{D}{}^{*0}$ molecule is proposed 
as very plausible interpretation for $X(3872)$ because of 
the very narrow width ($\Gamma < 1.2$ MeV) and the mass 
($M = 3871.69 \pm 0.17$ MeV/$c^2$) \cite{Agashe:2014kda}
being very close to the $D^0 \bar{D}{}^{*0}$ threshold.
Inspired by this argument, $B \to D^0\bar{D}{}^{*0} K$ decays are 
reconstructed to see $D^0 \bar{D}{}^{*0}$ invariant mass spectrum
and a clear excess has been observed
\cite{Aubert:2007rva,Adachi:2008sua} 
and Belle result is shown in Fig. \ref{fig_xtoddstar}.  
The observed mass peak is consistent with the one determined by 
the $J/\psi \pi^+\pi^-$ mode. 
The resultant ${\cal B}(X(3872) \to D^0 \overline{D}{}^{*0})$ is found 
to be 10 times as large as ${\cal B}(X(3872) \to J/\psi \pi^+ \pi^-)$. 

The difference between the observed mass and the threshold is regarded as 
the binding energy and it is small,
$0.11 \pm 0.21$ MeV \cite{Agashe:2014kda}.
This value would give an estimation of the $X(3872)$'s size, i.e. 
the distance between $D^0$ and $\bar{D}{}^{*0}$ mesons to be 10 fm
(see section~\ref{sec:theory_X3872} for details).
In this case, the entire volume of $X(3872)$ would become about 1000
times larger than that for the $J/\psi$ whose diameter is estimated 
to be 0.4 fm, thus the $D^0 \bar{D}{}^{*0}$ mode should have 
much larger branching fraction with respect to the $J/\psi \pi^+\pi^-$ 
mode.
Small binding energy raises another question in the molecular state
hypothesis, because the $X(3872)$ must be so fragile that it would be 
hard to explain the observed production rate in the high energy 
$p\bar{p}$ collisions at the Tevatron \cite{Bignamini:2009sk}.

Then it is plausible that $X(3872)$ would be an admixture of molecular state 
and the charmonium having the same quantum numbers (section \ref{sec:theory_X3872}).  
$\chi_{c1}(2P)$ is an unconfirmed charmonium with $J^{PC}=1^{++}$
and its original mass is expected to be about 3920 MeV/$c^2$.
In the admixture scenario, the $X(3872)$ is comprised by mainly 
the $D^0 \bar{D}{}^{*0}$ molecule and also contain sizable 
$\chi_{c1}(2P)$ component. 
The pure $\chi_{c1}(2P)$ state itself must be a compact object 
thus it is possible to explain the production in high energy collisions 
at either Tevatron or LHC. 
It also can explain the significant 
branching fractions to the final states containing a $J/\psi$.
The $\chi_{c1}(2P)$ has not yet been observed, the admixture hypothesis
also give an explanation; we can think the $X(3872)$ partner state which
is perpendicular to the $X(3872)$ by containing $\chi_{c1}(2P)$ as the 
main component and it would become several
hundreds MeV/$c^2$ heavier than the lower state. 
This larger mass allows a strong decay by emitting pion(s), 
consequently it may become so broad state that it is uneasy to recognize 
as a resonance in the currently available data statistics.
Hence, the admixture hypothesis has not been contradicting with the 
compilation of the experimental results.

The radiative decay of $X(3872) \to \psi(2S) \gamma$ would be suppressed 
in pure meson-meson molecule hypothesis because of small $Q$-value, 
while it can be significant if $X(3872)$ is the unconfirmed charmonium 
of $\chi_{c1}(2P)$. 
BaBar reported that 
${\cal B}(X(3872) \to \psi(2S) \gamma)$ is three times as large as 
$J/\psi \gamma$, 
$\frac{{\cal B}(X(3872) \to \psi(2S) \gamma)}
{{\cal B}(X(3872) \to J/\psi \gamma)}=3.4\pm1.4$
 \cite{Aubert:2008ae}.
But Belle found no evidence and gave an upper limit of
$\frac{{\cal B}(X(3872) \to \psi(2S) \gamma)}
{{\cal B}(X(3872) \to J/\psi \gamma)} < 2.1$ 
at 90\% C.L. \cite{Bhardwaj:2011dj}. 
Recently LHCb experiment also reported the radiative decays of 
$X(3872) \to J/\psi \gamma$ and $\psi(2S) \gamma$. 
Reconstruction of the $\psi$ which denotes a $J/\psi$ or $\psi(2S)$ meson is done by the $\mu^+ \mu^-$ mode only, because of the difficulty to achieve enough reconstruction efficiencies for other decay modes. 
Signal extraction is done by an extended maximum likelihood fit in the two-dimensional space of the invariant masses of $\psi \gamma K^{\pm}$ and $\psi \gamma$, where the former peaks at the $B^{\pm}$ mass and the latter forms a peak at the $X(3872)$ mass for the signal. 
It confirms the $X(3872)$ radiative decay in $J/\psi \gamma$ mode, at the same time, it finds an evidence of $X(3872) \to \psi(2S) \gamma$ with the significance of 4.4$\sigma$ \cite{Aaij:2014ala}. 
The ratio of the $X(3872) \to \psi(2S) \gamma$ to the $X(3872) \to J/\psi \gamma$ is obtained to be 
$\frac{{\cal B}(X(3872) \to \psi(2S) \gamma)}
{{\cal B}(X(3872) \to J/\psi \gamma)}=2.46\pm0.64\mbox{(stat)}\pm0.29\mbox{(syst)}$.
Neither BaBar nor Belle results contradicts with this LHCb result within the uncertainties and the resultant branching fraction disfavors an interpretation as a pure $D\bar{D}^*$ molecule and supports 
charmonium and molecular admixture state hypothesis.

All above functional hypotheses are the main progress in both theory and 
experiment in the recent  years to give an interpretation for the $X(3872)$ 
state, which will be discussed in some detail 
in Section~\ref{sec:theory_X3872}.

\subsection{$Y(4260)$ and some similar states in initial state radiation}
\label{sec:exp_y4260}

\begin{figure}[t]
\centering\includegraphics[width=3in]{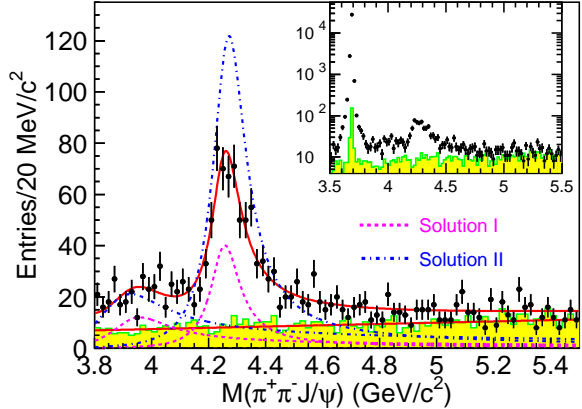}
\caption{Invariant mass distributions of the $J/\psi \pi^+\pi^-$ in ISR
events. Points with error mars are data and the shaded histograms are the 
normalized $J/\psi$ mass sidebands for background estimation.
The solid line shows the best fit with two coherent resonances and 
background while dashed and dot-dashed lines are solution I and II.
The distributions on a logarithmic vertical scale are shown in the inset
and the peak at 3686 MeV/$c^2$ is the $\psi(2S) \to J/\psi \pi^+\pi^-$
contribution \cite{Liu:2013dau}. 
}
\label{fig_y4260_final}
\end{figure}

In $e^+ e^-$ colliders, one of the beam particles may 
emit a photon and consequently $e^+ e^-$ annihilation takes place 
at lower effective center mass energy. 
This reaction is called the 
initial state radiation (ISR) or radiative return, and produces
$J^{PC}=1^{--}$ states via a virtual photon at any reachable invariant 
masses. 
In 2005, BaBar reported a resonance at 4260 MeV/$c^2$ in the
$J/\psi \pi^+ \pi^-$ final state \cite{Aubert:2005rm} and 
this state is named $Y(4260)$.
Belle confirmed the same resonance and another cluster of events 
at 4008 MeV/$c^2$ was seen \cite{Yuan:2007sj}.
Fig. \ref{fig_y4260_final} shows the $J/\psi \pi^+ \pi^-$ invariant mass 
distribution with the full data sample \cite{Liu:2013dau}. 
After these discoveries, the hadron which is created by ISR and 
contains $c\bar{c}$ constituent but is hardly interpreted  to be a 
charmonium is called "$Y$" as a custom.

In the analysis for $\psi(2S) \pi^+\pi^-$ channel, 
BaBar reported the relatively broad resonance having the mass 4324 MeV/$c^2$ 
and width 172 MeV/$c^2$ \cite{Aubert:2007zz}, while Belle found two peaks 
at 4361 MeV/$c^2$ and 4664 MeV/$c^2$ \cite{Wang:2007ea}. 
The last result brought by BaBar \cite{Lees:2012pv} confirmed those two peaks,
thus their first result is interpreted as the hypothesis that two resonances 
could not be separately recognized because of less statistics.
It is unusual that the $Y$ state decays to  $J/\psi \pi^+\pi^-$ or 
$\psi(2S) \pi^+\pi^-$ final states despite of the heavier mass than 
the $D\overline{D}$ threshold, contrary to the fact that $\psi(3770)$ decays 
predominantly to $D\overline{D}$ pair.
In addition, only one state is predicted among unconfirmed $J^{PC}=1^{--}$
charmonia above 4 GeV/$c^2$. 
Therefore these ``$Y$" states are thought to have different structure
from the conventional charmonium, as an orbital excitation of $c \bar c$.
Rather the large $\pi^+ \pi^-$ decay with its scalar and isoscalar correlations
would be possible from the flavor blind gluonic excitations, rendering 
its interpretation as a hybrid state~\cite{Close:2005iz,Kou:2005gt,Zhu:2005hp}.  
A short review of $Y(4260)$ with various interpretations are discussed in Section~\ref{sec:Y4260_theory}.  

The CLEO experiment also performed measurement of the $J/\psi \pi^+\pi^-$
production cross section and confirmed it exhibits a peak around 
$\sqrt{s}=4.26$ GeV/$c^2$ \cite{Coan:2006rv}, while BES measurement 
shows that there is a dip in the total hadronic cross section 
as a function of $\sqrt{s}$ in the same center-mass-energy region
\cite{Bai:2001ct}.
Thus, the fraction of $J/\psi \pi^+\pi^-$ production in the total 
hadronic cross section is estimated to be very high and the resultant 
partial width $\Gamma(Y(4260) \to J/\psi \pi^+\pi^-)$ should be more 
than 1.0 MeV (at 90\% C.L.) which is too large to interpret it as an 
ordinary charmonium. 
This $Y(4260)$ and related states are thought to be similar to the 
bottomonium-like exotics as described in section~\ref{sec:yb}. 
Such point of view lead us to search for the charged charmonium-like 
exotics as an intermediate state of $Y(4260)$ decays as discussed 
later.

\subsection{$Z_c(4430)^+$ and some similar states} 
\label{sec:exp_z4430}
\begin{figure}[t]
\centering\includegraphics[width=3.5in]{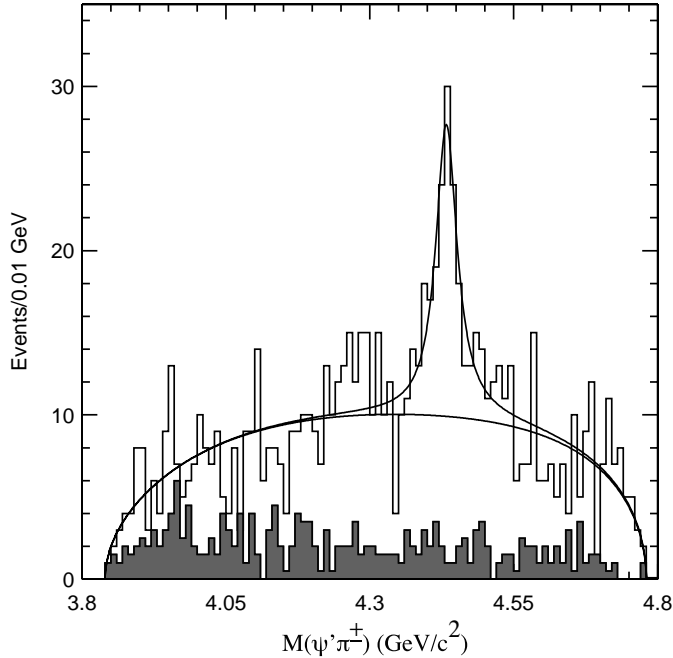}
\caption{The invariant mass spectrum for $\psi(2S) \pi^+$ in 
the $B \to \psi(2S) \pi^+ K$ decays. Significant peak is observed 
at 4430 MeV/$c^2$ and it is called $Z_c(4430)^+$. 
Superimposed lines show the Breit-Wigner lineshape for $Z_c(4430)^+$
signal and the phase space-like non-resonant $\psi(2S) \pi^+$
component. Shaded histogram shows the estimated background distribution 
by the $B$ decay signal sideband events \cite{Choi:2007wga}.
}
\label{fig_z4430}
\end{figure}

In either molecular state or tetraquark hypothesis, the number of 
constituent quarks is four. If a quarkonium-like state has an electric
charge, it becomes a firm signature to have four constituent quarks. 
The most straightforward way to search for the charged state is looking
for a peak in the invariant mass spectrum for a charmonium and a charged
pion system. In 2007, Belle reported a peak at 4430 MeV/$c^2$ in the 
$\psi(2S) \pi^+$ invariant mass spectrum for the 
$B \to \psi(2S) \pi^+ K$ three-body decays \cite{Choi:2007wga}
as shown in Fig. \ref{fig_z4430}.

BaBar analyzed the same $B$ decay process and found the effect
corresponding only to 1.9$\sigma$ statistical significance, 
thus an upper limit was presented \cite{Aubert:2008aa}, 
stating that two $B$-factories' results are not statistically 
contradicting each other.
In $B \to \psi(2S) \pi^+ K$ decays, major contributions are coming from
ordinary quasi-two-body $B$ decays, $B \to \psi(2S) K^*$ where 
$K^* \to K \pi^+$. In Belle first discovery paper \cite{Choi:2007wga},
$B \to \psi(2S) K^*(892)$ and $B \to \psi(2S) K^*(1430)$ dominant regions 
were removed according to the $K \pi^+$ invariant mass.
However, it is important to see if the potential interference between 
those $B \to \psi(2S) K^*$ decays and other featureless process such as 
non-resonant three-body $B$ decays to $\psi(2S) \pi^+ K$ may exhibit 
a peak-like projection in $\psi(2S) \pi^+$.
Hence, Belle reanalyzed the same data with a Dalitz analysis approach
which is a fit featuring all possible three-body $B$ decay amplitudes 
written as a function on the Dalitz plane \cite{Mizuk:2009da}. 
The result obtained by this advanced analysis technique turned out 
to be statistical significance of 6.4$\sigma$.
In addition, the most recent Belle study using a four dimensional full 
amplitude analysis of $\bar{B}^0 \to \psi(2S) \pi^+ K^-$ decays constrain 
the spin and parity of the $Z_c(4430)^+$. Here, the $\psi(2S)$ is reconstructed via its $e^+e^-$ or $\mu^+\mu^-$ decay channels and the amplitude of this $B$ decay is expressed by the helicity formalism in a four-dimensional parameter space of 
$M^2_{K\pi}, M^2_{\psi(2S)\pi}, \theta_{\psi(2S)}$ and $\phi$, where 
$M^2_{K\pi}$ ($M^2_{\psi(2S)\pi}$) is the squared mass of $K\pi$ 
($\psi(2S) \pi$) system, $\theta_{\psi(2S)}$ is the $\psi(2S)$ helicity angle defined as the angle between the daughter lepton's momentum in 
the $\psi(2S)$ rest frame and $\psi(2S)$ momentum in the mother $B$ rest frame and $\phi$ is the angle between the $\psi(2S)$ decay plane and the plane containing $K$ and $\pi$ momentum vectors. 
Each of different intermediate two-body states is represented by the 
Breit-Wigner and the sum of those contributions forms the total decay 
amplitude. 
In total, 6 possible $K^*$ resonances, non-resonant $K\pi$ component and 
a charged $Z_c$ state are taken into account.
The fit result favors $J^P = 1^+$ over the $0^-$, $1^-$ and $2^+$ hypothesis at 3.4$\sigma$ or more \cite{Chilikin:2013tch}.
At last, in 2014, LHCb reported confirmation of the resonant structure of 
$Z_c(4430)^+$ \cite{Aaij:2014jqa} by performing the same amplitude analysis 
as Belle. As a result, they rule out the $J^P=0^-$, $1^-$,
$2^+$ and $2^-$ hypotheses by more than 9$\sigma$ relative to $1^+$, thus 
now the quantum number is unambiguously determined.
Exploiting the large $B$ decay signal statistics brought by $pp$ collision environment, fit to obtain the $Z_c(4430)^+$ amplitude in six $M^2_{\psi(2S) \pi}$ bins has been performed and plotted in an Argand diagram. 
The phase is found to increase as a function of $M^2_{\psi(2S) \pi}$, 
it is consistent with the expectation from a Breit-Wigner formula, 
thus the $Z_c(4430)^+$ is confirmed to be a real resonance.
Its mass and width are determined to be $M = 4478^{+15}_{-18}$ MeV/$c^2$ 
and $\Gamma = 181\pm31$ MeV, respectively in the latest world average \cite{Agashe:2014kda}.

The discovery of $Z_c(4430)^+$ is striking to lead us to know 
that exotic hadrons can really be formed. 
However, it is not yet easy to reach convincing interpretation for the 
$Z_c(4430)^+$ for the following reasons; (1) its mass is so away from 
any charmed meson pair threshold that a molecular state hypothesis can
hardly recognized as a plausible scenario, (2) neither other decay mode
or a partner state have been reported. 
Further searches for other 
decay modes and partner states from experimental side 
are still necessary to settle the argument 
about this charged state's structure.
Various theoretical proposals are briefly discussed in 
Section~\ref{sec:z4430_theory}.

In $\bar{B}^0 \to \chi_{c1} \pi^+ K^-$ decay, Belle reported two resonance-like structures in the $\chi_{c1} \pi^+$ mass spectrum \cite{Mizuk:2008me}.
From a Dalitz plot analysis fit gives the two resonance-like objects' masses 
and widths of: 
$M_1 = 4051$ MeV/$c^2$, 
$\Gamma_1 = 82$ MeV, 
$M_2 = 4248$ MeV/$c^2$, 
$\Gamma_2 = 177$ MeV.
While, BaBar reported that the obtained distribution in the same $B$ decay mode
can be explained without these charged charmonium-like objects
\cite{Lees:2011ik}. 
Thus attempts to make a confirmation with higher statistics is necessary for 
these states.
Belle also visited the $\bar{B}^0 \to J/\psi \pi^+ K^-$ decay. 
In order to resolve possible intermediate states, an amplitude analysis which is the same approach as $Z_c(4300)^+ \to \psi(2S) \pi^+$ is 
performed with taking 10 possible $K^*$ resonances into account. 
The $Z_c(4430)^+ \to J/\psi \pi^+$ is found to be evident and in addition,
one more charged charmonium-like state of $Z_c(4200)^+ \to J/\psi \pi^+$
is observed with a significance of 6.2$\sigma$. 
Its mass and width are obtained to be $M = 4196^{+31}_{-29}{}^{+17}_{-13}$ MeV/$c^2$ and $\Gamma = 370 \pm 70 ^{+70}_{-132}$ MeV where the first and second errors are statistical and systematic, respectively, and $J^P = 1^+$ is the most favored quantum number. 

These results tell us that multi-body $B$ decays are functional as the 
source of the exotic charmonium-like states and extraction of information 
often requires Dalitz or more sophisticated amplitude analysis technique.

\subsection{$Z_c(3900)^+$ and some similar states}
\label{sec:z3900}
The large rate for $\Upsilon(nS) \pi^+\pi^-$ ($n=1,2,3)$ production at 
$\Upsilon(5S)$ led to the discovery of $Z_b(10610)^+$ and 
$Z_b(10650)^+$ in 
$\Upsilon(nS) \pi^+$ and $h_b(mP) \pi^+$ ($m=1,2$) 
final states as to be discussed in subsection \ref{sec:Zb}.
Since $Y(4260)$ has an anomalous high rate $J/\psi \pi^+\pi^-$ production
as described in subsection \ref{sec:exp_y4260}, it can be interpreted as an analog 
in charmonium(-like) sector. In this point of view, 
the charged bottomonium-like particles' appearance at $\Upsilon(5S)$ 
inspired us to visit the $J/\psi \pi^+$ intermediate state in 
$Y(4260) \to J/\psi \pi^+\pi^-$.

\begin{figure}[t]
\centering\includegraphics[width=3.5in]{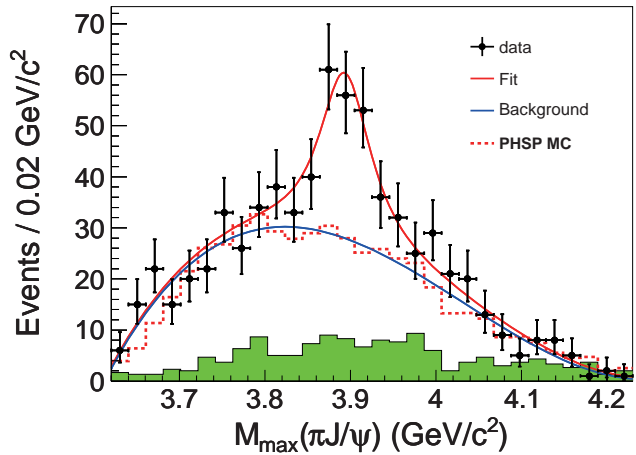}
\caption{The invariant mass spectrum for the larger one among two possible
combination of the $J/\psi$ and one charged pion ($M_{\rm max}(\pi J/\psi)$)
in the $Y(4260) \to J/\psi \pi^+ \pi^-$ candidates. 
Significant peak is seen around 3900 MeV/$c^2$ and it is called 
$Z_c(3900)^+$. 
Superimposed lines show the Breit-Wigner lineshape for $Z_c(3900)^+$
signal and the phase-space like background.
Shaded histogram shows the estimated non-$J/\psi$ background by the 
normalized sideband events \cite{Liu:2013dau}. 
}
\label{fig_z3900}
\end{figure}

In a $Y(4260) \to J/\psi \pi^+ \pi^-$ event, there are two combinations
of the $J/\psi$ and one charged pion. Among them, the one having 
the larger invariant mass, $M_{\rm max}(\pi J/\psi)$ is taken.
The $M_{\rm max}(\pi J/\psi)$ distribution is shown in 
Fig. \ref{fig_z3900}. We see a significant enhancement at 3900 MeV/$c^2$
with 5.2$\sigma$ statistical significance.
Its mass and width are obtained to be $M= 3895.5\pm6.6\pm4.5$ MeV/$c^2$
and $\Gamma = 63\pm24\pm26$ MeV, where first and second errors are 
statistical and systematic, respectively.
This structure can be interpreted as a new charged charmonium-like state.

BES III experiment at Beijing Electron Positron Collider performed 
the dedicated run 
with setting the center mass energy at 4260 MeV to collect 525 pb$^{-1}$. From this data sample, they also 
find the structure in $M_{\rm max}(\pi J/\psi)$ around 3.9 GeV/$c^2$
and get mass $M=3899.0\pm3.6\pm4.9$ MeV/$c^2$
and width $\Gamma = 46\pm10\pm20$ MeV \cite{Ablikim:2013mio}.
Hence this charged charmonium-like state $Z_c(3900)^+$ is confirmed.

Because of the similarity between $Y(4260) \to J/\psi \pi^+\pi^-$ and 
$\Upsilon(5S) \to \Upsilon(nS) \pi^+\pi^-$ as well as discovery of 
$Z_b(10610)^+$ and $Z_b(10650)^+$ states, one possible hypothesis is 
regarding $Z_c(3900)^+$ as a $(D \bar{D}{}^*)^+$ molecular state.
Using BES III data collected at $\sqrt{s} = 4.26$ GeV, 
$e^+e^- \to (D \bar{D}{}^*)^{\pm} \pi^{\mp}$ process is visited by
a partial reconstruction technique with requiring a charged pion and a $D$ meson to tag the candidate events. The $D^*$ signal is identified as the 
corresponding peak in the $\pi D$ recoil mass spectrum. 
In both $D^0 D^{*-}$ and $D^+ \bar{D}{}^{*0}$ cases, clear enhancement 
in $D \bar{D}{}^*$ mass spectrum just above the threshold is significant.
The extracted resonance parameters are 
the pole mass $M_{\rm pole} = 3883.9 \pm 1.5 \pm 4.2$ GeV/$c^2$ and 
width $\Gamma_{\rm pole} = 24.8 \pm 3.3 \pm 11.0$ MeV.
Here, $Z_c(3885)^+$ denotes this structure in the $(D \bar{D}{}^*)^+$ final state. Assuming that the $Z_c(3885)^+$ is same as the $Z_c(3900)^+$ in 
$J/\psi \pi^+$ mode, the ratio of partial decay widths is obtained to be
$\Gamma(Z_c(3885)^+ \to (D\bar{D}{}^*)^+)/\Gamma( Z_c(3900)^+ \to J/\psi \pi^+ ) = 6.2 \pm 1.1 \pm 2.7$, it is pretty much smaller than that for conventional charmonium states thus it implies very different mechanism in 
the $Y(4260)$-$Z_c^+$ system \cite{Ablikim:2013xfr}.

More searches for analogous reactions have been performed. Exploiting a partial reconstruction technique, $e^+e^- \to (D^* \bar{D}{}^*)^{\pm} \pi^{\mp}$ process is reconstructed and an enhancement just above 
$(D^*\bar{D}{}^*)^{\pm}$ threshold has been observed.
This enhancement is called $Z_c(4025)^+$ and the fit assuming that it is 
a resonance described by an $S$-wave Breit-Wigner lineshape gives the mass and width to be $M = 4026.3 \pm 2.6 \pm 3.7$ MeV/$c^2$ and 
$\Gamma = 24.8 \pm 5.6 \pm 7.7$ MeV, respectively. This can be regarded as a candidate for the charm sector counterpart of the $Z_b(10650)^+$ state \cite{Ablikim:2013emm}. 
$e^+e^- \to \pi^+\pi^- h_c$ process is also studied at 13 center-mass-energies from 3.900 GeV to 4.420 GeV by a full reconstruction approach by the radiative decay mode of $h_c \to \gamma \eta_c$ followed by 
the $\eta_c$ decays in 16 exclusive hadronic final states. 
Spectrum of the invariant mass of $\pi^+ h_c$ system, $M_{\pi^+ h_c}$, shows a distinct structure and it is referred to $Z_c(4020)^+$ with a mass 
of $4022.9\pm0.8\pm2.7$ MeV/$c^2$ and width of $7.9\pm2.7\pm2.6$ MeV \cite{Ablikim:2013wzq}.

These results suggests a similar special mechanism in the $J^{PC}=1^{--}$ 
state production followed by the decay to a quarkonium and a charged pion
pair to produce a charged object as an intermediate state in both 
charmonium and bottomonium cases.
However, as described in Section \ref{sec:zb}, all the 
$B\overline{B}{}^*$, $B^* \overline{B}{}^*$, $\Upsilon(nS) \pi^+$ 
($n=1,2,3$) and $h_b(mP)\pi^+$ ($m=1,2$) modes are commonly possessed 
by the two charged states of 
$Z_b(10610)^+$ and $Z_b(10650)^+$ at $\Upsilon(5S)$, while 
several different charged charmonium-like structures are found 
depending on the final state at $Y(4260)$ region. 
Proper theoretical approaches to explain such a difference between 
charmonium-like and bottomonium-like cases are still awaited. 
Current theoretical studies for $Z_c(3900)^+$ and related discussions are briefly given 
in Section~\ref{sec:z3900_theory}.

\subsection{$Y_b$ state} 
\label{sec:yb}

As described in section~\ref{sec:exp_y4260},
the discovery of $Y(4260)$ in initial state radiation 
process~\cite{Aubert:2005rm}
and following similar states~\cite{Yuan:2007sj,Aubert:2007zz,Wang:2007ea}
are thought to be candidates of exotic states.
The corresponding exotic hadrons where a charm and an anti-charm quark
are replaced by a bottom and an anti-bottom quark  is expected to
have masses around $\Upsilon(5S)$ region~\cite{Hou:2006it}~\cite{Abe:2007tk}.
The corresponding particle (denoted as $Y_b$ hereafter) has 
$J^{PC} = 1^{--}$ 
and large decay rate into $\Upsilon(nS) \pi^+\pi^-$ ($n = 1, 2, 3$)
final state is expected.  Therefore, Belle studied the decays 
$\Upsilon(5S) \to \Upsilon(nS) \pi^+\pi^-$ using 21.7 fb$^{-1}$ data
accumulated at $\Upsilon(5S)$ peak.

Assuming that $\Upsilon(5S)$ is an ordinary bottomonium above open bottom 
threshold, only a few events were expected with this amount of data.
However, 
Belle observed clear and anomalously large signal in
$\Upsilon(1S) \pi^+\pi^-$ and $\Upsilon(2S) \pi^+\pi^-$ channels.
The measured partial widths are about 100 times larger than those of 
$\Upsilon(n'S)$ ($n' = 2, 3, 4$) resonances.
The result was presented at Hadron 2007 conference~\cite{Hou:2007hd}.

\begin{figure}
\begin{center}
\resizebox{0.7\textwidth}{!}{\rotatebox{0}{\includegraphics{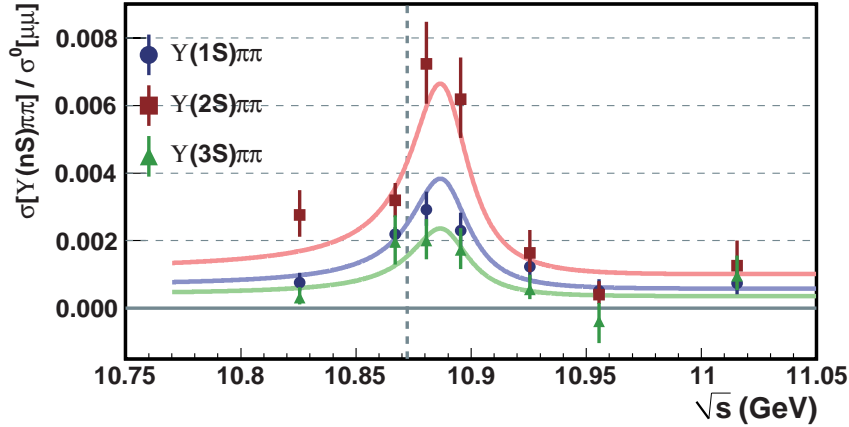}}}
\caption{The energy-dependent cross sections for 
$e^+e^- \to \Upsilon(nS)\pi^+\pi^-$ ($n=1, 2, 3$) processes normalized to the 
leading-order $e^+e^- \to \mu^+\mu^-$ cross sections. The results of the 
fits are shown as smooth curves. The vertical dashed line indicates the 
energy at which the hadronic cross section is maximal~\cite{Chen:2008xia}. }
\label{fig:yb_ypipi}
\end{center}
\end{figure}

\begin{figure}
\begin{center}
\resizebox{0.7\textwidth}{!}{\rotatebox{0}{\includegraphics{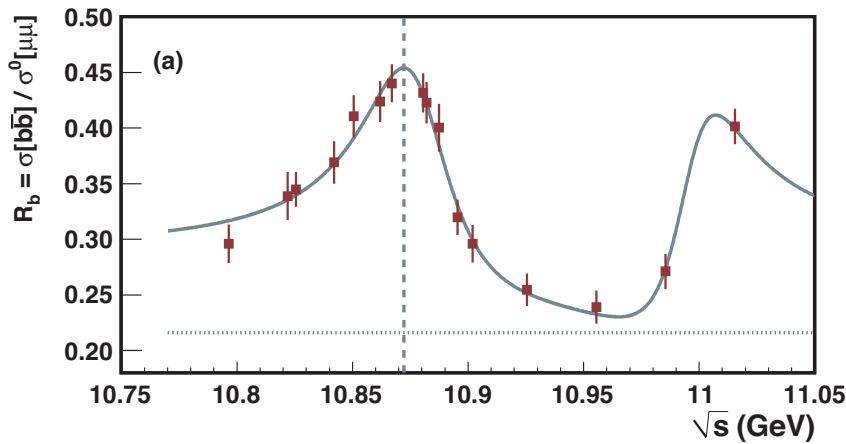}}}
\caption{ The energy-dependent cross sections for  $e^+e^- \to b\bar b$ 
processes normalized to the leading-order $e^+e^- \to \mu^+\mu^-$ 
cross sections. The horizontal dotted line in is the non-interfering
non-resonance contribution~\cite{Chen:2008xia}. } 
\label{fig:yb_rb}
\end{center}
\end{figure}

Although the inclusive energy dependence does not show a peak like structure, 
it showed a characteristic structure when exclusive process going to $J/\psi \pi \pi$
was measured at around $Y(4260)$.  
This lead to the idea that there could be a $Y_b$ state different from $\Upsilon(5S)$
with a different mass around 10.9 GeV/$c^2$.
In order to  check this hypothesis, an energy scan was proposed
and performed in December 2007.
The data were taken at 6 energy points from $\sqrt{s}$ = 10.83 GeV
to 11.02 GeV with a total integrated luminosity of 8.1 fb$^{-1}$.
In addition, nine scan points with an integrated luminosity of 
about 30 pb$^{-1}$ each, in order to measure hadronic line
shape with more data points.
The energy dependence of cross section $e^+e^- \to \Upsilon(nS)\pi^+\pi^-$
($n = 1, 2, 3$) are shown in Fig.~\ref{fig:yb_ypipi}~\cite{Chen:2008xia}.  
Peak structure are observed
for all three final states.
The line shapes of cross sections are fitted by a common Breit-Wigner 
function, and mass and width values are compared with those obtained
from the fit to hadronic cross section (Fig.~\ref{fig:yb_rb}).  
The differences of the  Breit-Wigner parameters determined from hadronic decay and those from  $\Upsilon(nS)\pi^+\pi^-$ decays are measured
to be $9 \pm 4$ MeV/$c^2$ in mass and $-15 ^{+11}_{-12}$ MeV/$c^2$ in
width, which correspond to 2.0$\sigma$ including systematic error.


In 2010 from May to June, which is a last run period of Belle experiment,
new energy scan was performed.  Data are taken at additional 16 energy 
points which were taken with about 1 fb$^{-1}$ each from 10.63 to 11.02 GeV.
Also, fine energy scan is made with 5 MeV interval with 50 pb$^{-1}$
at each of 61 points from 10.75 to 11.05 GeV.
The new data together with the previously taken data were analyzed.  
The results on the differences in the two Breit-Wigner
parameters are $9.2 \pm 3.9$ MeV/c$^2$ in mass and $5.2 \pm 8.1$ MeV/c$^2$ in
width, which corresponds to about 2$\sigma$ level~\cite{Santel:2015qga}.
Here, note that we could use only the relatively simple model comprised of 
sum of two Breit-Wigner functions and flat non-resonant component. 
Potentially its imperfect ability to describe hadronic cross sections 
may limit reliability of the obtained parameters.
We can not totally exclude such an issue as a reason why 
the resonance particle decaying to hadrons and that decaying
to $\Upsilon(nS)\pi^+\pi^-$ are not proved to be different particles.
However, still a mystery or exotic feature on anomalously large rates to  
$\Upsilon(nS)\pi^+\pi^-$ vitally remains.

\subsection{$Z_b(10610)^+$ and $Z_b(10650)^+$} 
\label{sec:zb}

\subsubsection{Observation of $h_b(1P)$ and $h_b(2P)$ }
\ \ \

\begin{figure}[t]
\begin{center}
\resizebox{0.5\textwidth}{!}{\rotatebox{0}{\includegraphics{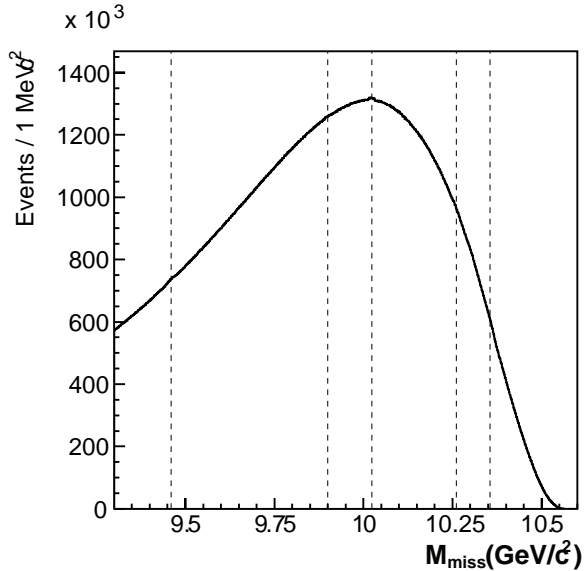}}}
\caption{ The missing mass distribution for the selected $\pi^+\pi^-$ pairs 
(solid histogram) 
Vertical lines indicate the locations of the $\Upsilon(1S)$, $h_b(1P)$, 
$\Upsilon(2S)$, $h_b(2P)$, and $\Upsilon(3S)$ signals~\cite{Adachi:2011ji}. }
\label{fig:hb_all}
\end{center}
\end{figure}

\begin{figure}[t]
\begin{center}
\resizebox{0.8\textwidth}{!}{\rotatebox{0}{\includegraphics{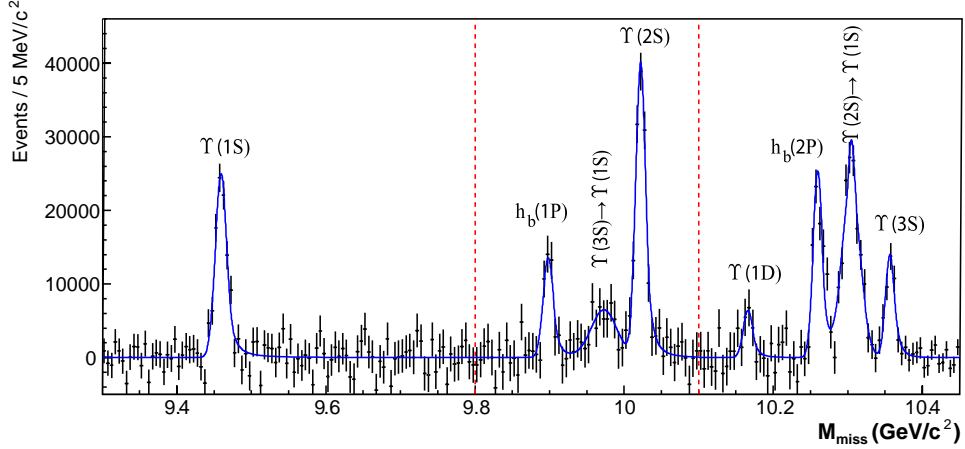}}}
\caption{ The missing mass spectrum with the background 
subtracted (points with errors) and signal component of the fit function 
overlaid (smooth curve). The vertical lines indicate boundaries of the fit
regions~\cite{Adachi:2011ji}. }
\label{fig:hb_fit}
\end{center}
\end{figure}

In addition to anomalously large rates to $\Upsilon(nS)\pi^+\pi^-$
($n = 1, 2, 3$)
at $\Upsilon(5S)$, another exotic feature was found, again motivated
from the analogy with $Y(4260)$.  
CLEO observed that the rates of $e^+e^- \to h_c \pi^+\pi^-$ also 
increase around $Y(4260)$ in a similar way as 
$e^+e^- \to J/\psi \pi^+\pi^-$ with comparable 
magnitudes~\cite{CLEO:2011aa}.
As mentioned in section \ref{sec:z3900}, BES III found a charged charmonium-like state at 4200 MeV/$c^2$ in the $\pi^+ h_c$ system of the 
$e^+e^- \to h_c \pi^+\pi^-$ process \cite{Ablikim:2013wzq}.
Indication of existence of $Y_b$ state also leads to an idea of large rates 
of $e^+e^- \to h_b(mP) \pi^+\pi^-$ ($m = 1,2$) at $\Upsilon(5S)$.
The $h_b(mP)$ had not been observed yet and was hoped to be
observed in this process.
Belle searched for $h_b(mP)$ states in missing mass distribution
against $\pi^+\pi^-$ using the 121.4 fb$^{-1}$ data sample accumulated
at $\Upsilon(5S)$ which includes about 100 fb$^{-1}$ taken in the period
from October 2008 to the end of 2009 in addition to the one used for 
discovery of anomalously large transition rates to 
$\Upsilon(1S) \pi^+\pi^-$ and $\Upsilon(2S) \pi^+\pi^-$ final states.
In order to suppress $e^+e^- \to q\bar q$ ($q = u,d,s,c$) continuum
background, ratio of the second to zeroth Fox-Wolfrum moments, $R_2$,
is required to be $< 0.3$.
The missing mass distribution shown in Fig.~\ref{fig:hb_all} is fitted 
with polynomial background function plus bottomonium resonance states
with their masses being free parameters. 
The widths of the resonance peaks are calibrated exclusively reconstructed 
$e^+e^- \to \pi^+\pi^- \Upsilon(nS) [\to \mu^+\mu^-]$ samples, where
the $\Upsilon(nS)$ mass is calculated as the missing mass against
$\pi^+\pi^-$.  The result is shown in Fig.~\ref{fig:hb_fit}. In addition to
clear peaks of $\Upsilon(1,2,3S)$ at the expected masses, clear peaks
of $h\b(1P)$ and $h_b(2P)$ were discovered~\cite{Adachi:2011ji}. 
As seen in Fig.~\ref{fig:hb_fit}, the rates of $h_b(mP) \pi^+\pi^-$ are similar
order as those of $\Upsilon(nS) \pi^+\pi^-$ in spite of the expectation
that  $h_b(mP) \pi^+\pi^-$ is suppressed because spin flip of bottom
quark is required.


\subsubsection{ Observation of $Z_b(10610)^+$ and $Z_b(10650)^+$ }
\label{sec:Zb}
\ \ \

\begin{figure}[t]
\begin{center}
\resizebox{0.3\textwidth}{!}{\rotatebox{0}{\includegraphics{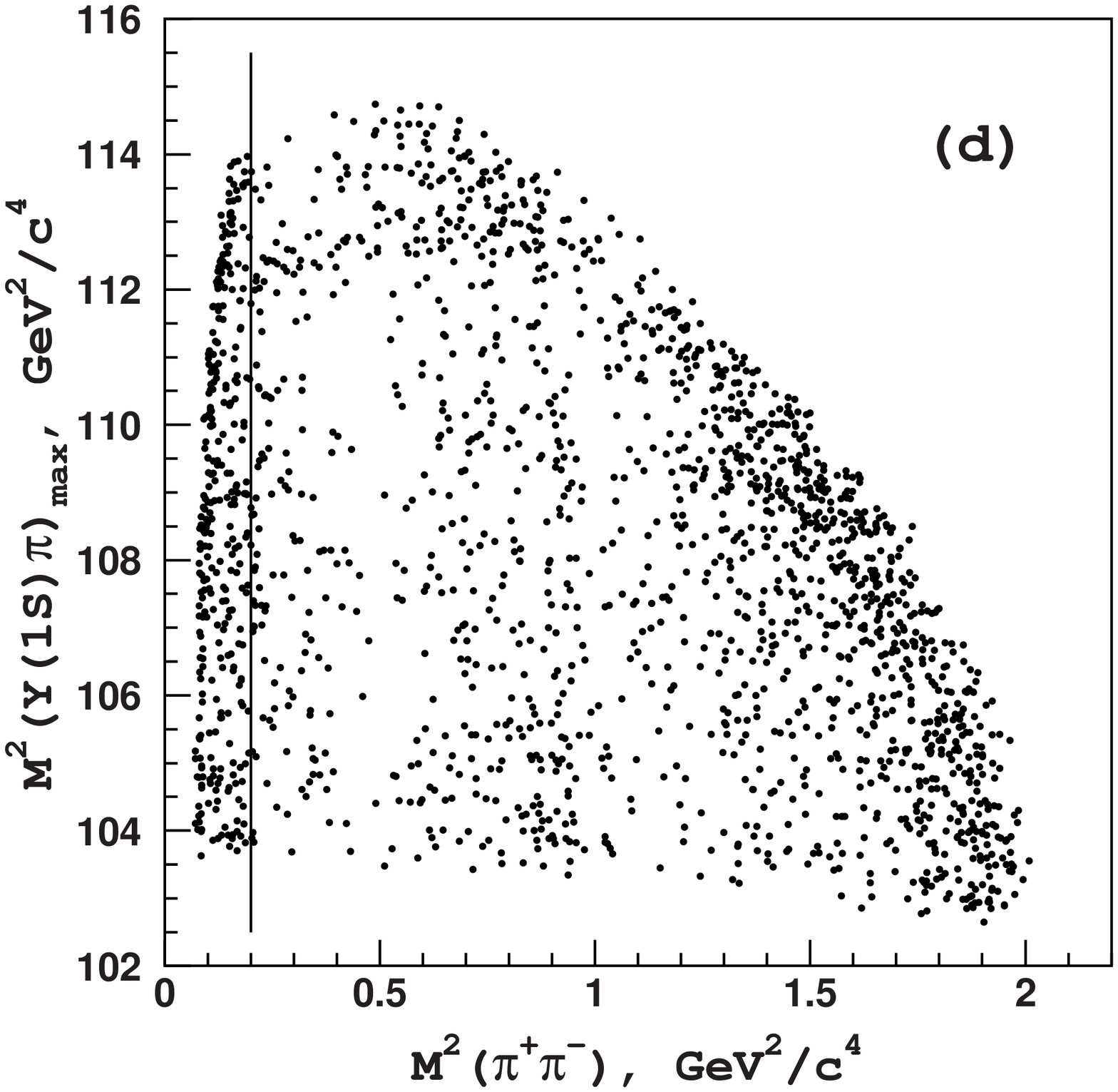}}}
\resizebox{0.3\textwidth}{!}{\rotatebox{0}{\includegraphics{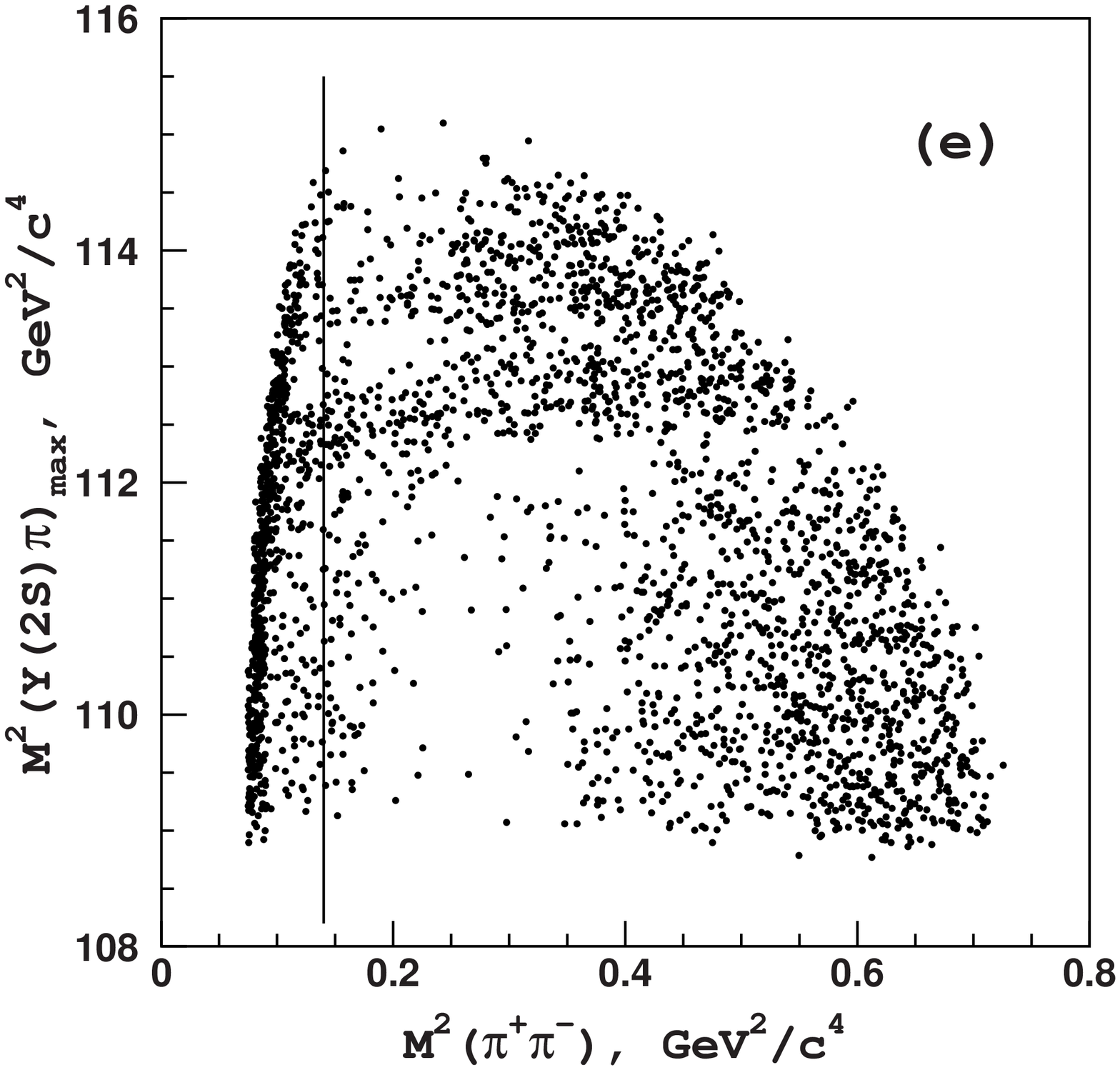}}}
\resizebox{0.3\textwidth}{!}{\rotatebox{0}{\includegraphics{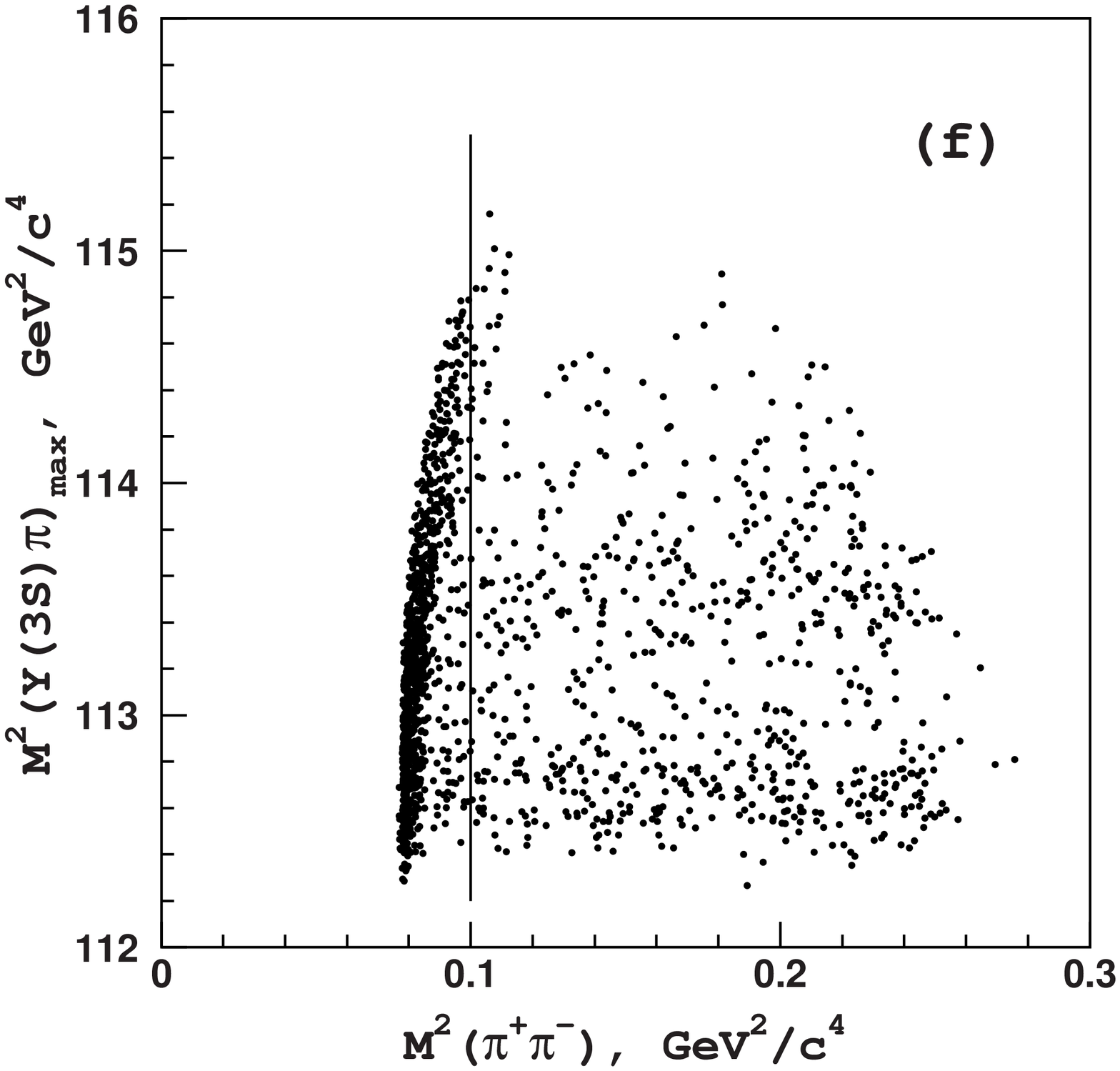}}}
\caption{ Dalitz plots for $\Upsilon(nS)\pi^+\pi^-$ events of the
(left) $\Upsilon(1S)$; (middle) $\Upsilon(2S)$; (right) $\Upsilon(3S)$~\cite{Belle:2011aa}. }
\label{fig:zb_dalitz}
\end{center}
\end{figure}

\begin{figure}[t]
\begin{center}
\resizebox{0.32\textwidth}{!}{\rotatebox{0}{\includegraphics{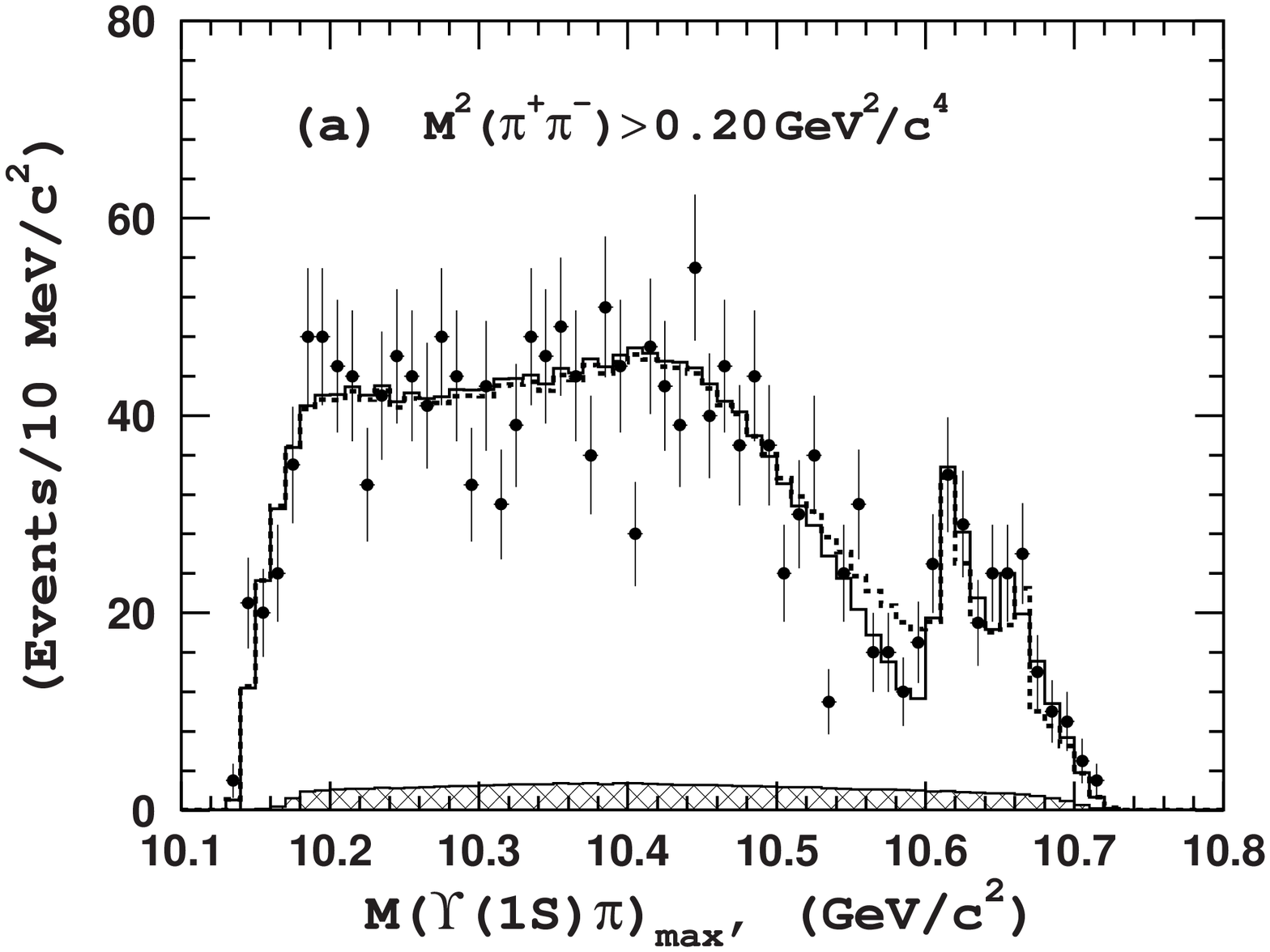}}}
\resizebox{0.32\textwidth}{!}{\rotatebox{0}{\includegraphics{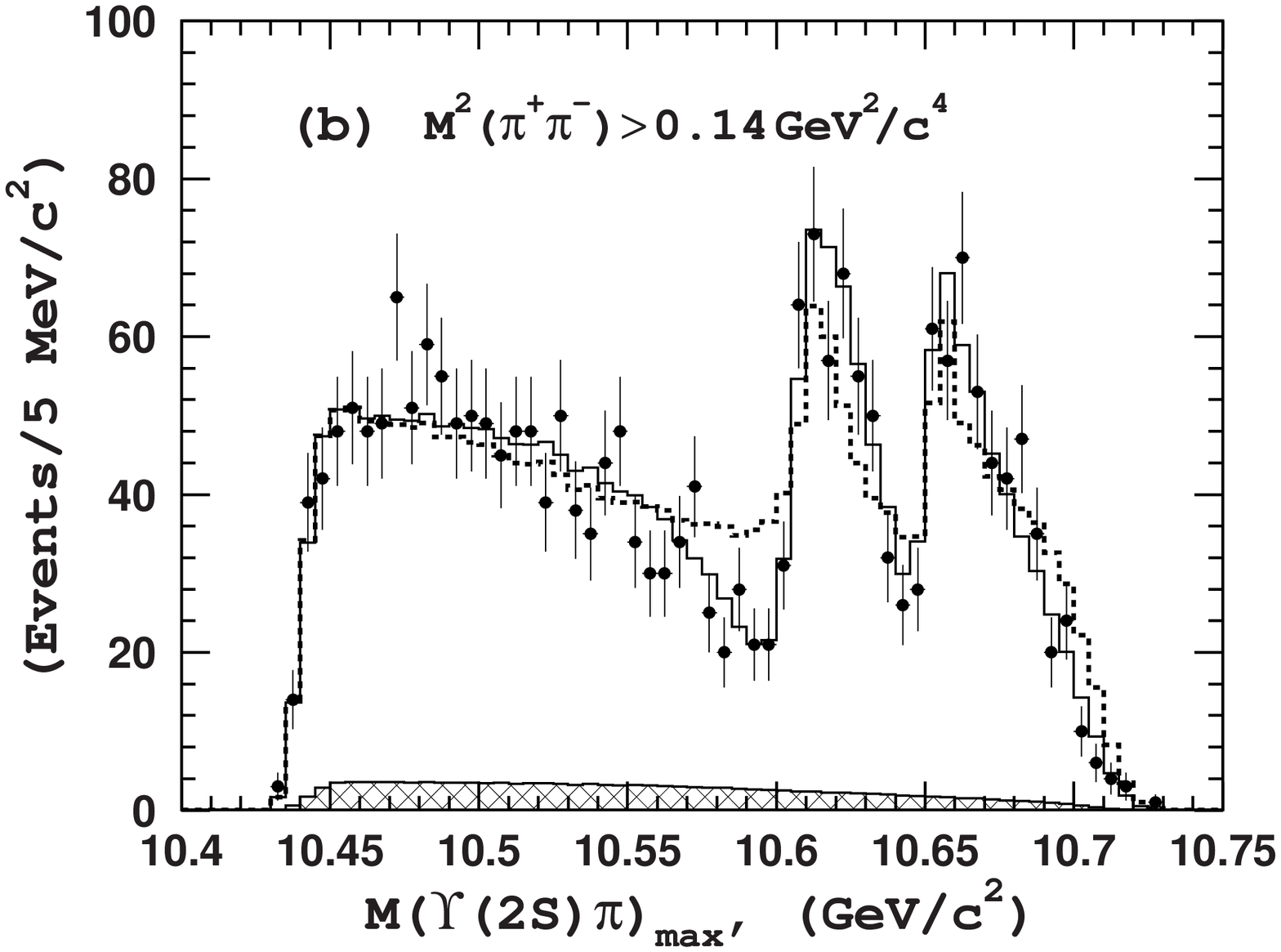}}}
\resizebox{0.32\textwidth}{!}{\rotatebox{0}{\includegraphics{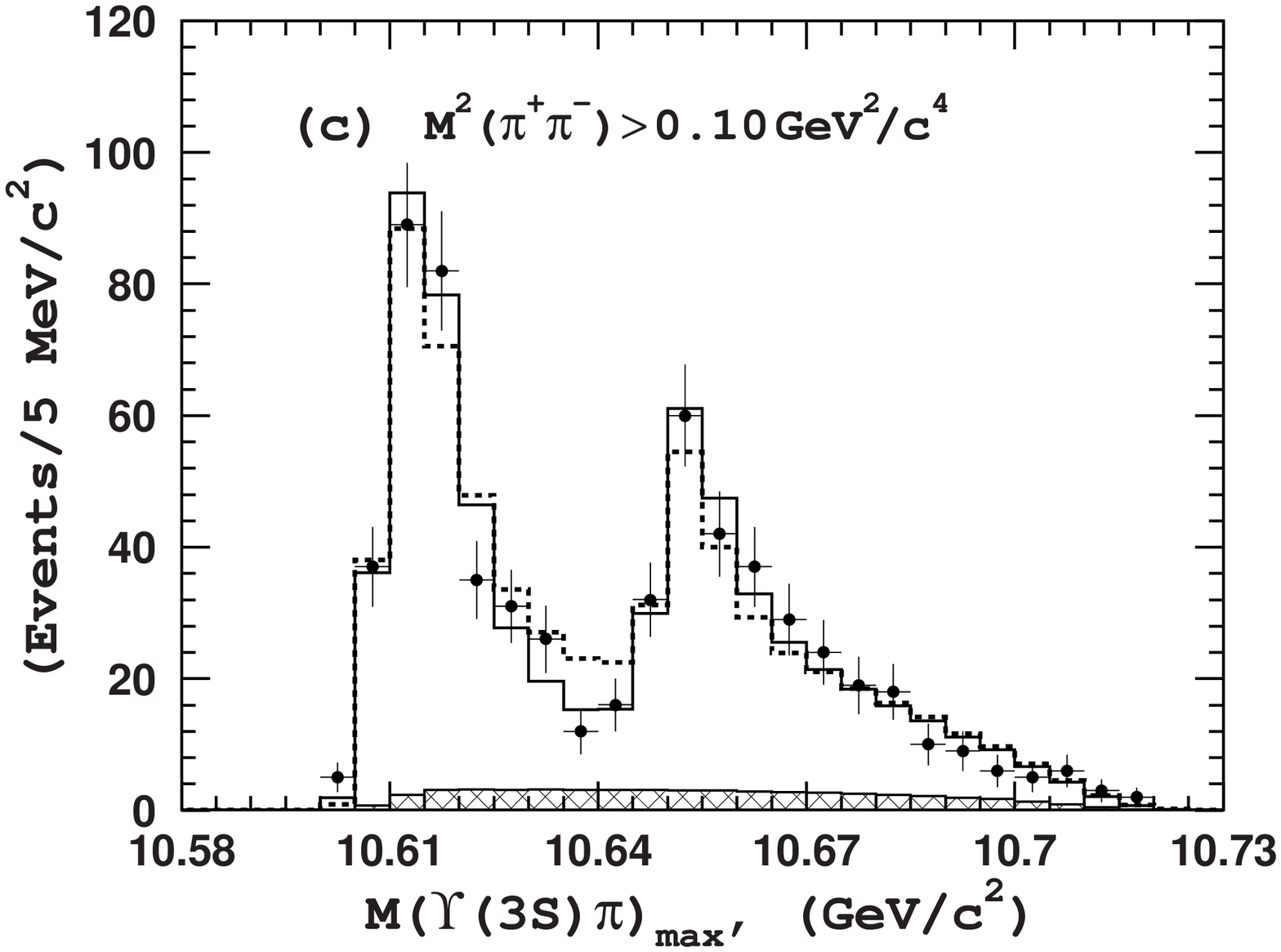}}}
\caption{ $\Upsilon(nS)\pi$ invariant mass distributions of 
$\Upsilon(nS)\pi^+\pi^-$ events for the
(left) $\Upsilon(1S)$; (middle) $\Upsilon(2S)$; (right) $\Upsilon(3S)$.
Points with error bars are data and hatched histograms show background.
Solid histograms are fits with $J^P = 1^+$ assigned to both $Z_b$ states
and dashed histograms with $J^P = 2^+$ (see text)~\cite{Belle:2011aa}. }
\label{fig:zb_mypi}
\end{center}
\end{figure}

In order to understand abnormal features in $e^+e^- \to \Upsilon(nS) \pi^+\pi^-$
and $e^+e^- \to h_b(mP) \pi^+\pi^-$ processes at $\Upsilon(5S)$, more 
detailed studies of final particles were performed. 
The invariant mass distributions intermediate states of three-body 
$\Upsilon(nS) \pi^+\pi^-$
system were checked when these process were first observed with 
21.7 fb$^{-1}$ data, but no clear structure was seen due to low statistics.
With five times larger data sample of 121.4 fb$^{-1}$, 
Dalitz plots of
$\Upsilon(nS) \pi^+\pi^-$ now show clear structures of two resonances in 
$\Upsilon(nS) \pi$ mass distribution as shown in 
Fig.~\ref{fig:zb_dalitz}~\cite{Belle:2011aa}.
Amplitude analysis of these Dalitz plot distributions revealed two
resonance states at masses of 10510 MeV/$c^2$ and 10560 MeV/$c^2$ in
all three transitions, $\Upsilon(5S) \to \Upsilon(nS)\pi^+\pi^-$ 
($n = 1, 2, 3$) with significances of well above 10$\sigma$ as shown in 
Fig.~\ref{fig:zb_mypi}.


\begin{figure}[t]
\begin{center}
\resizebox{0.4\textwidth}{!}{\rotatebox{0}{\includegraphics{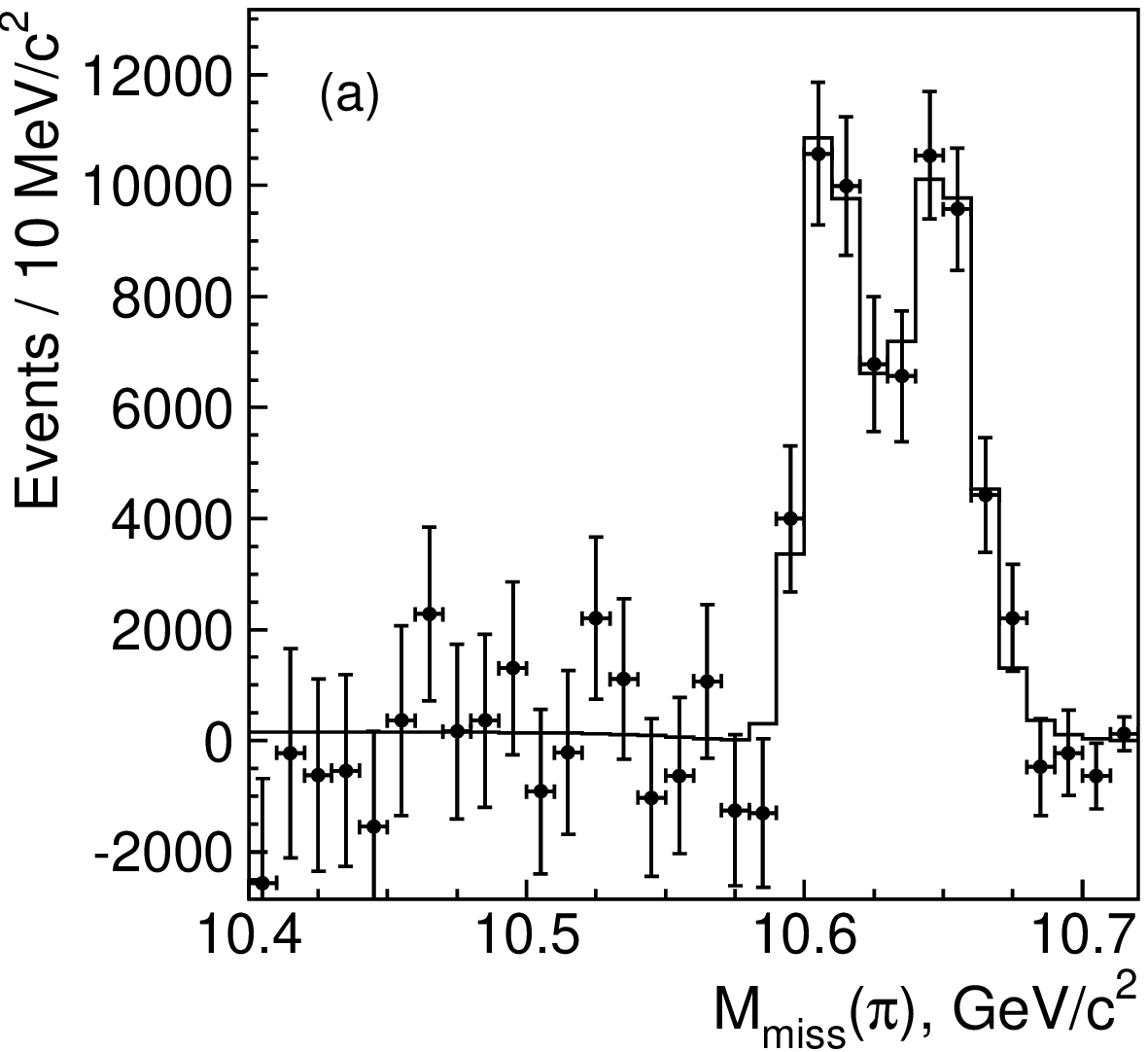}}}
\resizebox{0.4\textwidth}{!}{\rotatebox{0}{\includegraphics{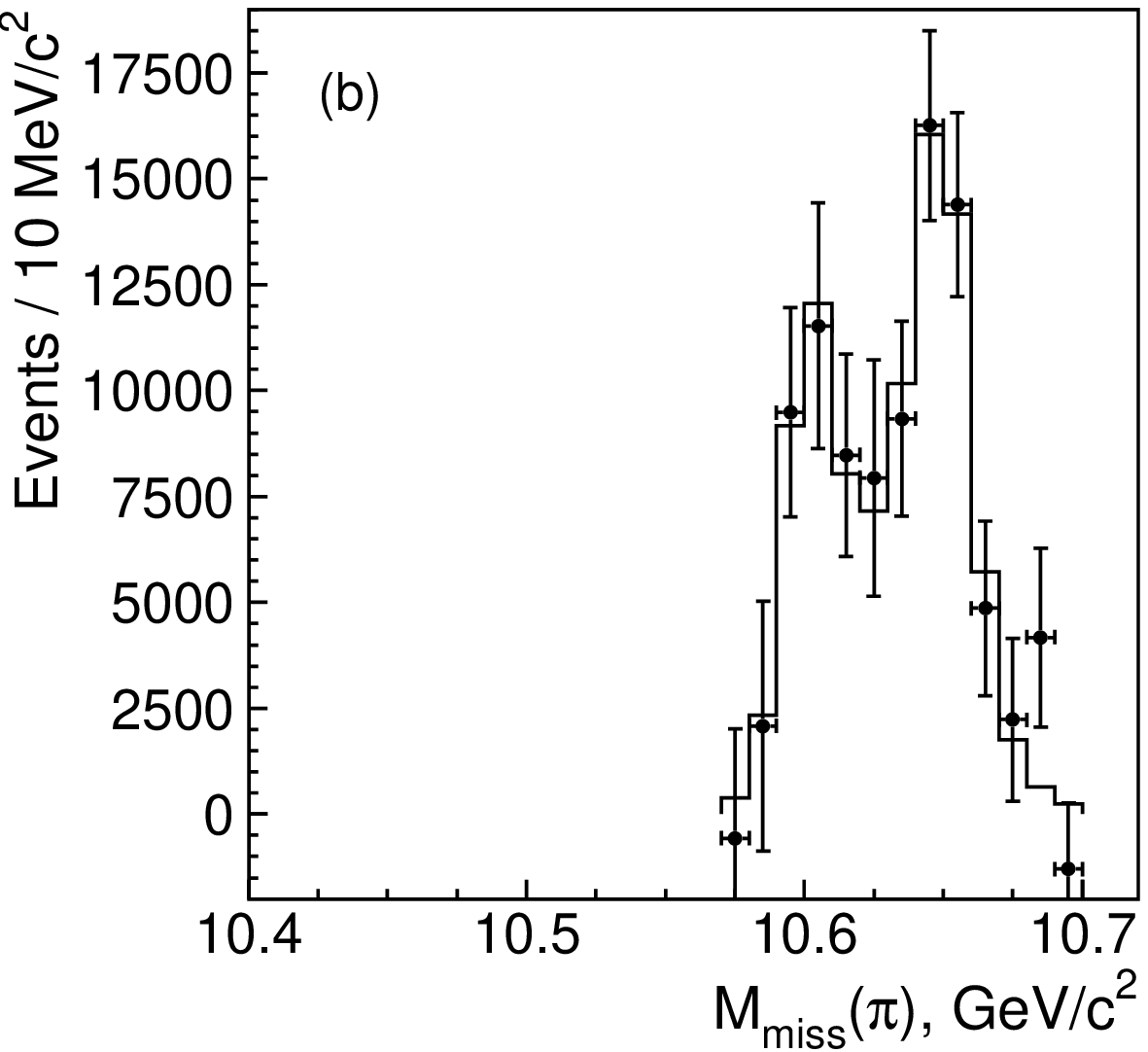}}}
\caption{ The (a) $h_b(1P)$ and (b) $h_b(2P)$ yields as a function of 
$M_{\rm miss}(\pi)$ (points with error bars) and results of the fit 
(histogram)~\cite{Belle:2011aa}. } 
\label{fig:zb_hbpi}
\end{center}
\end{figure}

The invariant mass distributions of $h_b(mP)\pi^+$ ($m =1,2$) are obtained
by the yields of $\Upsilon(5S) \to h_b(mP)\pi^+\pi^-$ signals in bins of
$h_b(mP)\pi^+$ mass, where $h_b(mP)\pi^+$ mass calculated as the missing mass
against $\pi^-$ ($M_{\rm miss}(\pi^-)$).  
For each bin, signal yield is obtained by fitting
the distribution of missing mass against $\pi^+\pi^-$ as described previously.
The resulting mass distributions of signal yields are shown in 
Fig.~\ref{fig:zb_hbpi}~\cite{Belle:2011aa}.
The two peaks of resonances are clearly seen and their masses and widths
are consistent with those obtained from 
$\Upsilon(5S) \to \Upsilon(nS)\pi^+\pi^-$ ($n = 1, 2, 3$).

The spin and parity of two $Z_b$ resonances are determined by the full amplitude
analysis with angular distributions in six dimensions for processes
$\Upsilon(5S) \to \Upsilon(nS)\pi^+\pi^-$ with 
$\Upsilon(nS) \to \mu^+\mu^-$~\cite{Garmash:2014dhx}.
The maximum likelihood fits with six-dimensional amplitude are performed
for all possible hypotheses of $J^P$ for $Z_b(10610)^+$ and $Z_b(10610)^+$,
each with $J^P = 1^+, 1^-, 2^+$ and $2^-$ ($0^\pm$ are forbidden because
$Z_b$ decays to both $\Upsilon(nS)\pi$ and $h_b(mP)\pi$).
The fit favors both $Z_b$'s with $1^+$ over all other combinations of
$J^P$ values with significances more than $6\sigma$.
The discrimination among different $J^P$ values mainly arises from 
an interference between $Z_b$ and other amplitude in Dalitz plot and
angular distributions as shown in Fig.~\ref{fig:zb_mypi}.

\begin{table}[t]
\caption{Summary of masses and widths of $Z_b$'s in 
$\Upsilon(nS)\pi^+$ $(n = 1, 2, 3)$ decays~\cite{Garmash:2014dhx} and 
$h_b(mP)\pi^+$ $(m = 1,2)$ decays~\cite{Belle:2011aa}. 
}
\label{table:zb_massw}
\centering
\begin{tabular}{|l| lll |}
\hline
Parameter & $\Upsilon(1S)\pi^+$ & $\Upsilon(2S)\pi^+$ &
$\Upsilon(3S)\pi^+$   \\
\hline
$Z_b(10610)^+~M$ MeV/c$^2$ & $10608.5 \pm 3.4 ^{+3.7}_{-1.4}$ &
$10608.1 \pm 1.2 ^{+1.5}_{-0.2}$ & $10607.4 \pm 1.5 ^{+0.8}_{-0.2}$  \\
~~~~~~~~~~~~~~$\Gamma$ MeV/c$^2$ & $18.5 \pm 5.3 ^{+6.1}_{-2.3}$ &
$20.8 \pm 2.5 ^{+0.3}_{-2.1}$ & $18.7 \pm 3.4 ^{+2.5}_{-1.3}$  \\
$Z_b(10650)^+~M$ MeV/c$^2$ & $10656.7 \pm 5.0 ^{+1.1}_{-3.1}$ &
$10650.7 \pm 1.5 ^{+0.5}_{-0.2}$ & $10651.2 \pm 1.0 ^{+0.4}_{-0.3}$  \\
~~~~~~~~~~~~~~$\Gamma$ MeV/c$^2$ & $12.1 ^{+11.3 +2.7}_{-4.8 -0.6}$ &
$14.2 \pm 3.7 ^{+0.9}_{-0.4}$ & $9.3 \pm 2.2 ^{+0.3}_{-0.5}$  \\
Relative phase (deg) & $67 \pm 36 ^{+24}_{-52}$ & $-10 \pm 13 ^{+34}_{-12}$ &
$-5 \pm 22 ^{+15}_{-33}$  \\
\hline
\hline
Parameter &  $h_b(1P)\pi^+$ &  $h_b(2P)\pi^+$  & \\
\hline
$Z_b(10610)^+~M$ MeV/c$^2$ & 
$10605 \pm 2 ^{+3}_{-1}$ & $10599 ^{+6 +5}_{-3 -4}$ & \\
~~~~~~~~~~~~~~$\Gamma$ MeV/c$^2$ & 
$11.4 \pm ^{+4.4 +2.1}_{-3.9 -1.2}$ & $13  ^{+10 +9}_{-3 -4}$ & \\
$Z_b(10650)^+~M$ MeV/c$^2$ & 
$10654 \pm 3 ^{+1}_{-2}$ & $10651 ^{+2 +3}_{-3 -2}$ & \\
~~~~~~~~~~~~~~$\Gamma$ MeV/c$^2$ & 
$20.9 \pm ^{+5.4 +2.1}_{-4.7 -5.7}$ & $19 \pm 7 ^{+11}_{-7}$ & \\
Relative phase (deg) & $187 ^{+44 +3}_{-57 -12}$ & $181 ^{+65 +74}_{-105-109}$ & \\
\hline
\end{tabular}
\end{table}

The masses, widths and other quantities for two $Z_b$ states obtained 
for five transitions are summarized in Table~\ref{table:zb_massw}.
The average masses and widths are 
$M_{Z_b(10610)^+} = 10607.2 \pm 1.1$ MeV/$c^2$,
$M_{Z_b(10650)^+} = 10651.3 \pm 0.8$ MeV/$c^2$,
$\Gamma_{Z_b(10610)^+} = 18.0 \pm 2.2$ MeV/$c^2$ and 
$\Gamma_{Z_b(10650)^+} = 11.3 \pm 1.8$ MeV/$c^2$.
The masses of $Z_b(10610)^+$ and $Z_b(10650)^+$ are very close to
sum of masses of $B\bar B^*$ and $B^* \bar B^*$, respectively. 
It should be noted that the relative phase between the two $Z_b$ amplitudes are
consistent with zero for $\Upsilon(5S) \to \Upsilon(nS)\pi^+\pi^-$ and
with $\pi$ for $\Upsilon(5S) \to h_b(mP)\pi^+\pi^-$ processes, which 
would be expected from $B^{(*)}\bar B^*$ molecular model~\cite{Bondar:2011ev}.

\subsubsection{$\Upsilon(5S) \to [B^{(*)}\bar B^*]^\pm\pi^\mp$}
\ \ \ \ 

\begin{figure}[t]
\begin{center}
\resizebox{0.5\textwidth}{!}{\rotatebox{0}{\includegraphics{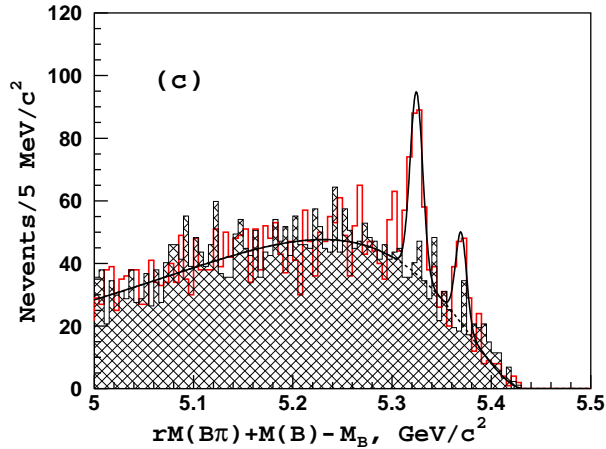}}}
\caption{ The distribution of missing mass against $B\pi$ ($rM(B\pi)$) 
for selected candidates (red histogram) and background 
(hatched histogram)~\cite{Adachi:2012cx}. }
\label{fig:zb_bbpi_mm}
\end{center}
\end{figure}

\begin{figure}[t]
\begin{center}
\resizebox{0.45\textwidth}{!}{\rotatebox{0}{\includegraphics{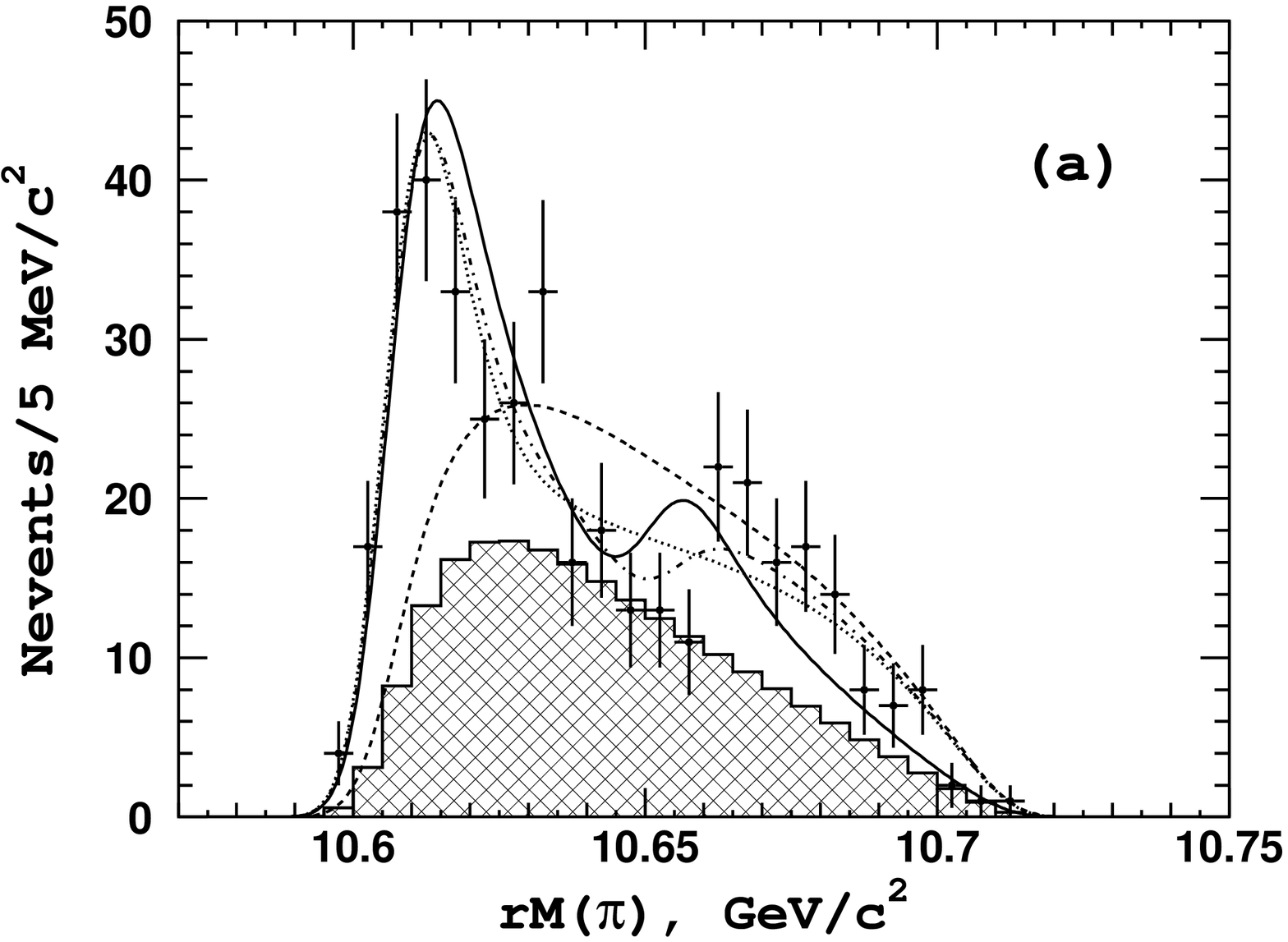}}}
\resizebox{0.45\textwidth}{!}{\rotatebox{0}{\includegraphics{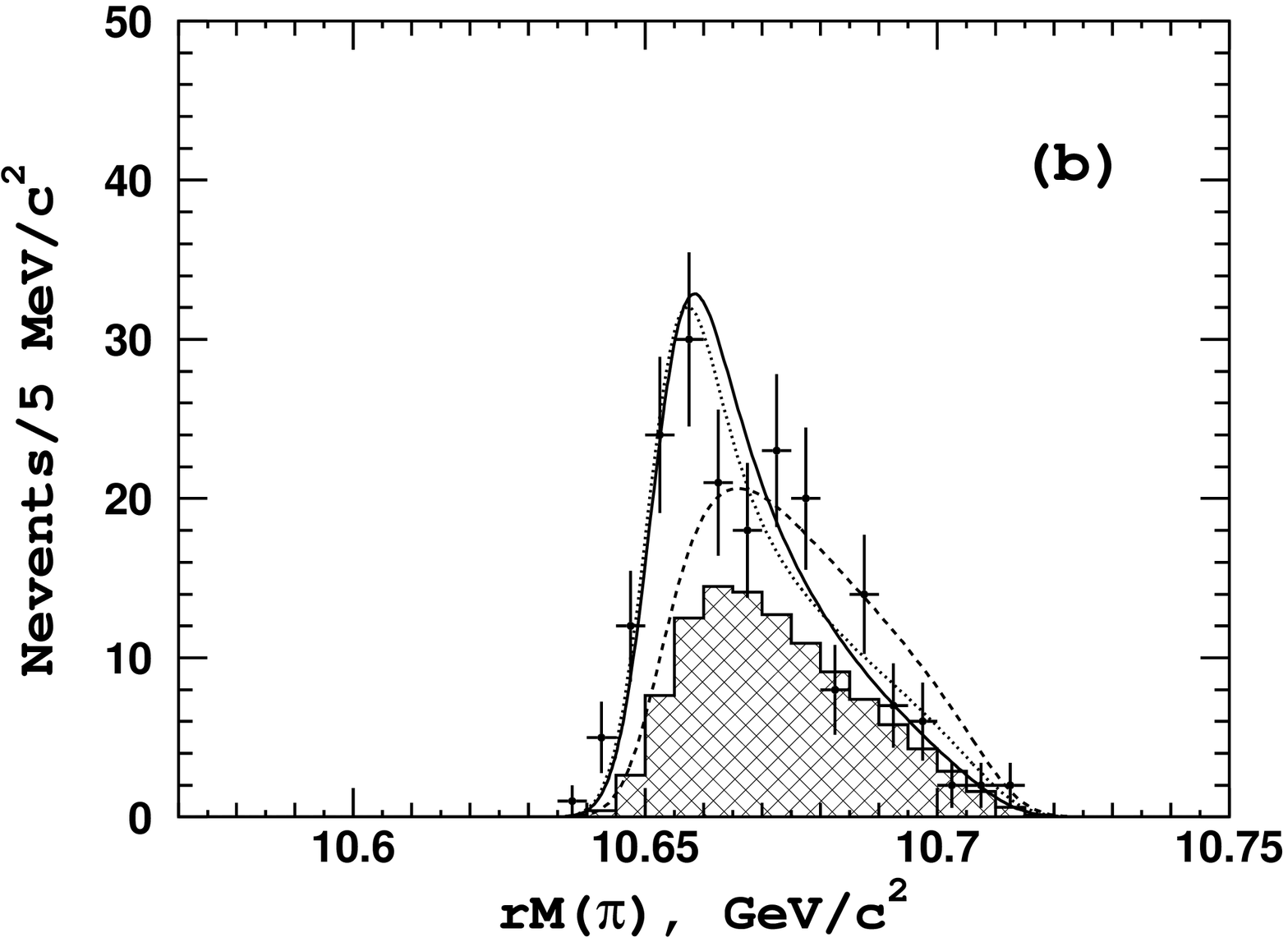}}}
\caption{ The distribution of missing mass against $\pi$ ($rM(\pi)$) for 
(a) $\Upsilon(5S) \to B\bar{B}^*\pi$ and 
(b) $\Upsilon(5S) \to B^*\bar{B}^*\pi$ 
candidate events. Points with error bars are data, the solid line is the 
result of the fit with $Z_b$'s only, the dashed line - fit to pure 
non-resonant amplitude, the dotted line - fit to a single $Z_b$
state plus a non-resonant amplitude, and the dash-dotted - two 
$Z_b$ states and a non-resonant amplitude. The hatched histogram
represents background component normalized to the estimated number
of background events~\cite{Adachi:2012cx}. }
\label{fig:zb_bbpi_bb}
\end{center}
\end{figure}

The proximity of $Z_b(10610)^+$ and $Z_b(10650)^+$ masses to the sum of 
$B^{(*)}\bar B^*$ masses indicates that these states would be
$B^{(*)}\bar B^*$ molecular states and decay to $B^{(*)}\bar B^*$ as will be discussed 
in section~\ref{sec:Zb}.
Rather large branching fractions in $\Upsilon(5S) \to B\bar B^*\pi$
decay was found with a data sample of 23.6 fb$^{-1}$~\cite{Drutskoy:2010an}.
The intermediate structure of $\Upsilon(5S) \to B^{(*)}\bar B^{(*)}\pi$ 
three-body decays are studied using 121.4 fb$^{-1}$ data 
sample~\cite{Adachi:2012cx}.
We reconstruct one $B$ meson with $B^+ \to J/\psi K^+$, 
$B^+ \to \bar D^0 \pi^+$, $B^0 \to J/\psi K^{*0}$, $B^0 \to D^- \pi^+$
and $B^0 \to D^{*-} \pi^+$.
Reconstructed $B^+$ or $B^0$ candidates are combined with a $\pi^-$
candidates and missing mass against $B\pi$ combination is calculated.
The missing mass distribution is shown in Fig.~\ref{fig:zb_bbpi_mm}.  
Two peaks correspond
to $B\bar{B}^*\pi$ and $B^*\bar{B}^*\pi$ signals with signal yields of 
$184 \pm 19$ and $82 \pm 11$ events, respectively.  
The $B^{(*)}\bar{B}^*$ invariant mass is calculated as a missing mass against
$\pi$ for events in $B\bar{B}^*\pi$ and $B^*\bar{B}^*\pi$ signal regions.
The distributions are shown in Fig.~\ref{fig:zb_bbpi_bb}.  
The fits to the distribution yield significant signal of $Z_b(10610)^+$
($8\sigma$) in $B\bar{B}^*\pi$ and $Z_b(10650)^+$ ($6.8\sigma$) 
in $B^*\bar{B}^*\pi$ transitions, 
while marginal signal of $Z_b(10650)^+$ is seen in $B\bar{B}^*\pi$ transition.


The relative branching fractions of $Z_b^+$ 
decaying to observed channels 
($\Upsilon(nS)\pi^+$, $h_b(mP)\pi^+$ and $B^{(*)}\bar{B}^*$) are calculated 
from the transition rates to three-body final
states and the fraction of intermediate $Z_b^+$ states.
Table~\ref{table:zb_bf} give the branching fractions assuming that sum 
of these modes
is 100\%.  The dominant decay modes are $B\bar{B}^*$ for $Z_b(10610)^+$
and $B^*\bar{B}^*$ for $Z_b(10650)^+$, respectively.

\begin{table}[t]
\caption{ Branching fractions (\%) of $Z_b^+$'s in different channels~\cite{Adachi:2012cx}.} 
\label{table:zb_bf}
\centering
\begin{tabular}{|l|cc|}
\hline
Channel  & $Z_b(10610)^+$  &  $Z_b(10650)^+$ \\
\hline
$\Upsilon(1S)\pi^+$   & $0.61 \pm 0.28$  &  $0.19 \pm 0.09$ \\
$\Upsilon(2S)\pi^+$   & $4.19 \pm 1.51$  &  $1.54 \pm 0.69$ \\
$\Upsilon(3S)\pi^+$   & $2.49 \pm 0.96$  &  $1.81 \pm 0.75$ \\
$h_b(1P)\pi^+$        & $4.40 \pm 2.17$  &  $10.3 \pm 5.5$ \\ 
$h_b(2P)\pi^+$        & $6.26 \pm 3.76$  &  $19.0 \pm 9.3$ \\  
$B^+\bar B^{*0} + \bar B^0B^{*+} $ & $82.0 \pm 3.5$  &  - \\  
$B^{*+}\bar B^{*0} $ &  -  &  $67.2 \pm 7.1$ \\  
\hline
\end{tabular}
\end{table}

\subsubsection{Neutral partner of $Z_b^+$ states}
\ \ \ \ 

\begin{figure}[t]
\begin{center}
\resizebox{0.32\textwidth}{!}{\rotatebox{0}{\includegraphics{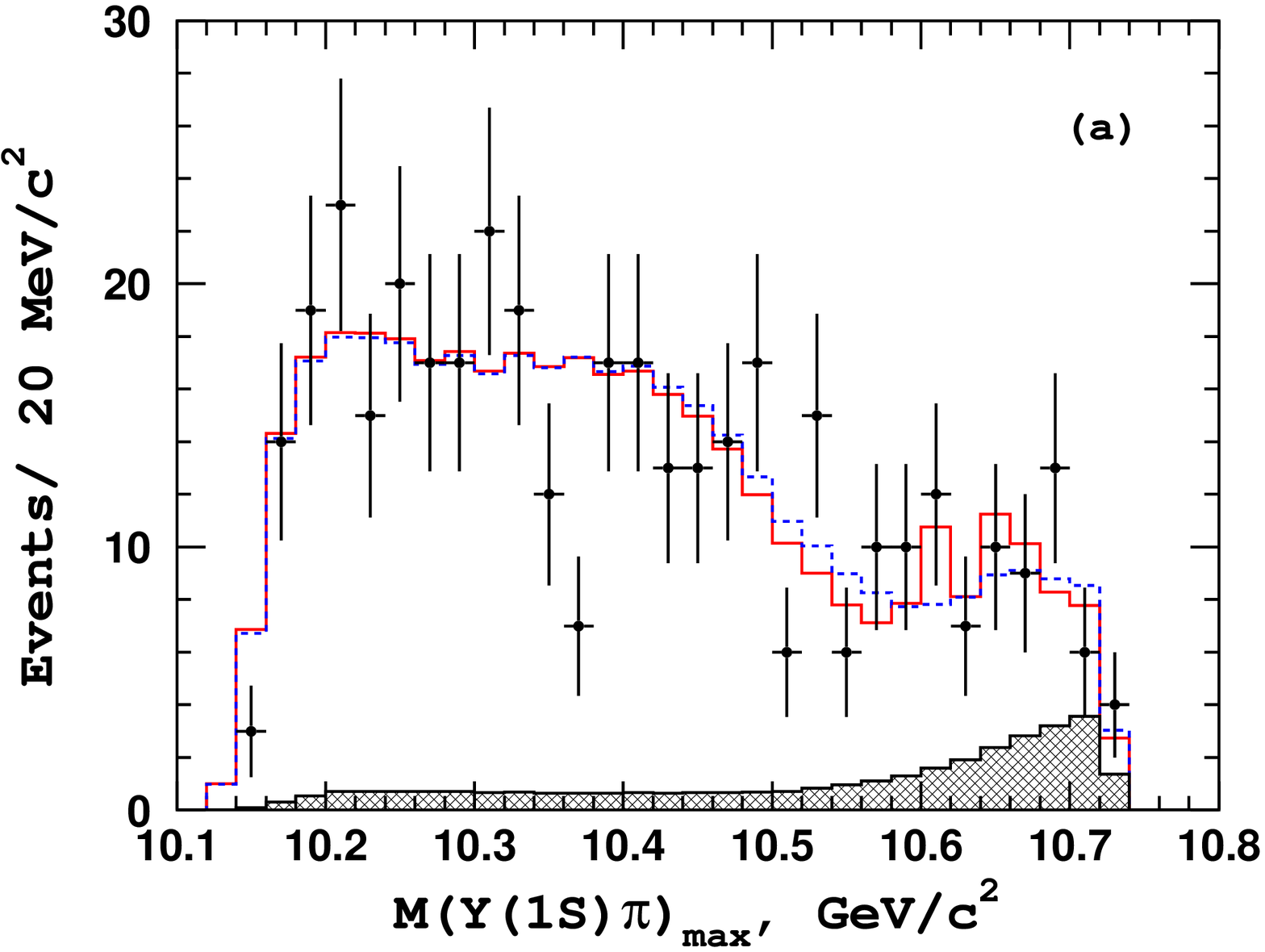}}}
\resizebox{0.32\textwidth}{!}{\rotatebox{0}{\includegraphics{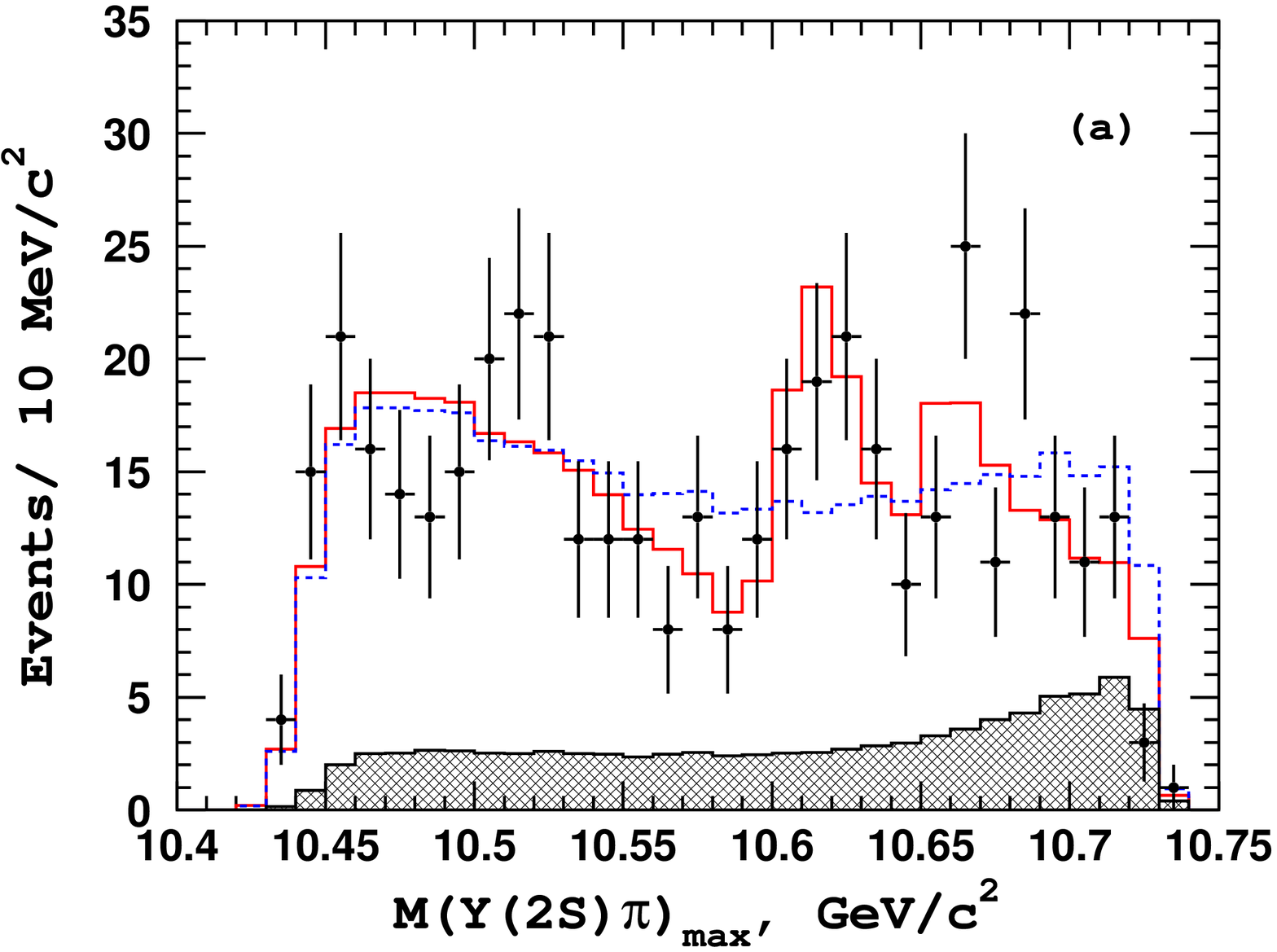}}}
\resizebox{0.32\textwidth}{!}{\rotatebox{0}{\includegraphics{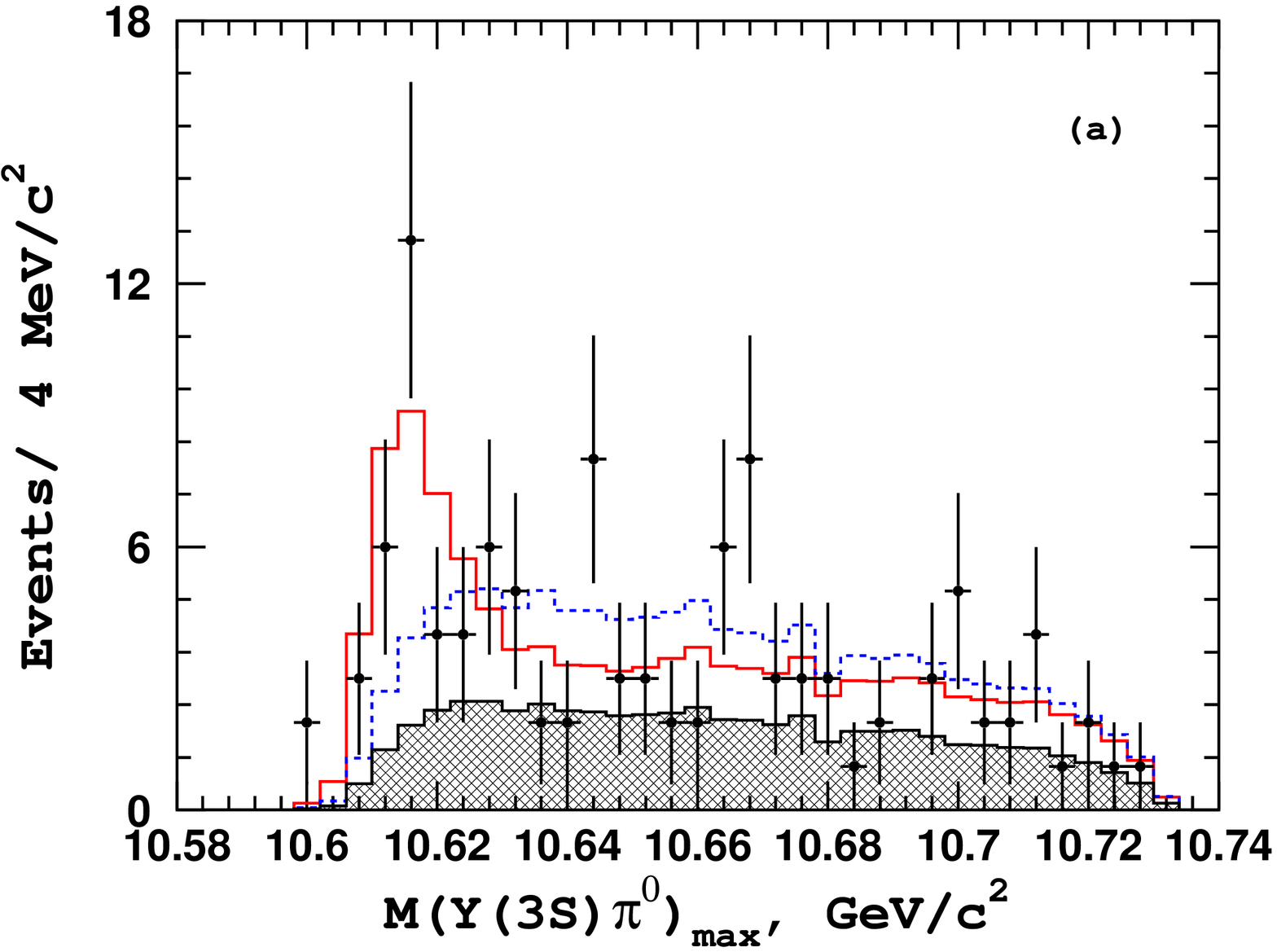}}}
\caption{ $\Upsilon(nS)\pi^0$ invariant mass distributions of 
$\Upsilon(nS)\pi^0\pi^0$ events for the
(left) $\Upsilon(1S)$; (middle) $\Upsilon(2S)$; (right) $\Upsilon(3S)$.
Points with error bars are data and hatched histograms show background.
Solid red and blue histograms show the fits with and without $Z_b^0$'s,
respectively~\cite{Krokovny:2013mgx}. }
\label{fig:zb0_mypi}
\end{center}
\end{figure}

Since $Z_b^+$ states decay to $\Upsilon(nS)\pi^+$, $h_b(mP)\pi^+$ and 
$B^{0(*)}\bar B^{*+}$, they are expected to be isovector states ($I = 1$).
Their neutral partners are searched for in 
$\Upsilon(5S) \to \Upsilon(nS)\pi^0\pi^0$ transitions.
The analysis is performed in a similar way as 
$\Upsilon(5S) \to \Upsilon(nS)\pi^+\pi^-$, replacing  
$\pi^\pm$ by $\pi^0$~\cite{Krokovny:2013mgx}.
The results have provided the first observation of 
$\Upsilon(5S) \to \Upsilon(nS)\pi^0\pi^0$ transitions with their rates
being consistent with half of 
$\Upsilon(5S) \to \Upsilon(nS)\pi^+\pi^-$ rates which are expected from
the isospin relation.
The amplitude analysis of Dalitz plot distributions has revealed $Z_b(10610)^0$
in $\Upsilon(2S)\pi^0\pi^0$ and $\Upsilon(3S)\pi^0\pi^0$ transitions
with a significance of $6.5\sigma$ as shown in Fig.~\ref{fig:zb0_mypi}.  
No significant signal is observed for $Z_b(10650)^0$ due to insufficient
statistics.  This result confirms that isospin of $Z_b$ state is one.

\subsection{Experiment summary}
\label{sec:summary_experiments}

Thanks to the world highest luminosity provided by the KEKB asymmetric-energy
$e^+e^-$ collider, the Belle experiment has recorded high statistics data
including 772 million $B\bar{B}$ pairs at $\Upsilon(4S)$ as well as 
the $\Upsilon(5S)$ data corresponding to the integrated luminosity of 
121.4 fb$^{-1}$. 
Accumulation of these excellent quality data is essential for all the 
discoveries of the $X$, $Y$, $Z$ states.
The molecular states turned out to be playing an important role near the 
thresholds. When the state is neutral and isosinglet, possible mixing with
the heavy quarkonia possessing same quantum number also affects a lot the 
properties of the observed states.
As already discussed, interpreting $X(3872)$ as an admixture of 
$D\bar{D^*}$ and $\chi_{c1}(2P)$ would show no conflict with all the 
available experimental results.
The $B^* \bar{B}^{(*)}$ molecule picture works well to describe the 
charged bottomonium-like states, $Z_b(10610)^+$ and $Z_b(10650)^+$;
the co-existence of $\Upsilon(nS) \pi^+\pi^-$ ($n=1, 2, 3$) and 
$h_b(mP) \pi^+ \pi^-$ ($m=1,2$) which usually does not take place because 
of the suppression to require heavy quark's spin flip. 
Large branching fraction to $B^* \bar{B}^{(*)}$ final state also supports 
the molecular hypothesis.

Reaching such very plausible interpretations would be a remarkable progress
made in recent years. In total, seven charged quarkonium-like states have been reported, they can not be interpreted as usual mesons but should have at least four constituents of $Q\bar{Q}u\bar{d}$, where $Q\bar{Q}$ detonates $c\bar{c}$ or $b\bar{b}$. Compilation of the relevant activities lead us to predictions of the partner states and their possible decay modes. 
Here, we also should not forget about many results brought by other experiments, BaBar, CLEO, BES III, CDF, D0 and LHCb. All those have also been greatly contributing to make clear existence of many states or point out necessity of further confirmation. All the available information as of today is summarized in Table \ref{tab_summary}.

\begin{table*}[tbp]
{\scriptsize
\caption{Newly found unconventional state candidates in the $c\bar{c}$ and
$b\bar{b}$ regions, ordered by mass. 
The mass and width values are as quoted in in PDG2014~\cite{Agashe:2014kda}.
$X(3945)$ and $Y(3940)$ have been subsumed under $X(3915)$ due to compatible properties and it is interpreted as $\chi_{c0}(2P)$ in PDG2014~\cite{Agashe:2014kda}.
The states known as $X(3823)$~\cite{Bhardwaj:2013rmw} and $Z(3930)$~\cite{Uehara:2005qd} are omitted because interpretations for these states are established as $\psi_2(1D)$ and $\chi_{c2}(2P)$, respectively shortly after their discoveries.
Original $Y(4008)$ measurement~\cite{Yuan:2007sj} is superseded by the latest one~\cite{Liu:2013dau} and its mass and width values are taken from the latter.
Since no public averrage values for $Y(4140)$ mass and width, CDF~\cite{Aaltonen:2009tz} values are written.
The $b\bar{b}$ in $Z_b(10610)^+$ and $Z_b(10650)^+$ decay modes represents $\Upsilon(nS)$ $(n=1,2,3)$ or $h_b(mP)$ $(m=1,2)$.
$Z_b(10610)^0$ width and $J^P$ are assumed to be same as $Z_b(10610)^+$.
 } 
\begin{center}
\label{tab_summary}
\hspace*{0cm}
\begin{tabular}{lcccll}
\hline\hline
\rule[10pt]{-1mm}{0mm}
 State & $m$~(MeV) & $\Gamma$~(MeV) & $J^{PC}$ & \ Process~(mode) & 
      \ References \\
\hline
\rule[10pt]{-1mm}{0mm}
$X(3872)$& 3871.69$\pm$0.17 & $<$1.2 &
    $1^{++}$
    & {\small $B\to K (\pi^+\pi^-J/\psi)$} &
    {\small Belle~\cite{Choi:2003ue,Choi:2011fc}, BaBar~\cite{Aubert:2008gu},} \\
& & & & & {\small LHCb~\cite{Aaij:2011sn, Aaij:2013zoa}} \\
& & & & {\small $p\bar p\to (\pi^+\pi^- J/\psi)+ ...$} &
    {\small CDF~\cite{Acosta:2003zx,Abulencia:2006ma,Aaltonen:2009vj}, D0~\cite{Abazov:2004kp}}  \\
& & &   & {\small $e^+ e^- \to \gamma (\pi^+ \pi^- J/\psi)$} &
    {\small BES III~\cite{Ablikim:2013dyn}} \\
& & &   & {\small $B\to K (\omega J/\psi)$} &
    {\small Belle~\cite{Abe:2005ix},
    BaBar~\cite{delAmoSanchez:2010jr}} \\
& & & & {\small $B\to K (D^{*0} \bar{D}^0)$} &
    {\small Belle~\cite{Gokhroo:2006bt,Adachi:2008sua}, 
    BaBar~\cite{Aubert:2007rva}} \\
& & & & {\small $B\to K (\gamma J/\psi)$ and } &
    {\small Belle~\cite{Bhardwaj:2011dj}, BaBar~\cite{Aubert:2008ae},}  \\
     &&&&{\small $B\to K (\gamma \psi(2S))$}& {\small LHCb~\cite{Aaij:2014ala}}   \\
$Z_c(3900)^+$ & $3888.7\pm3.4$ & 35$\pm$7 & $1^{+}$ &
    {\small $e^+e^- \to (J/\psi~\pi^+) \pi^-$} &
    {\small Belle~\cite{Liu:2013dau},
    BES III~\cite{Ablikim:2013mio}} \\ 
     & & & & {\small $e^+e^-\to (D \bar{D}^*)^+ \pi^-$} &
    {\small BES III~\cite{Ablikim:2013xfr}}  \\
$X(3915)$ & $3915.6\pm3.1$ & 28$\pm$10 & $0/2^{?+}$ &
    {\small $B\to K (\omega J/\psi)$} &
    {\small Belle~\cite{Abe:2004zs},
    BaBar~\cite{delAmoSanchez:2010jr}} \\ 
     & & & & {\small $e^+e^-\to e^+e^- (\omega J/\psi)$} &
    {\small Belle~\cite{Uehara:2009tx}, BaBar~\cite{Lees:2012me}}  \\
$X(3940)$ & $3942^{+9}_{-8}$ & $37^{+27}_{-17}$ & $?^{?+}$ &
     {\small $e^+e^-\to J/\psi(D D^*)$} &
     {\small Belle~\cite{Abe:2007sya}}  \\ 
&&&& {\small $e^+e^-\to J/\psi\ (...)$} &
     {\small Belle~\cite{Abe:2007jna}}  \\ 
$Y(4008)$ & $3891 \pm 42$ & 255$\pm$42 & $1^{--}$ &
     {\small $e^+e^-\to\gamma  (\pi^+\pi^-J/\psi)$} &
     {\small Belle~\cite{Yuan:2007sj,Liu:2013dau}}  \\
$Z_c(4050)^+$ & $4051^{+24}_{-43}$ & $82^{+51}_{-55}$ & ?&
     {\small $ B\to K (\pi^+\chi_{c1}(1P))$} &
     {\small Belle~\cite{Mizuk:2008me}, BaBar~\cite{Lees:2011ik}}  \\
$X(4050)^+$ & $4054 \pm 3$ & $45 $ & ? &
     {\small $ e^+ e^- \to  (\pi^+ \psi(2S)) \pi^-$} &
     {\small Belle~\cite{Wang:2014hta}}  \\
$Y(4140)$ & $4143.4\pm3.0 $ & $15^{+11}_{-\ 7}$ & $?^{?+}$ &
     {\small $B\to K (\phi J/\psi)$} &
     {\small CDF~\cite{Aaltonen:2009tz},D0~\cite{Abazov:2013xda}}  \\
$X(4160)$ & $4156^{+29}_{-25} $ & $139^{+113}_{-65}$ & $?^{?+}$ &
     {\small $e^+e^- \to J/\psi (D \bar{D}^*)$} &
     {\small Belle~\cite{Abe:2007sya}} \\
$Z_c(4200)^+$ & $4196^{+35}_{-32}$ &
     370$^{+99}_{-149}$ &?&
     {\small $ B\to K (\pi^+ J/\psi)$} &
     {\small Belle~\cite{Chilikin:2014bkk}}  \\
$Z_c(4250)^+$ & $4248^{+185}_{-\ 45}$ &
     177$^{+321}_{-\ 72}$ &?&
     {\small $ B\to K (\pi^+\chi_{c1}(1P))$} &
     {\small Belle~\cite{Mizuk:2008me}, BaBar~\cite{Lees:2011ik}}  \\
$Y(4260)$ & $4263\pm5$ & 108$\pm$14 & $1^{--}$ &
     {\small $e^+e^-\to\gamma  (\pi^+\pi^- J/\psi)$} &
     {\small BaBar~\cite{Aubert:2005rm,Lees:2012cn}, CLEO~\cite{He:2006kg},} \\ 
     &&&&& {\small Belle~\cite{Yuan:2007sj,Liu:2013dau}}  \\ 
& & & & {\small $e^+e^-\to (\pi^+\pi^- J/\psi)$} & {\small CLEO~\cite{Coan:2006rv}, BES III~\cite{Ablikim:2013xfr}} \\
& & & & {\small $e^+e^-\to (\pi^0\pi^0 J/\psi)$} & {\small CLEO~\cite{Coan:2006rv}} \\
$X(4350)$ & $4350.6^{+4.6}_{-5.1}$ & $13.3^{+18.4}_{-10.0}$ & ?$^{?+}$ &
     {\small $e^+e^-\to e^+e^- (\phi J/\psi)$} &
     {\small Belle~\cite{Shen:2009vs}} \\ 
$Y(4360)$ & $4361 \pm 13$ & 74$\pm$18 & $1^{--}$ &
     {\small $e^+e^-\to\gamma  (\pi^+\pi^- \psi(2S))$} &
     {\small BaBar~\cite{Aubert:2007zz}, Belle~\cite{Wang:2007ea,Wang:2014hta}}    \\ 
$Z_c(4430)^+$ & $4485^{+36}_{-25}$ & $200^{+49}_{-58}$ & $1^+$ &
     {\small $B\to K (\pi^+\psi(2S))$} &
     {\small Belle~\cite{Choi:2007wga,Mizuk:2009da,Chilikin:2013tch},} \\
& & & & & {\small BaBar~\cite{Aubert:2008aa}, LHCb~\cite{Aaij:2014jqa}} \\
& & & & {\small $B\to K (\pi^+ J/\psi)$} &
    {\small Belle~\cite{Chilikin:2014bkk}, BaBar~\cite{Aubert:2008aa}}  \\
$X(4630)$ & $4634^{+\ 9}_{-11}$ & $92^{+41}_{-32}$ & $1^{--}$ &
     {\small $e^+e^- \to \gamma (\Lambda_c^+ \Lambda_c^-)$} &
     {\small Belle~\cite{Pakhlova:2008vn}} \\
$Y(4660)$ & 4664$\pm$12 & 48$\pm$15 & $1^{--}$ &
     {\small $e^+e^-\to\gamma (\pi^+\pi^- \psi(2S))$} &
     {\small Belle~\cite{Wang:2007ea}} \\
$Z_b(10610)^+$ & 10607.2$\pm$2.0 & 18.4$\pm$2.4 & $1^+$ &
     {\small $e^+e^- \to (b\bar{b}~\pi^+) \pi^-$} &
     {\small Belle~\cite{Belle:2011aa}} \\
$Z_b(10610)^0$ & 10609$\pm$4$\pm$4 & N.A. & $1^{+-}$ &
     {\small $e^+e^- \to (\Upsilon(2,3S)  \pi^0) \pi^0$} &
     {\small Belle~\cite{Krokovny:2013mgx}} \\
$Z_b(10650)^+$ & 10652.2$\pm$1.5 & 11.5$\pm$2.2 & $1^+$ &
     {\small $e^+e^- \to (b\bar{b}~\pi^+) \pi^-$} &
     {\small Belle~\cite{Belle:2011aa}} \\
$Y_b(10888)$ & 10888.4$\pm$3.0 & 30.7$^{+8.9}_{-7.7}$ & $1^{--}$ &
      {\small $e^+e^-\to(\pi^+\pi^- \Upsilon(nS))$} &
      {\small Belle~\cite{Abe:2007tk,Chen:2008xia}} \\
\hline\hline
\end{tabular}
\end{center}
}
\end{table*}

Determination of still unsettled quantum numbers and various partner state searches require more statistics, continuing to 
exploit possibilities to access various production mechanisms as well as 
variety of decay modes with higher statistics data would be the upgrade 
$B$-factory project, Belle II experiment's mission in hadron spectroscopy, where friendly competition with other running experiments, BES III, LHCb and some other projects is expected to provide many interesting discoveries.

\section{Theory} 
\label{sec:theory}
%
\subsection{Overview} 
\label{sec:Introduction_theory}

In this chapter, we discuss theoretical methods from some fundamentals to  applications to the $XYZ$ exotic hadrons.
We emphasize the roles of  heavy quarks, which are essential to understand the properties of these states.    

Let us start with the quarkonium spectroscopy, which has been successfully described phenomenologically
by quark models.  
The method has been working well particularly in the mass region below the thresholds of open heavy mesons, such as $D\bar{D}$, $D\bar{D}^{\ast}$ and $D^{\ast}\bar{D}^{\ast}$ (see Fig.~\ref{fig_ccbar}).
There are two essential ingredients in the phenomenological quark model: 
(i) heavy quarks as effective degrees of freedom for constituents and 
(ii) a quark potential which has two important components of the color Coulomb force at short distances and of the confinement (linear) force at long distances~\cite{Barnes:2005pb}. 
Phenomenologically, the potential is supplemented by the spin-dependent and velocity-dependent forces.
The important static part of potential has been nicely reproduced by the lattice QCD calculations~\cite{Bali:2000gf}, and in this respect the model has a  correspondence to QCD.

Recent experimental observations of $XYZ$ exotic hadrons (section~\ref{sec:experiments}) have unveiled, however, 
that those successes are limited only below the thresholds of open heavy hadrons.
Near and above the thresholds, many exotic states have been found, 
whose properties cannot be well understood by the conventional quark model with the incorporation of (i) and (ii).  
Thus various ideas have been proposed; 
more degrees of freedom in addition to heavy quarks, such as light constituent quarks and constituent gluons.
In early stages, for instance,  theoretical studies were performed in Refs.~\cite{Manohar:1992nd,Lipkin:1986dw,Tornqvist:1993ng}, 
where the light degrees of freedom were incorporated in hadronic dynamics through 
$D^{(\ast)}\bar{D}^{(\ast)}$ mesons, 
and exotic hadrons  were predicted in the form of their hadronic molecules.
Due to the lack of experimental information at that time, however, 
much progresses were not made.  

We can understand at least in a qualitative manner how the introduction of light degrees of freedom is essential to understand the $XYZ$ exotic hadrons.
As for the conventional quarkonia below the thresholds, orbital motions of the constituent quarks carry (excitation) energy whose scale is much smaller than the heavy quark mass.
Hence we may regard them as light degrees of freedom (d.o.f.).
For example, the mass difference between $S$-wave quarkonia ($\eta_{c}$ and $J/\psi$) and $P$-wave ones ($h_{c}$ and $\chi_{cJ}$) is  about 300 MeV which is explained by the excitation energies of orbital motions.
With this energy the light quark-antiquark pair creation is still suppressed, or slightly more precisely, they may occur only virtually.
Near and above the thresholds with more excitation energies, however, 
light quarks can appear as explicit degrees of freedom.
They can be practically constituents of hadrons to form the so called  multiquark systems ($\bar{Q}Q\bar{q}q$) with compact structure or an extended structure of hadronic molecules through the rearrangement of the multiquarks into several hadrons.
The mechanism of how various kinds of hadrons are formed is dictated by QCD, in particular by 
those of quark confinement, mass generation and chiral symmetry breaking.  
In other words, by properly describing the  properties of exotic hadrons we hope to 
know better non-perturbative dynamics of QCD.  

Let us turn to some features associated with heavy particle nature.  
These are the essences of the potential non-relativistic QCD (pNRQCD)~\cite{Brambilla:1999xf} and 
heavy quark spin symmetry~\cite{Manohar:2000dt}. 
\begin{itemize}
\setlength{\itemsep}{1em}
\item
{\bf Constituent nature:}
For heavy quarks, particle number of constituent quarks is well conserved, where 
quark-antiquark pair creation is suppressed at the QCD energy scale.
Thus the concept of constituents works and, 
therefore, a non-relativistic treatment is justified not only kinematically but also 
dynamically with the heavy particle number fixed.  
\item
{\bf Potential:}
Heavy quarks move slowly. 
Therefore,  the potential approximation for the interaction by 
instantaneously propagating force works well.
Moreover, kinetic energy is suppressed for heavy particles, and 
even a small attraction can allow bound states or resonances.  
Their properties are sensitive to the details of the interaction.  
This feature may provide us with precise information about the interaction 
among the constituent particles.
The inter-quark potential can be supplied directly from QCD in the pNRQCD formalism \cite{Brambilla:1999xf}.
\item
{\bf Spin Symmetry:}
Heavy quark spins are conserved  in the heavy quark limit.
It is the exact symmetry of QCD for heavy quarks
where they interact through vector (gauge) gluons.  
Thus, just as in non-relativistic theories,  the spin of the heavy quark is conserved
in addition to the total hadron spin.
\end{itemize}

Because of the above features, more predictions about heavy hadrons have been made theoretically
 than so far observed.  
For example, tetraquark mesons $T_{cc}$ with charm number two ($|C|=2$) 
are considered to be stable states against the decays by strong 
interaction\footnote{See Refs.~\cite{Ader:1981db,Zouzou:1986qh,Heller:1986bt,Carlson:1987hh,SilvestreBrac:1993ss,SilvestreBrac:1993ry,Semay:1994ht,Pepin:1996id,SchaffnerBielich:1998ci,Brink:1998as,Janc:2004qn,Barnea:2006sd,Vijande:2007fc,Vijande:2007rf,Vijande:2007ix,Ebert:2007rn,Navarra:2007yw,Zhang:2007mu,Lee:2007tn,Lee:2009rt,Yang:2009zzp,Vijande:2009kj,Carames:2011zz,Vijande:2011zz} for compact tetraquark picture, Refs.~\cite{Manohar:1992nd,Tornqvist:1993ng,Ding:2009vj,Molina:2010tx,Ohkoda:2012hv} for extended hadronic molecule picture, Refs.~\cite{Bicudo:2012qt,Brown:2012tm,Ikeda:2013vwa} for lattice QCD result.}.
If they exist, they are the ground states of hadrons with charm number two and baryon number zero.  

In this section, we present theory discussions with some 
examples for exotic hadrons.  
In section~\ref{sec:basic_idea}, we describe the basic theoretical ideas for the study of heavy hadrons, 
starting from the QCD Lagrangian.  
First we briefly discuss chiral symmetry which is important for light quark dynamics.  
Then we turn to discussions of heavy quark dynamics.   
In particular, 
we present how the heavy quark (spin) symmetry shows up in normal hadrons as well as in exotic ones, with an 
emphasis that the heavy quark symmetry holds not only at the level of quark dynamics, 
but also at the level of hadron dynamics.
In section~\ref{sec:Theoretical_methods}, we discuss several theoretical methods  
including the quark model (section~\ref{sec:Quark_model}), 
heavy-hadron chiral effective theories (section~\ref{sec:HHChT}), 
models for hadronic molecules (section~\ref{sec:Hadron_model}), QCD sum rules (section~\ref{sec:QCDSR}) and 
lattice QCD studies (section~\ref{sec:LQCD}).
In section~\ref{sec:Examples}, we apply the idea of the hadronic molecule 
for the description of 
$X(3872)$ (section~\ref{sec:theory_X3872}), and $Z_{b}(10610)$ and $Z_{b}(10650)$ (section~\ref{sec:Zbtheory}).
We briefly overview current theoretical status for other states of $Y(4260), Z_c^+(4430)$ and $Z_c^+(3900)$,  and mention the recent studies based on the dynamical treatments (section~\ref{sec:quick_view}).  
As future prospects, in section~\ref{sec:future}, we discuss possible new hadronic and nuclear systems, 
which have been predicted theoretically, but not yet observed in experiments.
We consider the stability of $D^{(\ast)}D^{(\ast)}$ hadronic molecules (section~\ref{sec:Tcc_molecule}), and $T_{cc}$ state (section~\ref{sec:Tcc}).

\subsection{Basic idea} 
\label{sec:basic_idea}

\subsubsection{The QCD Lagrangian}
\label{sec:QCD}
\ \ \

The QCD Lagrangian is well-known and is given by
\begin{eqnarray}
{\cal L}_{\rm{QCD}} = -\frac{1}{4}F_{\mu\nu}^{a} F^{a\,\mu\nu} + \sum_{f} \bar{\psi}_{f} (iD\hspace{-0.6em}/ - m_{f}) \psi_{f}.
\label{eq_LQCD}
\end{eqnarray}
Here the sum $\sum_{f}$ goes over flavors $f$, 
$F^{a}_{\mu\nu}=\partial_{\mu} A^{a}_{\nu} - \partial_{\nu} A^{a}_{\mu} - g_{\rm s} f^{abc} A^{b}_{\mu} A^{c}_{\nu}$ is the gluon field tensor with the color index $a$ and Lorentz ones $\mu, \nu$, and $D_{\mu} = \partial_{\mu} - g_{\rm s} A_{\mu}^{a}T^{a}$ the covariant derivative with the gluon field $A^{a}_{\mu}$ 
($a=1,\dots,N_{\rm c}^{2}-1$; $\mu=0,\cdots,4$;
$N_c$, the number of colors).

The Lagrangian (\ref{eq_LQCD}) looks similar to that of QED, but is dynamically very much different.
In QCD, it is known that the  strong coupling constant $\alpha_{\mathrm s}(\mu^2)$
runs as a function of a momentum scale $\mu$  
as dictated by the following renormalization group equation 
with the beta function $\beta(\alpha_{\mathrm s})$
\be
\mu^2 \frac {d\alpha_{\mathrm s}}{d\mu^2} = \beta(\alpha_{\mathrm s}) \equiv
-(b_1 \alpha_{\mathrm s}^2 + b_2 \alpha_{\mathrm s}^3 + \cdots ), 
\label{RGE}
\ee
where the coefficients $b_n$ can be calculated 
at each order of $n$ loops.
For instance, at the one-loop level, 
$
b_1 = (33 - 2 N_f)/(12 \pi)
$
with the flavor number $N_f$.  
At this level, one can solve Eq.~(\ref{RGE}) analytically to find the solution
\be
\alpha_{\mathrm s}(\mu^2) = \frac{4 \pi}{\left( 11 - \frac{2}{3} N_f\right) \ln(\frac{\mu^2}{\Lambda_{\mathrm{QCD}}^2} ) }.
\label{eq_alpha_S}
\ee
Hence $\alpha_{\mathrm s}(\mu^2) $ diverges at the QCD scale $\mu^2 = \Lambda_{\rm QCD}^2$, 
where $\Lambda_{\rm QCD}$ is several hundred MeV.
The emergence of the dimensional scale brings the rich structure in the QCD vacuum\footnote{We should remind us that there is no scale parameter in the original QCD Lagrangian~(\ref{eq_LQCD}). The existence of $\Lambda_{\rm QCD}^2$ is a consequence of the quantum fluctuations in the QCD vacuum.}.
Physically, due to the self coupling of the gluons, the forces among the colored quarks and gluons increase as their distances increase. 
This causes non-trivial phenomena such as color confinement, mass generation of hadrons 
and spontaneous breaking of chiral symmetry.  
Thus at the hadronic scale, we need essentially a non-perturbative method.  

Now, six kinds of quark have very different bare (current) masses; u, d, s quarks are light while c, b, t quarks are heavy in comparison with $\Lambda_{\rm QCD}$; $m_u, m_d, m_s \ll \Lambda_{\rm QCD} \ll m_c, m_b, m_t$. In fact, these inequality may not be well satisfied for the strange quark with $m_s \sim$ 100 MeV. But for the discussion up to the bottom region (without top which is too far from the current discussions and is the scale where the weak interaction is not longer weak) $\Lambda_{\rm QCD} \simeq 200$-$300$ MeV, and so at least the inequality holds. The different mass scales for the light and heavy quarks provide us with different structures in hadron dynamics.
%
In fact, because of the hierarchy of the quark masses and $\Lambda_{\mathrm{QCD}}$, it is possible to introduce two different small parameters; 
(light mass)/$\Lambda_{\mathrm{QCD}}$ and 
$\Lambda_{\mathrm{QCD}}$/(heavy mass).  
These two parameters may be used as a small parameter for perturbation series in QCD.
For light hadrons, chiral perturbation is the relevant one associated with chiral symmetry 
with its spontaneous breaking, and for heavy hadrons spin symmetry is the one 
due to heavy quark symmetry.  
We will see that, for the discussion of exotic hadrons, both symmetries are useful and important.  

\subsubsection{Chiral symmetry}
\label{sec:ChS}
\ \ \

Let us consider the massless limit ($m_{f} \rightarrow 0$)
for up, down and strange quarks ($f=u,d,s$).
In this case, chiral fermions of right- and left-handed chiralities 
\begin{eqnarray}
\psi_{{\rm R}f} = \frac{1+\gamma_{5}}{2} \psi_{f}, \; \; \; 
\psi_{{\rm L}f} = \frac{1-\gamma_{5}}{2} \psi_{f},
\end{eqnarray}
may decouple.  
This is shown explicitly by the QCD Lagrangian, which, by ignoring the mass term, can be written 
as the sum of the right- and left-handed terms;
\begin{eqnarray}
 \bar{\psi}_{f} (i D \hspace{-0.6em}/ - m_{f}) \psi_{f}
\rightarrow
\bar{\psi}_{{\rm R}f} i D \hspace{-0.6em}/ \hspace{0.3em} \psi_{{\rm R}f} +
\bar{\psi}_{{\rm L}f} i D \hspace{-0.6em}/ \hspace{0.3em} \psi_{{\rm L}f}.
\end{eqnarray}
This is invariant under separate flavor transformations for the right and left components,  
${\rm SU}(N_{\rm F})_{\rm R} \times {\rm SU}(N_{\rm F})_{\rm L}$;
\begin{eqnarray}
\psi_{\rm R} \rightarrow e^{i \theta_{\rm R}^{a}T^{a}} \psi_{\rm R}, \; \; \; 
\psi_{\rm L} \rightarrow e^{i \theta_{\rm L}^{a}T^{a}} \psi_{\rm L},
\end{eqnarray}
with $\psi_{\rm R/L} = (u_{\rm R/L}, d_{\rm R/L}, s_{\rm R/L})^{\rm t}$ and with $T^{a}$ ($a=1,\dots,N_{\rm F}^{2}-1$) being generators of the flavor ${\rm SU}(N_{\rm F})_{\rm R/L}$ symmetry for the right- and left-handed components, respectively.

Chiral symmetry manifests for the QCD Lagrangian of the light flavor sector, while 
it may not so for the  vacuum.    
It is known that non-trivial configurations of the gluons field, instantons, may provide interactions
which mix  the left- and right-handed quarks.
When the interaction is sufficiently strong, the left and right 
quark-antiquark pairs condensate
\begin{eqnarray}
\langle \bar{\psi}_{f} \psi_{f} \rangle 
= \langle \bar \psi_{\rm L} \psi_{\rm R} \rangle
+ \langle \bar \psi_{\rm R} \psi_{\rm L} \rangle \neq 0,
\end{eqnarray}
causing the breaking of the chiral (left- and right-handed) symmetry of the vacuum.  
This resolves the degeneracy of hadrons of opposite parities, leading to rich/complicated spectrum of hadrons.
For instance, 
the Nambu-Goldstone bosons of negative parity ($\pi$, $K$, $\eta$ mesons) become (approximately) 
massless, while their chiral symmetry partners, scalar mesons ($f_0$'s, $\kappa$ and $a_{0}$ mesons), 
remain massive.  
The ground state nucleon is of positive parity, while its negative parity partner 
appear at  about 600 MeV above the nucleon~\cite{Jido:2001nt}.  

An interesting feature of hadron dynamics is that mesons can mediate interactions among colorless hadrons. 
Such an  interaction is regarded as a residual interaction after saturating the color charges, 
and is considered to be relatively weak as compared to the force among the colored quarks and gluons.  
Among various meson exchanges, of particular importance is the pion exchange interaction.  
There are two features in pion-exchange interaction; 
\begin{itemize}
\item It is a long distance force, because of the small mass as the Nambu-Goldstone boson.
\item It generates the strong tensor force that mixes different angular momenta, $L$ and $L\pm2$.
\end{itemize}
Those two features lead to rich structures in the many-body hadron systems. 
A typical example is the atomic nucleus, where a variety of modes are realized, e.g., 
single particle motions, collective motions such as surface vibrations and rotations, alpha-clustering, 
spin- and isospin-correlations, and so on. 
In the early developments of the nuclear physics, these modes have been separately studied in a phenomenological manner~\cite{bohr1969nuclear}.
Recently, however, there are attempts for systematic description in a unified manner, 
where the pion exchange force plays crucial roles~\cite{Otsuka:2005zz,Myo:2009ru,Pieper:2001mp,Ogawa:2011zzb,Myo:2011ci,Myo:2012pv}.
Here what we are interested in are the hadrons that are many-body systems of quarks and gluons,
in particular, exotic states containing both heavy and light quarks.  
Once light quarks exist, they can couple to the pion, and cause the strong tensor force.
The possible pion exchange force as well as the presence of heavy quarks may 
give us rich structure as in atomic nuclei, which is one of the main issues in the present discussions.

\subsubsection{Heavy-quark symmetry}
\label{sec:HQS}
\ \ \

Let us ask what will happen when a quark becomes very heavy as compared to the QCD scale of several hundred MeV.
For sufficiently large mass $m_{\rm Q}$, 
the parameter $\Lambda_{\mathrm{QCD}}/m_{\mathrm Q}$ is regarded as a small parameter for 
a systematical expansion of  the strong interaction dynamics in powers of it.  
In this subsection, we briefly discuss the usefulness of such expansions for the charm and bottom sectors, 
and apply to the study of heavy hadron phenomenology  (see Refs.~\cite{Manohar:2000dt,Neubert:1993mb} for more precise information).

Let us consider a system which contains one heavy quark.  
In the limit $m_{\rm Q} \to \infty$, disturbance from the QCD interaction of order 
$\Lambda_{\mathrm{QCD}}$ may be neglected, and so
the heavy quark can remain almost on-shell though it is confined.   
Thus we introduce the on-shell momentum $m_Q v^\mu$ by the four-velocity $v^\mu, v^2 = 1$. 
In reality, finitely heavy quark is disturbed by the QCD interaction and the momentum is changed from its on-shell value to, 
\begin{eqnarray}
p^{\mu} = m_{\rm Q} v^{\mu} + k^{\mu},
\label{eq:momentum}
\end{eqnarray}
where $k^{\mu}$ is a residual momentum whose scale is much smaller than $m_{\rm Q}$, 
characterizing the offsell-ness of the heavy quark.  
In many cases of practical applications,  
the four-velocity may be chosen as $v^{\mu}=(1,\vec{0}\,)$, the one in the rest frame of the heavy quark.  

According to the decomposition of the momentum, the effective field for the heavy quark with four-velocity $v^{\mu}$ is defined as
\begin{eqnarray}
Q_{v}(x) = e^{im_{\rm Q}v\cdot x} \frac{1+v\hspace{-0.5em}/}{2} Q(x),
\label{eq:heavy_quark_effective_field_1}
\end{eqnarray}
for the original heavy quark field $Q(x)$.
The factor $e^{im_{\rm Q}v\cdot x}$ indicates that the momentum scale $m_{\rm Q} v^{\mu}$ is extracted from $Q(x)$, so that the energy of the effective heavy quark $Q_{v}$
is defined by subtracting the mass.
As a result, the effective field has only a residual momentum $k^{\mu}$.
The operator $(1+v\hspace{-0.5em}/)/2$ is the projection operator to select the positive-energy component of $Q(x)$.

Similarly  we can define the effective field for the heavy {\it antiquark} as
\begin{eqnarray}
{\cal Q}_{v}(x) = e^{im_{\rm Q}v\cdot x} \frac{1-v\hspace{-0.5em}/}{2} Q(x),
\label{eq:heavy_quark_effective_field_2}
\end{eqnarray}
in the same frame, by applying the projection operator $(1-v\hspace{-0.5em}/)/2$.
We note also that ${\cal Q}_{v}(x)$ corresponds to the excitation with mass $2m_{\rm Q}$, 
as shown soon.

With the above setup, we separate the heavy quark part from the light quark and gluon part in the QCD Lagrangian, as a sum of the heavy quark part and the light quark and gluon part,
\begin{eqnarray}
{\cal L}_{\rm{QCD}} = {\cal L}_{\mathrm{heavy}}  + {\cal L}_{\mathrm{light}}.
\label{eq:QCD}
\end{eqnarray}
where the explicit form of ${\cal L}_{\mathrm{heavy}} $ is
\begin{eqnarray}
{\cal L}_{\mathrm{heavy}} 
&=&
\bar{Q} ( iD\hspace{-0.7em}/ -m_{\rm Q}) Q, 
\end{eqnarray}
and the explicit form of ${\cal L}_{\rm{light}}$ is omitted, because the contributions from light quarks and gluons are irrelevant to the heavy quark spin symmetry.
As for the heavy part, in the first term, a summation over heavy quark flavors can be taken, if multiple number of heavy flavors (e.g. charm and bottom) are considered.
In the present cases, we consider only a single heavy flavor.
By using the positive-energy field operator $Q_{v}$ in Eq.~(\ref{eq:heavy_quark_effective_field_2}), let us rewrite the heavy quark part as
\begin{eqnarray}
{\cal L}_{\mathrm{heavy}} 
&=&
\bar{Q}_{v} v \! \cdot\! i D Q_{v}
-\bar{{\cal Q}}_{v} (v \! \cdot\! i D + 2m_{\rm Q}) {\cal Q}_{v}
+\bar{Q}_{v} iD_{\perp}\hspace{-1.3em}/\hspace{0.7em} {\cal Q}_{v}
+\bar{{\cal Q}}_{v} iD_{\perp}\hspace{-1.3em}/\hspace{0.7em} Q_{v},
\end{eqnarray}
The term of $2m_Q$ in the second term indicates that ${\cal Q}_{v}$ corresponds to the excitation of the mass $2m_{\rm Q}$, hence the component ${\cal Q}_{v}$  should be suppressed for large $m_{\rm Q}$, so long as we are interested in the heavy quark limit.
In fact, using the equation-of-motion for ${\cal Q}_{v}$,
\begin{eqnarray}
(v \! \cdot\! i D + 2m_{\rm Q}) {\cal Q}_{v} = iD_{\perp}\hspace{-1.3em}/\hspace{0.7em}{ Q}_{v},
\end{eqnarray}
and eliminating ${\cal Q}_{v}$, we obtain
\begin{eqnarray}
{\cal L}_{\mathrm{heavy}} 
&=&
\bar{Q}_{v} \left( v \! \cdot\! i D + iD_{\perp}\hspace{-1.3em}/\hspace{0.7em} \frac{1}{2m_{\rm Q} + v \! \cdot\! i D} iD_{\perp}\hspace{-1.3em}/\hspace{0.7em} \right) Q_{v} \nonumber \\
&=&
\bar{Q}_{v} v \! \cdot\! i D Q_{v}
+ \bar{Q}_{v} \frac{(i D_{\perp})^2}{2m_{\rm Q}} Q_{v}
-  g_{\rm s} \bar{Q}_{v} 
    \frac{\sigma_{\mu \nu}G^{\mu \nu}}{4m_{\rm Q}} Q_{v}
+ {\cal O}(1/m_{\rm Q}^{2}),
\label{eq:HQET_Lagrangian}
\end{eqnarray}
with $D_{\perp}^{\mu}=D^{\mu} -v^{\mu} \, v \!\cdot\! D$~\cite{Manohar:2000dt}.
This is the result at the tree level, because we have used the equation-of-motion for ${\cal Q}_{v}(x)$.
In order to reach the final results consistent with the original QCD at a given order of $1/m_{\rm Q}$, we need to include the quantum corrections given by the Wilson coefficients.
As a result, we obtain the effective Lagrangian in the $1/m_{\rm Q}$ expansion for heavy quarks as
\begin{eqnarray}
    {\cal L}_{\rm{HQET}} &=& \sum_{\rm Q} \left[ \bar{Q}_{v} v \! \cdot\! i D Q_{v} + \bar{Q}_{v} \frac{(i D_{\perp})^2}{2m_{\rm Q}} Q_{v} 
    - c(\mu) g_{\mathrm{s}} \bar{Q}_{v} 
    \frac{\sigma_{\mu \nu}G^{\mu \nu}}{4m_{\rm Q}} Q_{v} 
    + {\cal O}(1/m_{\rm Q}^{2}) \right] \nonumber \\
    && + {\cal L}_{\mathrm{light}},
    \label{eq:HQET}
\end{eqnarray}
with 
 $\sigma^{\mu\nu}=i[\gamma^{\mu},\gamma^{\nu}]/2$.
In the third term in the square brackets in Eq.~(\ref{eq:HQET}), $c(\mu)$ the a Wilson coefficient which is determined by matching to QCD at an energy scale $\mu$ due to quantum corrections.
On the other hand, the first and second terms are not affected by the quantum corrections, because the invariance under the velocity-rearrangement (i.e. the Lorentz boost valid up to ${\cal O}(1/m_{\rm Q})$ ) is maintained.
In other words, the quantum corrections are protected from the Lorentz symmetry up to ${\cal O}(1/m_{\rm Q})$, as explained later.

The effective theory given by Eq.~(\ref{eq:HQET}) is called the heavy quark effective theory (HQET). 
The important feature is that the Lagrangian for the heavy quark is given as a series of the power expansion by $1/m_{\rm Q}$.
This feature enables us to discuss systematically at each ${\cal O}(1/m_{\rm Q}^{n})$ ($n=0,1,\dots$) for sufficiently large $m_{\rm Q}$.

Let us consider the leading order (LO) as the dominant term in the $1/m_{\rm Q}$ expansion.
The term for the heavy quark at LO, 
\begin{eqnarray}
\sum_{\rm Q} \bar{Q}_{v} v \!\cdot\! i D Q_{v}.
 \label{eq:HQET0}
\end{eqnarray}
has two important symmetries; the spin symmetry and the heavy-flavor symmetry.
The spin symmetry is understood from the invariance under the spin transformation for $Q_{v}$, $Q_{v} \mapsto S Q_{v}$ 
where $S \in \rm{SU}(2)_{\rm{spin}}$ is a spin transformation operator~\cite{Manohar:2000dt}).
This is clear because the term $v \!\cdot\! i D$ does not contain the Dirac (and hence spin) matrices.
The heavy-flavor symmetry, a unitary transformation for mixing the different heavy-flavors, is also understood from the fact that the term $v \!\cdot\! i D$ does not depend on the species of the heavy quark $Q$ and hence it is regarded as a unit matrix in the heavy-flavor space.
Consequently, the Lagrangian (\ref{eq:HQET0}) has ${\rm SU}(2N_{\rm h})$ symmetry including both the spin symmetry and the flavor symmetry, where $N_{\rm h}$ is a number of heavy flavors.
$N_{\rm h}=2$ for two flavors (e.g. charm and bottom), and $N_{\rm h}=1$ for single flavor (e.g. charm or bottom).
The heavy-flavor symmetry is used for the studies of the weak interaction, for instance for the process $b \rightarrow c\,e\,\bar{\nu}_{e}$~\cite{Manohar:2000dt}, but this is not covered in the present article.

As discussed above, the heavy-quark symmetry is exact in the heavy quark limit ($m_{\rm Q} \to \infty$).
In realistic situations, however, the heavy quark masses are still finite, and the heavy-quark symmetry is violated at ${\cal O}(1/m_{\rm Q})$.
To see the violation, we consider the next-to-leading order (NLO) in the $1/m_{\rm Q}$ expansion in ${\cal L}_{\rm{HQET}}$;
\begin{eqnarray}
\sum_{\rm Q} \! \left[ \bar{Q}_{v} \frac{(i D_{\perp})^2}{2m_{\rm Q}} Q_{v} 
    - c(\mu) g_{\mathrm{s}} \bar{Q}_{v} 
    \frac{\sigma_{\mu \nu}G^{\mu \nu}}{4m_{\rm Q}} Q_{v} 
 \right].
 \label{eq:HQET1}
\end{eqnarray}
Because $m_{\rm Q}$ differs for each heavy flavor, the first term does not provide a unit matrix in heavy-flavor symmetry; the heavy-flavor symmetry is violated.
Moreover, the second term is not invariant under both the flavor-symmetry and the spin transformation ($Q_{v} \mapsto SQ_{v}$) 
The latter violation is understood, because $\sigma^{\mu\nu}$ contains Dirac matrices proportional to the Pauli matrices in the rest frame.
The violation of the spin symmetry is known in the spin-interaction by the magnetic moment of an electron in the quantum electrodynamics (QED).
The present case is the version of the non-Abelian gauge theory.
It is worthwhile to note that the spin symmetry violation occurs at the NLO of the heavy mass expansion
 for heavy fermion whenever they couple to gauge fields (either Abelian or non-Abelian).

In the following discussion, we will consider only a single flavor ($N_{\rm h}=1$), and concentrate on the strong interaction, and consider only the spin symmetry.
This is the heavy-quark spin symmetry, or the heavy-quark symmetry (HQS) in short.

Now, let us come back again to the heavy quark limit ($m_{\mathrm{Q}} \rightarrow \infty$), where only the LO term in ${\cal L}_{\rm HQET}$ plays the role.
In this limit, the spin of the heavy quark is a well-conserved quantity.
Let us define the total spin (including simultaneously the angular momenta and spins) for hadrons as
\begin{eqnarray}
 \vec{J} = \vec{S} + \vec{j},
\end{eqnarray}
where $\vec{S}$ is a heavy quark spin (excluding the angular momentum), and $\vec{j}$ is the total spin of the light components ($u$, $d$, $s$ quarks, their antiquarks and gluons).
Because $\vec{J}$ and $\vec{S}$ are conserved quantities, $\vec{j}$ must be also conserved.
Hereafter we call $\vec{j}$ the brown-muck spin.
Originally, the word ``muck", implying  garbage or scattered useless things, has been used, because the structure of the light component is not important, but only the conserved total light spin $j$ is essential for the discussion of heavy quark symmetry~\cite{Neubert:1993mb,Manohar:2000dt}.
In the hadron dynamics at low energy, however, the properties of the brown muck light components turn to be important.  
For example, let us consider a $Q\bar{q}$ meson composed of a heavy quark $Q$ and a light antiquark $\bar{q}$, provided that the light component contains not only a $\bar{q}$ but also a multiple number of light quark-antiquark pairs ($\bar{q}q$) and gluons ($g$).
Hence the $Q\bar{q}$ should be regarded as multi-particle objects whose structure may be denoted by $Q\bar{q}+Q\bar{q}\bar{q}q+Q\bar{q}\bar{q}qg+\dots$ schematically\footnote{Here $q$ stands for the current quark with bare mass, which is different from the constituent quark in the quark model in section~\ref{sec:Quark_model}.}.
We emphasize that, regardless of the complexity of the light components,
 the $\vec{j}$ is a well-conserved quantity, because  $\vec{S}$ is conserved in the heavy quark limit and  $\vec{J}$ is also conserved.
Thus, $\vec{j}$ provides us with a good quantum number for the classification of the non-perturbative object composed of the light components~\cite{Manohar:2000dt,Neubert:1993mb}.

\begin{figure}[tb]
  \begin{center}
   \includegraphics[angle=0,width=60mm]{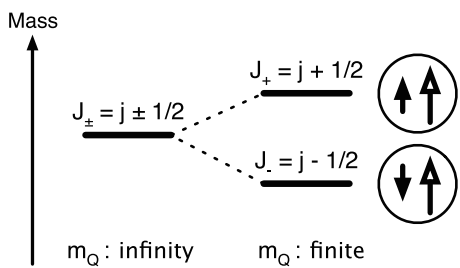}
  \end{center}
 \caption{A schematic picture of the spin degeneracy of states with total spin $J_{\pm} = j \pm 1/2$ in heavy quark mass limit (left) and the small mass splitting in finite heavy quark mass (right). The short and long arrows indicate the spins of heavy quarks and light components, respectively.}
\label{fig:Fig_spin_degeneracy_2}
\end{figure}

As an application of the heavy quark spin-conservation to the hadron spectroscopy, we find that the two states of heavy hadrons with brown-muck spins $j \ge 1/2$ and $J_{\pm}=j\pm1/2$ are degenerate in mass for $j \neq 0$, as shown in Fig.~\ref{fig:Fig_spin_degeneracy_2}.
This is because the spin-dependent interaction is suppressed by $1/m_{\mathrm{Q}}$ due to the HQS.
We will call those two paired states as the HQS doublet.
We note that $j$ takes a half-integer value for mesons, and an integer value for baryons.
For $j=0$ for a baryon, there is only a single state with $J_{+}= 1/2$.
We call this state the HQS singlet.

\begin{figure}[tb]
  \begin{center}
   \includegraphics[angle=0,width=90mm]{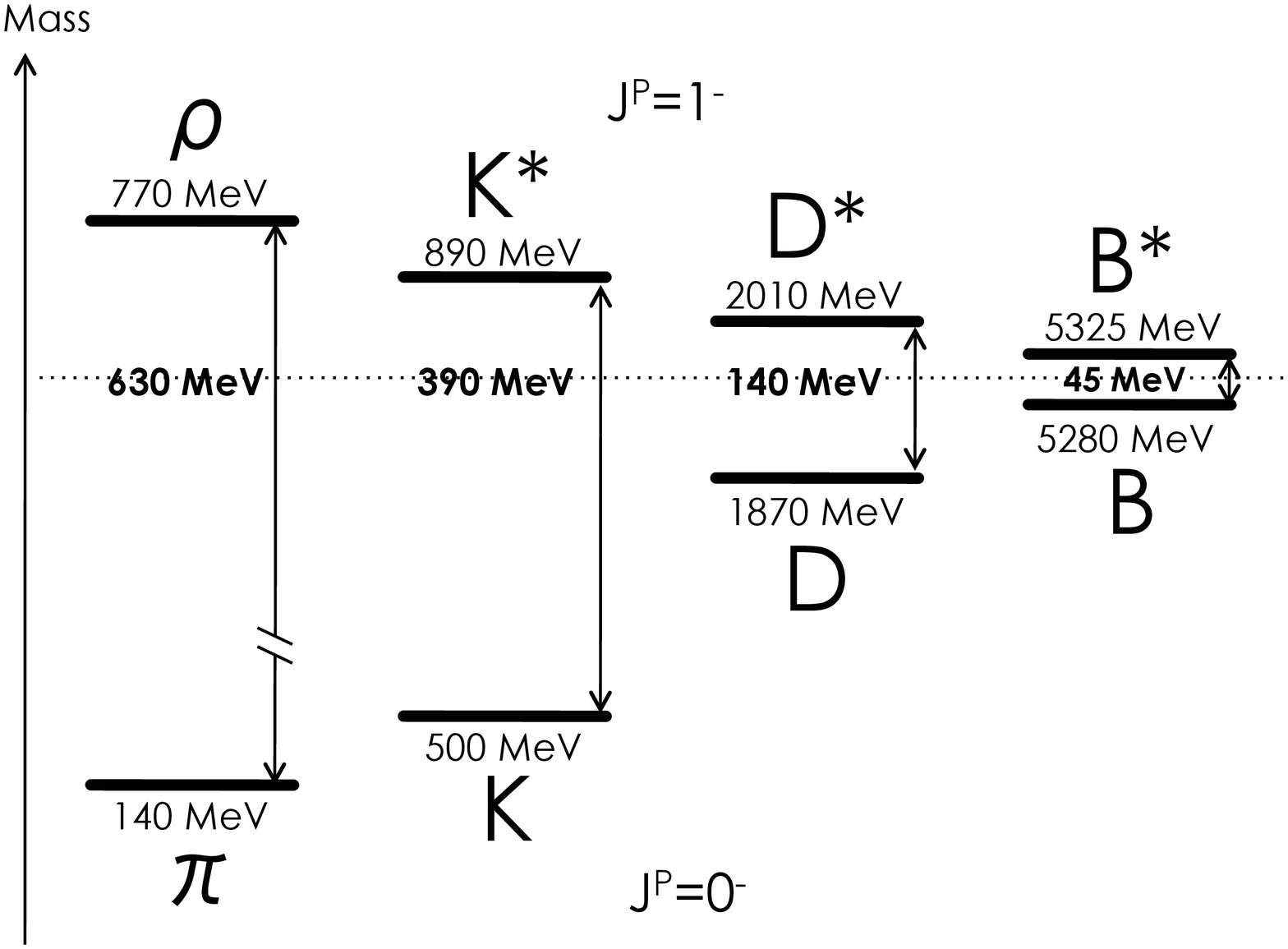}
  \end{center}
 \caption{A schematic picture of the mass splitting between the $J^{P}=0^{-}$ and $1^{-}$ mesons, $\pi$-$\rho$, $K$-$K^{\ast}$, $D$-$D^{\ast}$ and $B$-$B^{\ast}$.}
\label{fig:Fig_mass_splitting}
\end{figure}

Such classifications according to the spin symmetry is found in several heavy hadrons.
For examples, we consider $D$ and $D^{*}$ mesons for charm and $B$ and $B^{*}$ mesons for bottom, whose mass splittings are small; about 140 MeV for the former and about 45 MeV for the latter, as shown in Fig.~\ref{fig:Fig_mass_splitting}.
It may be worthwhile to compare those values with the large mass splittings in light hadrons; about 630 MeV for $\pi$ and $\rho$ mesons, and about 390 MeV for $K$ and $K^{\ast}$ mesons.
Those small mass splittings suggest that $(D^*,D)$ mesons and $(B^*,B)$ mesons, respectively, are approximately degenerate and are regarded as a HQS doublet with $j=1/2$.
Examples for the baryon sector are 
$\Sigma_{c}$ and $\Sigma_{c}^{\ast}$ baryons for charm and $\Sigma_{b}$ and $\Sigma_{b}^{\ast}$ baryons for bottom.
Their mass splittings are about 65 MeV and 20 MeV, respectively\footnote{We note that the corresponding hadrons in light flavors are nucleon and $\Delta$, and $\Sigma$ and $\Sigma^{\ast}$, whose mass splittings are about 290 MeV and 195 MeV, respectively.}.
Those small values suggest that $(\Sigma_{c}, \Sigma_{c}^{\ast})$ baryons and $(\Sigma_{b}, \Sigma_{b}^{\ast})$ baryons, respectively, are approximately degenerate, and hence they are regarded as the HQS doublets with $j=1$.
We note that the ground state baryons, $\Lambda_{c}$ for charm and $\Lambda_{b}$ for bottom, are regarded as the HQS singlet, because the light component would have $j=0$ in the corresponding state in the heavy limit.

\begin{figure}[t]
  \begin{center}
   \includegraphics[angle=0,width=90mm]{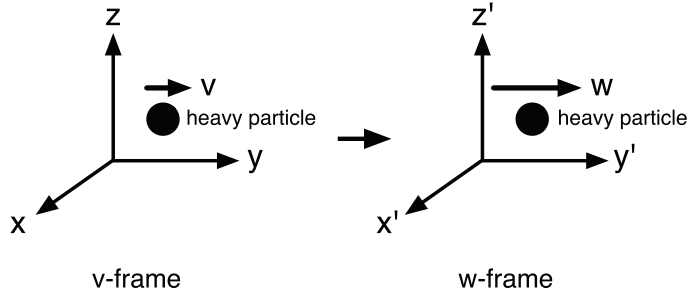}
  \end{center}
 \caption{The Lorentz boost from the $v$-frame to the $w$-frame. In each frame, the heavy particle has four-velocity $v^{\mu}$ and $w^{\mu}$, respectively. The different lengths of the arrows indicate the different spatial components for $v^{\mu}$ and $w^{\mu}$, respectively.}
\label{fig:Fig_velocity_rearrangement}
\end{figure}

Next let us consider the NLO terms in the $1/m_{\rm Q}$ expansion in the HQET Lagrangian (\ref{eq:HQET1}).
The contributions at NLO are roughly estimated as follows.
Employing the typical energy scale $\Lambda_{\rm QCD} \simeq 200$ - $300$ MeV, we find $\Lambda_{\rm QCD}/m_{\rm c}=0.15$ - $0.23$ ($m_{\rm c}=1.3$ GeV) for charm and $\Lambda_{\rm QCD}/m_{\rm b}=0.04$ - $0.06$ ($m_{\rm b}=4.7$ GeV) for bottom.
To discuss a general formulation for $1/M$ corrections, 
let us remember that the coordinate frame with four-velocity $v$ ($v$-frame) 
has been defined for the heavy particle with momentum $p^{\mu} = M v^{\mu} + k^{\mu}$,  
the sum of the on-shell part ($Mv^{\mu}$ with $v^{2}=1$) and the off-shell part ($k^{\mu}$).
The latter is a quantity of order ${\cal O}(\Lambda_{\rm QCD})$ (cf. Eq.~(\ref{eq:momentum})).
An important observation is, however, that those two parts are separated uniquely only in the heavy quark limit.
When the contribution from ${\cal O}(1/M)$ is taken into account, the choice of the frame is not unique.
Applying a small Lorentz boost, 
instead of the $v$-frame, we can consider another frame with four-velocity $w^{\mu}=v^{\mu}+q^{\mu}/M$ ($w$-frame), where $q^{\mu}$ is a small momentum which is much smaller than $M$ and satisfies ${\cal O}(v\cdot q) = q^{2}/M$ (Fig.~\ref{fig:Fig_velocity_rearrangement}).
Here, because of $w^{2}=(v+q/M)^2=v^{2}+{\cal O}(1/M^2)$, $w^{\mu}$ is properly normalized as $w^{2}=1$ up to ${\cal O}(1/M)$.
In the $w$-frame, the momentum $p^{\mu}$ is given by $p^{\mu} = M w^{\mu}+l^{\mu}$ with $l^{\mu}$ being defined by $l^{\mu} = k^{\mu}-q^{\mu}$.
Such transformation is called the ``velocity-rearrangement (VR)"~\cite{Luke:1992cs,Kitazawa:1993bk}.

The velocity-rearrangement can be applied to any heavy particles with an arbitrary spin.
For example, the heavy quark field $Q_{v}(x)$ defined in the $v$-frame
 is related by the Lorentz transformation to $Q_{w}(x)$ 
 in the $w$-frame with $w^{\mu}=v^{\mu}+q^{\mu}/m_{\mathrm{Q}}$. 
Any Lagrangian of the heavy particles has to be constrained by the invariance under the velocity-rearrangement, which may be called the ``velocity-rearrangement invariance (VRI)".
This gives useful constraints in constructing the effective Lagrangians at NLO, as presented explicitly in section~\ref{sec:HHChT}.
In fact, the Lagrangian of the HQET~(\ref{eq:HQET}) is  invariant under the velocity-rearrangement up to ${\cal O}(1/m_Q)$. 

Lastly we comment on the parametrization of the masses of heavy hadrons.
In the $1/m_{\rm Q}$ expansion, the mass of a heavy hadron is given by
\begin{eqnarray}
M_{H} &=& m_{Q} + \bar{\Lambda} - \frac{\lambda_{1}}{2m_{Q}} + 4\vec{S}_{Q} \!\cdot\! \vec{j} \, \frac{\lambda_{2}(\mu)}{2m_{Q}} + \mathcal{O}(1/m_{Q}^2),
\label{eq:mass_formula}
\end{eqnarray}
where we have defined the low-energy parameters, in the rest frame with $v_{\mathrm{r}}=(1,\vec{0}\,)$,
\begin{eqnarray}
 \bar{\Lambda} &=& \frac{1}{2M_{\mathrm{H}}} \langle \tilde{\mathrm{H}}_{v_{\mathrm{r}}} | \frac{\beta(\alpha_{\mathrm{s}})}{4\alpha_{\mathrm{s}}} G^{2} | \tilde{\mathrm{H}}_{v_{\mathrm{r}}} \rangle, 
\label{eq:lambda} \\
 \frac{\lambda_{1}}{m_{Q}} &=& -\langle \mathrm{H}_{v_{\mathrm{r}}} | \bar{Q}_{v_{\mathrm{r}}} g_{\mathrm{s}} \vec{x} \!\cdot\! \vec{E} \, Q_{v_{\mathrm{r}}} | \mathrm{H}_{v_{\mathrm{r}}} \rangle, 
\label{eq:chromoelectric} \\
 8 \vec{S}_{Q} \!\cdot\! \vec{j} \lambda_{2}(\mu) &=& \frac{1}{2} c(\mu) \langle \mathrm{H}_{v_{\mathrm{r}}} | \bar{Q}_{v_{\mathrm{r}}} g_{\mathrm{s}} \vec{\sigma} \!\cdot\! \vec{B} Q_{v_{\mathrm{r}}} | \mathrm{H}_{v_{\mathrm{r}}} \rangle, \label{eq:chromomagnetic}
\end{eqnarray}
with denoting the hadron state by $| \mathrm{H}_{v_{\mathrm{r}}} \rangle$~\cite{Neubert:1993zc,Bigi:1994ga,Bigi:1997fj,Yasui:2013iga} \footnote{In Ref.~\cite{Yasui:2013iga}, $| \tilde{\mathrm{H}}_{v_{\mathrm{r}}} \rangle$ is introduced as the hadron state whose normalization factor is consistent with those in Refs.~\cite{Bigi:1994ga,Bigi:1997fj}.}.
$\vec{S}_{Q}=\vec{\sigma}/2$ and $\vec{j}$ are the spin operators for the heavy quark and the brown muck respectively.
We have also introduced  the Gell-Mann--Low function $\beta(\alpha_{\mathrm{s}})=\mu \mathrm{d}\alpha_{\mathrm{s}}(\mu)/\mathrm{d}\mu$,
the chromoelectric gluon field $E^{i}=-G^{a\,0i}T^{a}$ and the chromomagnetic gluon field $B^{i}=\varepsilon^{ijk}G^{a\,jk}T^{a}$ ($i,j,k=1,2,3$; $a=1,\dots,8$).
In Eq.~(\ref{eq:chromomagnetic}), the Wilson coefficient $c(\mu)$ is determined by the matching to QCD at energy scale $\mu \simeq m_{\mathrm{Q}}$.
The first equation (\ref{eq:lambda}) originates from the scale anomaly in the trace of the energy-momentum tensor in QCD~\cite{Bigi:1994ga,Bigi:1997fj}.
The parametrization (\ref{eq:mass_formula}) holds generally in any hadronic states with a heavy quark, provided that the values $\bar{\Lambda}$, $\lambda_{1}$ and $\lambda_{2}(\mu)$ depend on each hadronic state.

\subsection{Theoretical methods} 
\label{sec:Theoretical_methods}

In this section, we discuss several theoretical methods for analyzing the exotic hadrons;
the quark model (section~\ref{sec:Quark_model}), the heavy-hadron chiral effective theory (section~\ref{sec:HHChT}), the hadronic molecule model (section~\ref{sec:Hadron_model}), the QCD sum rules (section~\ref{sec:QCDSR}) and the lattice QCD simulations (section~\ref{sec:LQCD}).

\subsubsection{Quark model} 
\label{sec:Quark_model}
\ \ \

The quark model was introduced by Gell-Mann and Zweig independently to explain the new particles found around 1960's in addition to the nucleons and pions.  In his famous paper entitled ``A schematic model of baryons and mesons" in 1964~\cite{GellMann:1964nj}, Gell-Mann showed that hadrons are classified systematically by a fewer constituent particles, named quarks ``$q$".  Baryons and mesons are then composed
of $qqq$ and $\bar{q}q$, respectively, and the light hadrons are explained by multiplets of SU(3) flavor symmetry for $u$, $d$ and $s$ quarks.  
In QCD, the SU(3) flavor symmetry is considered to be a consequence of spontaneous breaking of chiral symmetry where 
light current quarks acquire masses approximately similar and of order $\Lambda_{\rm QCD}$.  
In reality the $s$ quark has a heavier mass than the $u$, $d$ quarks.
But the small breaking of flavor symmetry was successfully applied to explain mass splittings among an SU(3) multiplet, which is known as the Gell-Mann--Okubo mass formula.  

Here we emphasize that the ``quark" in the quark model is not the same one as the quark in the QCD Lagrangian~(\ref{eq_LQCD}).
The ``quark" in the former is the constituent particles with dressed masses ($m_{u,d} \simeq 400$ MeV and $m_{s} \simeq 500$ MeV) which are dynamically generated as a result of the spontaneous breaking of chiral symmetry in the QCD vacuum~\footnote{There are studies for dynamical quarks from QCD, see Ref.~\cite{Roberts:1994dr} in the Schwinger-Dyson approach, and in lattice QCD simulation.}.
In this sense, the constituent quarks are composite particles with some structure.

Although the relationship between the quark model and QCD is not clear, it works well in a semi-quantitative manner, not only for the ground, 
but also for the classification of low-lying excited states.
Some of the excited states are described by orbital motions of the constituent quarks and/or by spin-spin interaction, and so on.
Thus, the quark model provides us with a simple picture for the internal structures of hadrons.

The quark model carries important aspects of low energy QCD dynamics, mostly based on flavor and spin symmetries~\cite{Donoghue:1992dd,Hosaka:2001ux}.
Both of them are, however, not exact symmetries of QCD,  but empirically they work well by 
approximate degeneracy of baryon octet states of spin 1/2 and decuplet states of spin 3/2.  
These symmetries may be explained due to spontaneous breaking of chiral symmetry generating constituent masses for $u$, $d$, $s$ quarks with almost equal amount of order $\Lambda_{\mathrm{QCD}}$.  
The quark model is often used also for studies of new hadron states. 
The deviations from quark model predictions often trigger new ideas for dynamics which are not incorporated by the quark model.  

In the quark model the wave functions are given as  products of  color, flavor, spin and orbital parts~\footnote{More details of the discussion are found in Refs.~\cite{Donoghue:1992dd,Nagashima1998,Hosaka:2001ux}.};
\begin{eqnarray}
(\mbox{color}) \times (\mbox{flavor}) \times (\mbox{spin}) \times (\mbox{orbital motion}).
\end{eqnarray}
Let us consider the ground states where all quarks and antiquarks are in the lowest $S$-wave states.
Then, because hadrons are color singlets, there are $3\times2=6$ remaining degrees of freedom by flavor ($u$, $d$, $s$) and spin ($\uparrow$ and $\downarrow$), whose components are denoted by $(u\!\uparrow,u\!\downarrow,d\!\uparrow,d\!\downarrow,s\!\uparrow,s\!\downarrow)$.
An approximate flavor and spin symmetries imply that these six states are regarded as components of the extended 
flavor-spin symmetry SU(6).  

In this symmetry, quarks belong to the fundamental representation ${\bf 6}=(3,1/2)$ (the latter notation means, 3 for flavor triplet and 1/2 for spin doublet) in the SU(6) symmetry.
Then hadron states are given as higher dimensional representations (multiplets) of the SU(6) symmetry.
The mesons are
\begin{eqnarray}
{\bf 6} \times {\bf 6}^{\ast} = {\bf 1} + {\bf 35} = (1,0) + (1,1) + (8,0) + (8,1),
\end{eqnarray}
where ${\bf 6}^{\ast}$ is the complex conjugate of ${\bf 6}$, corresponding to an antiquark.
We obtain the flavor singlet and octet states both for spin 0 and 1.
The baryons are given by
\begin{eqnarray}
{\bf 6} \times {\bf 6} \times {\bf 6} = {\bf 20}_{\rm A} + {\bf 70}_{\rm MA} + {\bf 70}_{\rm MS} + {\bf 56}_{\rm S}.
\end{eqnarray}
However, because the color d.o.f. are anti-symmetric in color singlet states, the flavor and spin d.o.f. must be symmetric and only ${\bf 56}_{\rm S}=(8,1/2)+(10,3/2)$ is allowed for the ground state baryons.
These are the baryons of flavor octet with spin 1/2 and flavor decuplet with spin 3/2, as anticipated.  
As another static property, the magnetic moments of hadrons can be explained.
The excited hadrons with $P$-wave, $D$-wave, $F$-wave and so on are classified according to the $\mbox{SU(6)} \times \mbox{O(3)}$ symmetry with orbital excitations incorporated by $\mbox{O(3)}$.

The above classification of the hadrons may be regarded as the ``taxonomy" of multiplets based on the SU(6) symmetry.
To understand quantitative properties of hadrons, we need further to consider the quark dynamics.
The study by R\'ujula, Georgi and Glashow in their paper ``{\it Hadron masses in gauge theory}" in 1975 played a seminal role to connect the quark model with QCD~\cite{DeRujula:1975ge}.
Considering the one-gluon exchange between two quarks, they proposed a potential of the form
\begin{eqnarray}
V = \alpha_{\rm s} \sum_{i>j} V^{\rm(S)}_{ij} + V^{\rm(L)}.
\label{eq:quark_model}
\end{eqnarray}
The short range part $V^{\rm(S)}_{ij}$ for quarks $i$ and $j$ is given by
\begin{eqnarray}
V^{\rm(S)}_{ij} &=& \vec{F}_{i} \!\cdot\! \vec{F}_{j}  \left[ \frac{1}{r} - \frac{1}{2m_{i}m_{j}} \left( \frac{\vec{p}_{i} \!\cdot\! \vec{p}_{j}}{r} + \frac{(\vec{r} \!\cdot\! \vec{p}_{i})(\vec{r} \!\cdot\! \vec{p}_{j})}{r^{3}} \right)
- \frac{\pi}{2} \delta^{(3)}(r) \left( \frac{1}{m_{i}^{2}} + \frac{1}{m_{j}^{2}} \right) \right. \nonumber \\
&&
- \frac{1}{m_{i}m_{j}} \left( \frac{8\pi}{3} \vec{s}_{i} \!\cdot\! \vec{s}_{j} \delta^{(3)}(r) + \frac{1}{r^{3}} S_{ij}  \right) \nonumber \\
&&
\left.
- \frac{1}{2r^{2}} \left( \frac{1}{m_{i}^{2}} \vec{l}_{i} \!\cdot\! \vec{s}_{i} - \frac{1}{m_{j}^{2}} \vec{l}_{j} \!\cdot\! \vec{s}_{j} + \frac{2}{m_{i}m_{j}} \left( \vec{l}_{i} \!\cdot\! \vec{s}_{i} - \vec{l}_{j} \!\cdot\! \vec{s}_{j} \right) \right)
\right],
\label{eq:QM_potential_short}
\end{eqnarray}
where $\alpha_{\rm s}$ is the strong coupling constant, $m_{i}$ are  quark masses, $\vec{s}_{i}=\vec{\sigma}_{i}/2$  spin operators ($\vec{\sigma}_{i}$ are the Pauli matrices), $\vec{F}_{i}=\vec{\lambda}_{i}/2$  color SU(3) operators ($\vec{\lambda}_{i}$ are the Gell-Mann matrices), $S_{ij} = 3 (\vec{s}_{i} \!\cdot\! \hat{r}) (\vec{s}_{j} \!\cdot\! \hat{r}) - \vec{s}_{i} \!\cdot\! \vec{s}_{j}$ the tensor operators with $\hat{r} = \vec{r}/|\vec{r}\,|$, and $\vec{l}_{i}$ the angular momentum.  
The long range part $V^{\rm(L)}$ is responsible for confinement and is given by a linear potential, such as the Cornel potential~\cite{Donoghue:1992dd}.

Among several terms, the color-spin dependent potential in the short range part
\begin{eqnarray}
V_{{\rm color}-{\rm spin}} = - \frac{8\pi}{3} \sum_{i,j} \frac{\alpha_{\rm s}}{m_{i}m_{j}} \vec{F}_{i} \!\cdot\! \vec{F}_{j} \, \vec{s}_{i} \!\cdot\! \vec{s}_{j} \, \delta^{(3)}(r),
\label{eq:color-spin_interaction}
\end{eqnarray}
plays an important role in the mass splittings in $S$-wave hadrons.
We note that $\vec{F}_{i} \!\cdot\! \vec{F}_{j}$ takes $-4/3$ for $\bar{q}q$ pair (color singlet) and $-2/3$ for $qq$ pair (color-antitriplet)~\footnote{
We define $\vec{F}_{i}=\vec{\lambda}_{i}/2$ and $\vec{\tau}_{i}=\vec{s}_{i}/2$ for the Gell-Mann matrices $\vec{\lambda}_{i}$ and the Pauli matrices $\vec{\sigma}_{i}$.
 We summarize the values of $\vec{\lambda}_{i} \!\cdot\! \vec{\lambda}_{j}$ as
 \begin{eqnarray}
   \vec{\lambda}_{i} \!\cdot\! \vec{\lambda}_{j}=
\left\{
\begin{array}{c}
 -\cfrac{8}{3} \hspace{0.5em} {\rm for} \hspace{0.5em} \bar{\bf 3}_{c} \\
 \cfrac{4}{3} \hspace{0.5em} {\rm for} \hspace{0.5em} {\bf 6}_{c}
\end{array}
\right.
\hspace{0.5em} {\rm in} \hspace{0.5em} {qq,} \hspace{0.5em} {\rm and} \hspace{0.5em}
\left\{
\begin{array}{c}
 -\cfrac{16}{3} \hspace{0.5em} {\rm for} \hspace{0.5em} {\bf 1}_{c} \\
 \cfrac{2}{3} \hspace{0.5em} {\rm for} \hspace{0.5em} {\bf 8}_{c}
\end{array}
\right.
\hspace{0.5em} {\rm in} \hspace{0.5em} q\bar{q}.
\label{eq:lambda_lambda}
 \end{eqnarray}
 and the ones of $\vec{\sigma}_{i} \!\cdot\! \vec{\sigma}_{j}$ as
  \begin{eqnarray}
   \vec{\sigma}_{i} \!\cdot\! \vec{\sigma}_{j}=
\left\{
\begin{array}{c}
 -3 \hspace{0.5em} {\rm for} \hspace{0.5em} S=0 \\
 1 \hspace{0.5em} {\rm for} \hspace{0.5em} S=1
\end{array}
\right..
\label{eq:sigma_sigma}
 \end{eqnarray}
 }.
Similarly, $\vec{s}_{i} \!\cdot\! \vec{s}_{j}$ takes $-3/4$ for spin 0 and $1/4$ for spin 1
 in $\bar{qq}$ ($qq$) pairs.
These spin-dependent interaction orders the masses as $M(S=1)>M(S=0)$ for mesons and $M(S=3/2)>M(S=1/2)$ for baryons.
Moreover, due to the color-dependent part, the mass differences in mesons and baryons are related as
$M(1)-M(0)>M(3/2)-M(1/2)$,
which is consistent with the experimental data.
We  note that in this inequality the color factor is important; 
without  it, the inequality would become reversed to be inconsistent with experimental data.

Next, we consider excited hadrons with various orbital motions.
In this case, we need to consider interactions depending on orbital states, in particular $\vec{l} \cdot \vec{s}$ and tensor interactions as shown in   
Eq.~(\ref{eq:quark_model}). 
Then, 
 the quark model has been shown to reproduce well properties of many excited hadrons only with a few input parameters.

Several comments are in order.
First, the coupling constant $\alpha_{\rm s}$ used in the quark model takes a value closed to one.
Hence, the short range part $V^{\rm(S)}_{ij}$ may not be regarded simply as a result of perturbation of QCD.
Second, the quark mass used in this model is a dynamical mass (constituent mass) which is of order $\Lambda_{\mathrm{QCD}} \sim$ a few hundred MeV.
Those masses are heavier than those of current quarks in the QCD Lagrangian, a few MeV.
Although it is generally believed that the large dynamical mass is due to spontaneous breaking of chiral symmetry, 
the relevance of ``quarks" used in the quark model to ``quarks" in the QCD Lagrangian is not clear yet.
Third, the long range part $V^{(\mathrm{L})}$ indicates  confinement through linearly increasing potential.
Such a behavior has been confirmed by the lattice QCD simulations~\cite{Bali:2000gf}.
The origin of the confinement potential and chiral symmetry breaking has been studied in lattice QCD by analyzing the quark (Dirac) spectral modes~\cite{Gongyo:2012vx}.  
They pointed out that the low energy modes contribute to chiral symmetry breaking while for the confinement higher modes seem to play a role.

In spite of the phenomenological model setting, whose relationship to QCD is not necessarily clear, the quark model has been successfully applied to hadron spectroscopy, 
especially in the non-relativistic quark model~\cite{Isgur:1979be,Isgur:1979wd,Isgur:1978xi,Isgur:1978xj,Isgur:1978xb,Isgur:1977ef}.  
In the present context, we refer to~\cite{Godfrey:1985xj} for charmonium mass spectrum which is used to draw Fig.~\ref{fig_ccbar}.  

However, the quark model has still left unresolved questions in hadron spectroscopy.
For example, a longstanding problem is the mass of $\Lambda(1405)$, which is the lowest state among negative parity baryons, but is overestimated by about 100 MeV in the quark model.  
Thus the  $\Lambda(1405)$ state has been investigated extensively as a candidate of the $\bar KN$ molecule (for a recent review, see~Ref.~\cite{Hyodo:2011ur})~\footnote{$\Lambda(1405)$ as $\bar{K}N$ bound state was in early days suggested by Dalitz and Tuan~\cite{Dalitz:1959dn}.}.
Another example is the mass ordering of scalar mesons.
In the quark model, the scalar mesons are naively considered to be $P$-wave states composed of quark and antiquark ($q\bar{q}$).
As proposed by Jaffe, however, the tetraquark picture ($qq\bar{q}\bar{q}$) can naturally explain the mass ordering, where the $qq=ud$, $ds$, $su$ diquarks are the building-blocks instead of the constituent quarks~\cite{Jaffe:1976ig,Jaffe:1976ih}.
For example, let us consider the light scalar mesons; $f_{0}(600)$, $\kappa(800)$, $f_{0}(980)$ and $a_{0}(980)$.
The mass ordering observed in experiments is
\begin{eqnarray}
M(f_{0}(600)) < M(\kappa(800)) < M(f_{0}(980)) \simeq M(a_{0}(980)).
\label{eq:scalar_meson_obs}
\end{eqnarray}
When two quarks ($q\bar{q}$; quark and antiquark) form the light scalar mesons, the mass ordering is naively given as
\begin{eqnarray}
M(u\bar{u}+d\bar{d}) \simeq M(u\bar{d})  < M(u\bar{s}) < M(s\bar{s}),
\end{eqnarray}
by using isospin symmetry and the larger mass of strange quark larger than that of up and down quarks.
However, this is clearly in contradiction to (\ref{eq:scalar_meson_obs}).
On the other hand, when four quarks with diquark pair ($[qq][\bar{q}\bar{q}]$; diquark $[qq]=[ud], [ds], [su]$ and antidiquark $[\bar{q}\bar{q}]=[\bar{u}\bar{d}], [\bar{d}\bar{s}], [\bar{s}\bar{u}]$ with spin 0 and antitriplet in SU(3) flavor symmetry) are considered as fundamental blocks, the mass ordering becomes
\begin{eqnarray}
M([ud][\bar{u}\bar{d}]) < M([ud][\bar{d}\bar{s}]) < M([su][\bar{s}\bar{u}]+[ds][\bar{d}\bar{s}]) \simeq M([su][\bar{s}\bar{u}]-[ds][\bar{d}\bar{s}]),
\end{eqnarray}
and can naturally explain the mass ordering in (\ref{eq:scalar_meson_obs}).

\begin{figure}[tb]
  \begin{center}
   \includegraphics[angle=0,width=100mm]{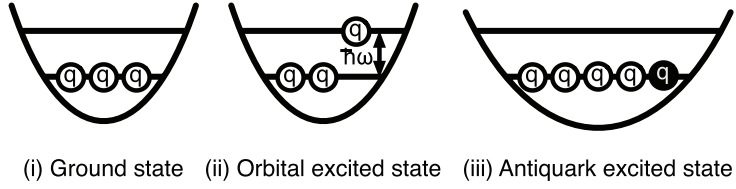}
  \end{center}
 \caption{A schematic picture of the quark model. (i) The ground state. (ii) Orbital excited state. (iii) Antiquark excited state.}
\label{fig:Fig_quark_model}
\end{figure}

Now in the quark model, excitation energies are brought either by orbital motions or by additional $q \bar q$ pair, where the quark ``q" is the constituent quark (Fig.~\ref{fig:Fig_quark_model}).  
In general,  hadrons are superpositions of the minimal quark configuration and other multi-quark configurations, such as $q\bar{q}+qq\bar{q}\bar{q}$~\footnote{
It is interesting to note that Gell-Mann has already suggested 
that $qqqq\bar{q}$ components can exist in addition to the normal $qqq$ in baryons, and $qq\bar{q}\bar{q}$ components could exist in addition to the normal $q\bar{q}$ in mesons~\cite{GellMann:1964nj}.}.
Multi-quark components suggest variety of hadron structures including  
hadronic molecules which are dynamically generated through hadron interactions.  
Whether excited states are described as an orbital excitations or as multi-quarks is one of relevant questions for hadron structure, especially for exotic hadrons, 
which should be eventually solved by QCD.  

Turning to quarkonium states of heavy flavors ($c\bar{c}$ and $b\bar{b}$), an interesting feature 
 is that many low lying excited states have rather small decay widths below the open flavor threshold ($\bar{D}^{(\ast)}D^{(\ast)}$ and $\bar{B}^{(\ast)}B^{(\ast)}$).
The decay modes are indeed restricted due to isosinglet nature of the charmonia and bottomonia.
For example, a single pion emission can occur only by small isospin breaking.
Therefore, the excited 
charmonia and bottomonia are well described by $Q\bar{Q}$ component with either node excitations, orbital excitations 
without additional light (anti)quarks and gluons like $q\bar{q}$, $q\bar{q}g$  below the open flavor threshold.  
This situation is very much different from that of light hadrons.
For instance, we remind that there is an evidence that the light scalar mesons are likely to be $qq\bar{q}\bar{q}$, 
while in the standard quark model, they are assigned as $^{3}P_{0}$ states of $P$-wave excitation.  
This is one of reasons that charmonia and bottomonia brings us with the precise mass spectroscopy. 
Recently, they are investigated by the QCD effective theories, such as the potential non-relativistic QCD (pNRQCD),
where the interaction between heavy quarks is introduced by the velocity-expansion~\cite{Brambilla:2004jw}~\footnote{It is important to distinguish the pNRQCD from the HQET (section~\ref{sec:HQS}). In the former, the relative motion of the two heavy quarks is given in the velocity-expansion, and, in the latter, the hadron dynamics is treated by the $1/m_{\mathrm{Q}}$-expansion in the given coordinate frame.}.
This formalism enables us to relate systematically the phenomenological quark model for quarkonia to QCD.  

Recent experimental discoveries of $XYZ$ exotic hadrons, however, have opened the question about the applicability of the quark model around and above the thresholds.
In addition to the conventional $Q\bar{Q}$ picture, a multi-quark  and/or the hadronic molecule configurations should be taken into account.  
These are the issues which will be discussed in the followings.

\subsubsection{Heavy hadron chiral effective theory} 
\label{sec:HHChT}
\ \ \

In section~\ref{sec:HQS}, we have discussed that the spin of a heavy quark is conserved in the heavy quark limit, and the total spin of the brown muck, namely the light component in the heavy hadron, is also conserved.
In this subsection, we  discuss the heavy hadrons effective theory in terms of the heavy quark symmetry for the heavy quark and chiral symmetry for the brown muck, the light degrees of freedom.
In order to present the concrete form of the equations, we concentrate on the heavy meson effective theory (HMET)~\cite{Casalbuoni:1996pg,Manohar:2000dt}.
The basic idea will be straightforwardly applicable to the heavy baryon effective theory.

We define the effective field $H_{\gamma}$ for the heavy meson with spin $\gamma$, which is composed by the heavy quark ($Q_{\alpha}$) with spin 1/2 ($\alpha= \pm 1/2$) and the brown muck ($\bar{q}_{\beta}$) with a spin component $\beta$ as introduced in section~\ref{sec:HQS}.
This is schematically expressed as
\begin{eqnarray}
H_{\gamma} = \sum_{\alpha,\beta} C_{\alpha \beta; \gamma} Q_{\alpha} \bar{q}_{\beta},
\label{eq:CG_simple}
\end{eqnarray}
where the sum is taken over the spin combinations with appropriate Clebsh-Goldan coefficients, denoted by $C_{\alpha\beta;\gamma}$.
For example, we consider the mesons composed of a heavy quark $Q$ and a light anti-quark $\bar{q}$ with conserved spin and parity $j^{\cal P}=1/2^{-}$, baryon number $-1/3$ and color antitriplet, which are denoted by $P_{v}(x)$ with spin 0 and $P_{v}^{\ast \, \mu}(x)$ with spin 1.
Here $v$ is the conserved four-velocity of the heavy meson, which is regarded as that of the heavy quark in the heavy quark limit.
Moreover, in this limit $P_{v}(x)$ and $P_{v}^{\ast \, \mu}(x)$ are degenerate in mass, and they are treated as heavy quark spin multiplets
\begin{eqnarray}
H_{v}(x) = \frac{1+v\hspace{-0.5em}/}{2} \left( P^{\ast}_{v}\hspace{-1em}/\hspace{0.5em}(x) + i \gamma_{5} P_{v}(x) \right),
\label{eq:Hv}
\end{eqnarray}
where $(1+v\hspace{-0.5em}/)/2$ is the projection operator 
as introduced in Eq.~(\ref{eq:heavy_quark_effective_field_1}) 
for the positive-energy state of the heavy quark $Q$
in the $Q \bar q$ meson.
More explicitly, the meson fields $P_{v}^{\ast\,\mu}$ and $P_{v}$ have the matrix structure
\begin{eqnarray}
P_{v}^{\ast\,\mu} \sim Q_{v}\gamma^{\mu}\bar{q}, \hspace{1em}
P_{v} \sim Q_{v} \gamma_{5} \bar{q},
\end{eqnarray}
where $Q_{v}$ is the heavy quark field in the $v$-frame as defined by Eq.~(\ref{eq:heavy_quark_effective_field_1}) and $\bar{q}$ is the brown muck field.
The transformation properties are important in the construction of the effective Lagrangian.
The heavy-quark-spin for the heavy quark and chiral transformation for the brown muck leads to the change of the field as
\begin{eqnarray}
H_{v}(x) \rightarrow S H_{v}(x) U_{V}^{\dag},
\label{eq:Hv_transformation}
\end{eqnarray}
where $S$ is the heavy-quark-spin operator and $U_{V}$ is the one for chiral transformation.
The effective Lagrangian for heavy hadrons is constructed order by order according to both heavy quark mass expansion and chiral order expansion.

First, we consider the leading order in both of them.
For the effective field $H_{v}(x)$ defined in Eq.~(\ref{eq:Hv}),
the effective Lagrangian at leading-order (LO) is given as
\begin{eqnarray}
{\cal L}_{\rm{HMET}}^{{\rm LO}} = - {\rm Tr}\, \bar{H}_{v}(x) i v\!\cdot\! D H_{v}(x) + g \, {\rm Tr} \, \bar{H}_{v}(x) H_{v}(x) \gamma_{\mu} \gamma_{5} A^{\mu}(x),
\label{eq:L_HMET_LO}
\end{eqnarray}
where the trace is taken over the Dirac matrices (see Eq.~(\ref{eq:Hv})), $D^{\mu}$ is the chiral covariant derivative, $D^{\mu} = \partial^{\mu} - V^{\mu}(x)$, and $V^{\mu}(x)$ and $A^{\mu}(x)$ are the vector and axial-vector currents of pions, respectively~\cite{Neubert:1993mb,Casalbuoni:1996pg,Manohar:2000dt}.
The coupling constant $g$ is a free parameter which should be determined from experimental data or by the non-perturbative methods for QCD.
We can verify that Eq.~(\ref{eq:L_HMET_LO}) is invariant under the transformation (\ref{eq:Hv_transformation}).

Second, we consider the effective Lagrangian at the next-to-leading order (NLO).
To include the $1/M$ corrections with $M$ being the mass of the heavy particle,
 we recall the velocity-rearrangement invariance in section~\ref{sec:HQS}.
In this scheme, we define the effective fields $H_{v}(x)$ and $H_{w}(x)$ in the two different frames with four-velocities $v$ and $w$.
Because the Lorentz boost between the $v$-frame and the $w$-frame is taken into account up to ${\cal O}(1/M)$,
the Lorentz transformation between $H_{v}(x)$ and $H_{w}(x)$ is also taken into account up to this order.
Then, we construct the effective Lagrangian invariant under the velocity-rearrangement. 
However, it is not easy to find out practically the velocity-rearrangement invariant terms, because the form of the Lorentz transformation between $H_{v}(x)$ and $H_{w}(x)$ is not so simple in general.
For $Q\bar{q}$ mesons, for example, the heavy field $H_{v}(x)$ and $H_{w}(x)$ are related as~\cite{Luke:1992cs,Kitazawa:1993bk}
\begin{eqnarray}
H_{v}(x) = \left( H_{w}(x) - \frac{1}{2M} \left[ q\hspace{-0.5em}/,H_{w}(x) \right] \right) e^{-iq\cdot x} + {\cal O}(1/M^2),
\end{eqnarray}
with $q/M=w-v$.
There are, not only the phase $e^{-iq\cdot x}$, but also the additional term in the parentheses, in which the latter causes the difficulty in finding the invariant effective Lagrangian.
To avoid this difficulty, instead of $H_{v}(x)$, we introduce the modified field as~\cite{Luke:1992cs,Kitazawa:1993bk,Yasui:2013iga}
\begin{eqnarray}
{\mathcal H}_{v} &=& H_{v} + \frac{1}{2M} \left( i\vecl{D}\hspace{-0.7em}/ H_{v} - H_{v} i\cev{D}\hspace{-0.7em}/ - 2v\!\cdot \!iD H_{v} \right) + \mathcal{O}(1/M^2),
\end{eqnarray}
which is given by the Lorentz boost from the $v$-frame to $(v+q/M)$-frame up to ${\cal O}(1/M)$, where $q^{\mu}$ is supposed to be the residual momentum in $H_{v}(x)$.
This has an interesting property
\begin{eqnarray}
{\mathcal H}_{v} = e^{-i q\cdot x} {\mathcal H}_{w} + \mathcal{O}(1/M^2),
\end{eqnarray}
namely a change only by the phase factor $e^{-i q\cdot x}$. 
This enables us to construct easily the effective Lagrangian with the velocity-rearrangement invariance~\cite{Luke:1992cs,Kitazawa:1993bk}.
Another $1/M$ correction is provided by the heavy-quark spin violation,
independent of the velocity-rearrangement.

Including all the $1/M$ contributions, we obtain the form of the effective Lagrangian up to ${\cal O}(1/M)$~\cite{Luke:1992cs,Kitazawa:1993bk,Yasui:2013iga}
\begin{eqnarray}
{\mathcal L}_{\mathrm{HMET}}^{{\rm LO}+{\rm NLO}}  
 &=& - {\mathrm{Tr}}\, \overline{H}_{v} v \!\cdot\! iD H_{v} - {\mathrm{Tr}}\, \overline{H}_{v} \frac{(iD)^2}{2M} H_{v} \nonumber + \frac{\lambda}{M} \mathrm{Tr}\, \overline{H}_{v} \sigma^{\mu\nu} H_{v} \sigma_{\mu\nu} \\
&&+ \left( g+\frac{g_1}{M} \right) {\mathrm{Tr}}\, \overline{H}_{v} H_{v} \gamma_{\mu} \gamma_{5} {\cal A}^{\mu} \nonumber \\
&&+ \frac{g}{2M} \left( {\mathrm{Tr}}\, v \!\cdot\! iD \overline{H}_{v} H_{v} \gamma_{\mu} \gamma_{5} {\cal A}^{\mu} - {\mathrm{Tr}}\,  \overline{H}_{v} v \!\cdot\! iD H_{v} \gamma_{\mu} \gamma_{5} {\cal A}^{\mu} \right) \nonumber \\
&&+ \frac{g}{4M} \varepsilon_{\mu\nu\rho\sigma} \left( {\mathrm{Tr}}\, iD^{\nu} \overline{H}_{v}H_{v} \sigma^{\rho\sigma} {\cal A}^{\mu} - {\mathrm{Tr}}\, \overline{H}_{v} iD^{\nu} H_{v} \sigma^{\rho\sigma} {\cal A}^{\mu} \right) \nonumber \\
&& + \frac{g_2}{M} {\mathrm{Tr}}\, \overline{H}_{v} \gamma_{\mu} \gamma_{5} H_{v} {\cal A}^{\mu} + {\mathcal O}(1/M^2),
\label{eq:Lagrangian_HMET_1/M}
\end{eqnarray}
where $\lambda$, $g_{1}$ and $g_{2}$ are new coupling constants at ${\cal O}(1/M)$\footnote{We consider only the contributions to the axial-vector coupling. For the vector-coupling, see Ref.~\cite{Kitazawa:1993bk}.}.
We note that fifth and sixth terms in the right-hand side are proportional to $g$.
It indicates that they are connected with the terms at LO due to the velocity-rearrangement invariance.
We note, however, that this is not the case in general.
For instance,
the terms proportional to $g_{1}$ and $g_{2}$ are not constrained by the velocity-rearrangement invariance.
More importantly, the term proportional to $g_{2}$ breaks the heavy-quark spin symmetry, due to 
the factor $\gamma_{\mu} \gamma_{5}$ which plays the role of the spin for the heavy $Q$.
Therefore, we have the heavy-quark-spin conserving part ($g$ and $g_{1}$) and the heavy-quark-spin non-conserving part ($g_{2}$).

Lastly we comment on the relation between the heavy hadron effective theory, such as the HMET, and the heavy-quark effective theory (HQET).
It is interesting to note that the $1/M$-expansion for heavy hadrons has a correspondence to the $1/m_{\rm Q}$-expansion for heavy quarks, when we retain the expansions up to NLO~\cite{Cheng:1993gc}.
This is because the relation $M=m_{\rm Q}+\Lambda$ ($\Lambda$ the energy from the light d.o.f., an order of $\Lambda_{\rm QCD}$) is inverted to
$
1/M = 1/m_{\rm Q} + {\cal O}(1/m_{\rm Q}^{2})
$.
From the HQET Lagrangian (\ref{eq:HQET_Lagrangian}), we remind us that there are two terms at ${\cal O}(1/m_{\rm Q})$; the heavy-quark-spin conserving and non-conserving terms\footnote{They are the second and third terms, respectively, in the last line in Eq.~(\ref{eq:HQET_Lagrangian}).}. 
It was shown in the HQET that, at ${\cal O}(1/m_{\rm Q})$, the former is related with color-electric gluon fields, and the latter with color-magnetic gluons~\cite{Bigi:1994ga,Bigi:1997fj,Neubert:1993zc} (cf. section~\ref{sec:HQS}).
Thus, the terms at ${\cal O}(1/M)$ in heavy hadrons are also related to the color-electric and -magnetic gluons.
This property is useful to study the role of the gluon dynamics in the vacuum and in finite temperature and density medium~\cite{Yasui:2013iga}.

\subsubsection{Hadronic molecule model} 
\label{sec:Hadron_model}
\ \ \

\begin{figure}[tb]
  \begin{center}
   \includegraphics[angle=0,width=50mm]{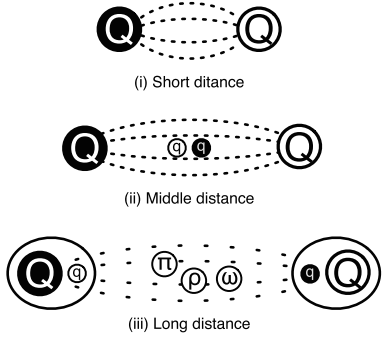}
  \end{center}
 \caption{A schematic picture for interaction between heavy quarks. (i) At short distance, gluon exchange is a dominant force. (ii) At middle distance, light quark-antiquark pairs appear as effective degrees of freedom around threshold energy. (iii) At long distance, the constituent becomes heavy hadrons and the meson exchange becomes a dominant force.}
\label{fig:Fig_interaction_heavy_quark}
\end{figure}

The quarkonium is a simple system composed of a heavy quark and a heavy antiquark.
Let us consider how the interquark interaction changes at different distance scales (Fig.~\ref{fig:Fig_interaction_heavy_quark}).
We have seen that the gluon exchange is a dominant force at short distances, as shown in Eq.~(\ref{eq:QM_potential_short}).
When the distance between heavy quarks become larger, light (constituent) quark-antiquark pairs can be created virtually from vacuum, and the light quark components can appear as effective degrees of freedom around the thresholds.
One can treat light components by hadron models, where the open heavy-hadrons are regarded as the effective degrees of freedom.
They are, for instance, $D$ and $\bar D$ mesons and their interactions are provided by meson exchanges.  
Here we discuss in detail the one-pion-exchange force as the most dominant interaction at long distances.
It is straightforward to include further massive mesons such as scalar ($\sigma$) and vector ($\rho$, $\omega$) mesons~\cite{Yamaguchi:2011xb,Yamaguchi:2011qw}.

{Let us consider the $P^{(\ast)}\bar{P}^{(\ast)}$ systems, where $P$ and $P^{\ast}$ ($\bar{P}$ and $\bar{P}^{\ast}$) are $(Q\bar{q})_{\rm spin \, 0}$ and $(Q\bar{q})_{\rm spin \, 1}$ mesons ($(\bar{Q}q)_{\rm spin \, 0}$ and $(\bar{Q}q)_{\rm spin \, 1}$) with $S$-wave.
The interaction Lagrangian for $P^{(\ast)}$ meson with axial-vector current induced by pions was constructed in section~\ref{sec:HHChT}.
It naturally gives the interaction vertices for pions and $P^{(\ast)}$ mesons.
From Eq.~(\ref{eq:L_HMET_LO}), we obtain the $\pi PP^*$ and $\pi P^* P^*$
vertices
\begin{eqnarray}
 {\cal L}_{\pi PP^*} &=& 
 2 \frac{g}{f_\pi}(P^\dagger_a P^\ast_{b\,\mu}+P^{\ast\,\dagger}_{a\,\mu}P_b)\partial^\mu\hat{\pi}_{ab} \, , 
 \label{eq:piPP*}\\
 {\cal L}_{\pi P^*P^*} &=& 
 2 i \frac{g}{f_\pi}\epsilon^{\alpha \beta \mu \nu} v_{\alpha}
 P^{\ast\,\dagger}_{a\,\beta}P^\ast_{b\,\mu}\partial_{\nu}
 \hat{\pi}_{ab} \, .
\label{eq:piP*P*}
\end{eqnarray}
The $\pi \bar{P} \bar{P}^{\ast}$ and $\pi \bar{P}^{\ast} \bar{P}^{\ast}$
vertices are obtained by changing the sign of the $\pi P P^{\ast}$ and
$\pi P^{\ast} P^{\ast}$ vertices in Eqs.~(\ref{eq:piPP*}) and
(\ref{eq:piP*P*}). 
It is important to note that the strength of the interaction vertex is the same for $\pi P P^{\ast}$ and $\pi P^{\ast} P^{\ast}$.
This is the consequence of the heavy quark symmetry.

The one-pion-exchange potential (OPEP) for 
$P\bar{P}^{\ast} \rightarrow
P^{\ast}\bar{P}$ and
$P^{\ast}\bar{P}^{\ast} \rightarrow
P^{\ast}\bar{P}^{\ast}$ is given from the vertices (\ref{eq:piPP*}) and (\ref{eq:piP*P*}) as 
\begin{eqnarray}
V^{\pi}_{P_{1}\bar{P}_{2}^{\ast} \rightarrow P_{1}^{\ast}\bar{P}_{2}} &\!=\!&
 -\left( \sqrt{2} \frac{g}{f_{\pi}} \right)^{2} \frac{1}{3} \left[ \vec{\varepsilon}_{1}^{\,\ast} \!\cdot\! \vec{\varepsilon}_{2} \, C_{\pi}(r) \!+\! S_{\varepsilon_{1}^{\ast},\varepsilon_{2}} \, T_{\pi}(r) \right]  \vec{\tau}_{1}  \!\cdot\! \vec{\tau}_{2}, \label{eq:pot_BB*B*B}  \\
V^{\pi}_{P_{1}^{\ast}\bar{P}_{2}^{\ast} \rightarrow P_{1}^{\ast}\bar{P}_{2}^{\ast}} &\!=\!&
 -\left( \sqrt{2} \frac{g}{f_{\pi}} \right)^{2} \frac{1}{3} \left[ \vec{T}_{1} \!\cdot\! \vec{T}_{2} \, C_{\pi}(r) \!+\! S_{T_{1},T_{2}} \, T_{\pi}(r) \right]  \vec{\tau}_{1}  \!\cdot\! \vec{\tau}_{2}, \label{eq:pot_B*B*B*B*}
\end{eqnarray}
and the OPEP for  $P\bar{P} \rightarrow P^{\ast}\bar{P}^{\ast}$ and $P\bar{P}^{\ast} \rightarrow P^{\ast}\bar{P}^{\ast}$ is as
\begin{eqnarray}
V^{\pi}_{P_{1}\bar{P}_{2} \rightarrow P_{1}^{\ast}\bar{P}_{2}^{\ast}} &\!=\!&
 -\left( \sqrt{2} \frac{g}{f_{\pi}} \right)^{2} \frac{1}{3} \left[ \vec{\varepsilon}_{1}^{\,\ast} \!\cdot\! \vec{\varepsilon}_{2}^{\, \ast} \, C_{\pi}(r) \!+\! S_{\varepsilon_{1}^{\ast},\varepsilon_{2}^{\ast}} \, T_{\pi}(r) \right] \vec{\tau}_{1}  \!\cdot\! \vec{\tau}_{2}, \label{eq:pot_BBB*B*} \\
V^{\pi}_{P_{1}\bar{P}_{2}^{\ast} \rightarrow P_{1}^{\ast}\bar{P}_{2}^{\ast}} &\!=\!&
 \left( \sqrt{2} \frac{g}{f_{\pi}} \right)^{2} \frac{1}{3} \left[ \vec{\varepsilon}_{1}^{\,\ast} \!\cdot\! \vec{T}_{2} \, C_{\pi}(r) \!+\! S_{\varepsilon_{1}^{\ast},T_{2}} \, T_{\pi}(r) \right]  \vec{\tau}_{1}  \!\cdot\! \vec{\tau}_{2}. \label{eq:pot_BB*B*B*}
\end{eqnarray}
Here three polarizations are possible for $P^{*}$ as defined by
$\vec{\varepsilon}^{\hspace{0.2em}(\pm)} \!=\! \left(\mp 1/\sqrt{2}, \pm i/\sqrt{2}, 0 \right)$ and
$\vec{\varepsilon}^{\hspace{0.2em}(0)} \!=\! \left(0, 0, 1\right)$,
and the spin-one operator $\vec{T}$ is defined by $T_{\lambda' \lambda}^{i}=i \varepsilon^{ijk} \varepsilon_{j}^{(\lambda')\dag} \varepsilon_{k}^{(\lambda)}$.
As a convention, we assign $\vec{\varepsilon}^{\,(\lambda)}$ for an
incoming vector particle and $\vec{\varepsilon}^{\,(\lambda)\ast}$ for
an outgoing vector particle.
The tensor operators are defined by
\begin{eqnarray}
S_{\varepsilon_{1}^{\ast},\varepsilon_{2}} &=& 3 ( \vec{\varepsilon}^{\,(\lambda_{1})\ast} \!\cdot\!\hat{r} ) ( \vec{\varepsilon}^{\,(\lambda_{2})} \!\cdot\!\hat{r} ) -  \vec{\varepsilon}^{\,(\lambda_{1})\ast} \!\cdot\! \vec{\varepsilon}^{\,(\lambda_{2})}, \\
S_{T_{1},T_{2}} &=& 3 ( \vec{T}_{1} \!\cdot\!\hat{r} ) ( \vec{T}_{2} \!\cdot\!\hat{r} ) - \vec{T}_{1} \!\cdot\! \vec{T}_{2}, \\
S_{\varepsilon_{1}^{\ast},\varepsilon_{2}^{\ast}} &=& 3 ( \vec{\varepsilon}^{\,(\lambda_{1})\ast} \!\cdot\!\hat{r} ) ( \vec{\varepsilon}^{\,(\lambda_{2})\ast} \!\cdot\!\hat{r} ) -  \vec{\varepsilon}^{\,(\lambda_{1})\ast} \!\cdot\! \vec{\varepsilon}^{\,(\lambda_{2})\ast}, \\
S_{\varepsilon_{1}^{\ast},T_{2}} &=& 3 ( \vec{\varepsilon}^{\,(\lambda_{1})\ast} \!\cdot\!\hat{r} ) ( \vec{T}_{2} \!\cdot\!\hat{r} ) -  \vec{\varepsilon}^{\,(\lambda_{1})\ast} \!\cdot\! \vec{T}_{2}.
\end{eqnarray} 
The functions $C_{\pi}(r)$ and $T_{\pi}(r)$ in the central and tensor potentials, respectively, have asymptotic behaviors at long distance;
\begin{eqnarray}
C_{\pi}(r) &\simeq& \frac{1}{r} e^{-m_{\pi}r}, \label{eq:C_function} \\
T_{\pi}(r) &\simeq& \left( 1 + \frac{3}{m_{\pi}r} + \frac{3}{(m_{\pi}r)^{2}} \right) \frac{1}{r} e^{-m_{\pi}r}.
\label{eq:T_function}
\end{eqnarray}
They are modified by form factor at short distance~\cite{Yasui:2009bz,Yamaguchi:2011xb,Yamaguchi:2011qw}.
$\vec{\tau}_{1}$ and $\vec{\tau}_{2}$ are isospin operators for $P^{(\ast)}_{1}$ and $\bar{P}^{(\ast)}_{2}$. 

To clarify the role of the OPEP, as an example,
we consider $J^{PC}=0^{+-}$ state shown in Fig.~\ref{fig:Fig_meson_meson_int}.
In the figure, we show $P\bar{P}(^{1}S_{0})$ for the initial state, $P^{\ast}\bar{P}^{\ast}(^{5}D_{0})$ for the intermediate state, and $P\bar{P}(^{1}S_{0})$ for  the final state\footnote{To be precise, more channels can couple to the state of this quantum number. See Table~\ref{tbl:BB_channels} for the $B^{(\ast)}\bar{B}^{(\ast)}$ mesons case.}.
We note that the spins of the heavy (anti)quarks $Q$ and $\bar{Q}$ are not changed in the heavy quark limit, while the spins of light quarks can be changed. 
We therefore concentrate on the alignment of spins of light (anti)quarks $\bar{q}$ and $q$ (called the light spin). 
In the process in Fig.~\ref{fig:Fig_meson_meson_int}, the light spins and orbital angular momentum are changed by the  tensor force from the OPEP in such a way that $P\bar{P}(^{1}S_{0}) \leftrightarrow P^{\ast}\bar{P}^{\ast}(^{5}D_{0})$.
The tensor force in the OPEP provides a strong attraction.
We note that, in order to switch on the tensor force, $P$ and $P^{\ast}$ (or $\bar{P}$ and $\bar{P}^{\ast}$) must be mixed in the process.
In the present case, the mass difference between $P$ and $P^{\ast}$ (or $\bar{P}$ and $\bar{P}^{\ast}$) are zero in the heavy quark limit, and hence they can be energetically easily mixed.
Thus, the $P$-$P^{\ast}$ ($\bar{P}$-$\bar{P}^{\ast}$) mixing accompanying with the OPEP is important in this system.

We emphasize that the $P$-$P^{\ast}$ ($\bar{P}$-$\bar{P}^{\ast}$) mixing is a characteristic phenomena seen in the heavy quark systems.
In the light quark systems, the mixing between the pseudoscalar meson and the vector meson is not important, because their mass splittings are larger, as presented Fig.~\ref{fig:Fig_mass_splitting}.
We also note that the $P$-$P^{\ast}$ ($\bar{P}$-$\bar{P}^{\ast}$) mixing is realized by the OPEP with strong tensor force.
The existence of the long range force in the heavy systems may be contrasted with that the interaction among the light Nambu-Goldstone bosons are supplied by the short range force, such as the Tomozawa-Weinberg interaction~\cite{Tomozawa:1993aa,Weinberg:1966kf}.
Thus, the simultaneous appearance of the heavy-quark-spin symmetry and the OPEP is a unique feature in the heavy quark systems.

In literature, there are pioneering works based on the hadron molecules where the importance of the pion exchange was pointed out for meson-meson systems \cite{Tornqvist:1991ks,Tornqvist:1993ng}.
It was presented that the S-wave and D-wave mixing in the tensor force plays an important role to induce the strong attraction to make the systems bound, as in the case of the deuteron.
Such meson-meson systems were called ``deuson".
It is important to note that the existence of $D\bar{D}^{\ast}$ molecules was predicted in Ref.~\cite{Tornqvist:1993ng}, whose mass is very close to X(3872).
In the present work, the importance of the OPEP is discussed for 
$Z_b(10610)$ and $Z_b(10650)$ in section~\ref{sec:Examples}.

\begin{figure}
\begin{center}
\resizebox{0.3\textwidth}{!}{\rotatebox{0}{\includegraphics{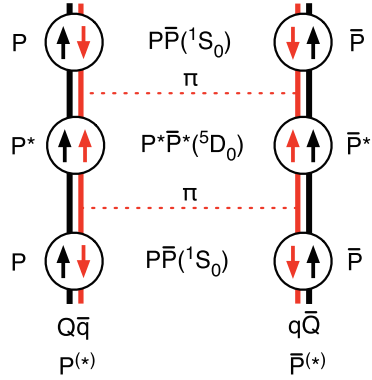}}}
\caption{A schematic picture of an interaction by one-pion exchange potential between $P^{(\ast)}$ and $\bar{P}^{(\ast)}$ mesons in $J^{PC}=0^{+-}$ (cf. Table~\ref{tbl:BB_channels} for $B^{(\ast)}\bar{B}^{(\ast)}$ mesons). The up and down spins for light component (red) and heavy component (black) are denoted in the circles.}
\label{fig:Fig_meson_meson_int}
\end{center}
\end{figure}

For more realistic studies, we should also consider short range forces by heavier meson exchanges like scalar and vector mesons, 
or by short distance mechanisms such as quark exchanges.  
For example, the $\omega$ meson exchange contributes as an attraction for $P^{(\ast)}\bar{P}^{(\ast)}$.
It has been argued that such contributions could be minor in comparison to the strong tensor force by the OPEP~\cite{Yamaguchi:2011xb,Yamaguchi:2011qw}.
On the other hand, there are opposite discussions which assert that the contributions from the OPEP is rather small~\cite{Nieves:2012tt}.
To analyze the interplays by long and short range forces will be left for future works.

\subsubsection{QCD sum rules} 
\label{sec:QCDSR}
\ \ \

In the following two sections, we would like to briefly discuss two theoretical methods, 
QCD sum rule and lattice simulations.  
Both of them are based on QCD to extract hadron properties directly.   
So far, their applicability especially to exotic hadrons is rather limited.   
However, they provide useful information on their properties.  

The idea of the QCD sum rule is to extract physical quantities by comparing  
the correlation function computed in two different 
methods~\cite{Shifman:1978bx,Shifman:1978by,Reinders:1984sr} .
In the QCD side, one calculates
a correlation function of currents 
\be
\Pi (p) = \int d^4x\ e^{ipx} \bra 0| T(A(x) \bar B(0)) |0\ket,
 \label{eq_def_correlationAB}
 \ee
in the asymptotic (deep Euclidean) region, where the perturbation method is available. 
The operators $A$ and $B$ are the currents which  create 
physical states  that we are interested  in.  
For instance,  the $\rho$ meson operator can be chosen as 
$A, B = \bar q \gamma_\mu \vec \tau q$, where $\vec \tau$ is an isospin matrix.  
The computation can be done by using the Wilson's operator product expansion (OPE) at short distances~\cite{Wilson:1969zs}, 
\be
\lim_{x \to y} A(x) \bar B(y)  \to \sum_i C_i (x-y) {\cal O}_i ((x+y)/2)  , 
\label{eq_WilsonsOPE}
\ee
where ${\cal O}_i$ are local operators and $C_i (x-y) $
the Wilson's coefficients  which can be calculated perturbatively.  
The indices summed over $i$ are  ordered by the dimensions of the  operators ${\cal O}_i$.  
Examples of such operators for hadrons written in terms of the quark and gluon fields 
are given in Ref.~\cite{Reinders:1984sr}.
In this decomposition, the singularities at short distances, $x \to y$ are isolated by the 
Wilson coefficients $C_i (x-y)$, and the operators ${\cal O}_i$ are regular.  
After taking the vacuum matrix elements, due to translational invariance, 
they reduce to constant values which characterizes the non-perturbative nature of the QCD vacuum
at long distances.

Now it is known that the asymptotic behaviors of 
$C_i$'s are determined up to logarithmic factor by the canonical 
dimensions of the operators 
$d_A, d_B$ and $d_i$, 
\be
C_i(x-y) 
\xrightarrow[|x-y| \ll 1/\Lambda]{}
|x-y|^{d_i - d_A - d_B} 
(1 + {\cal O}(x\Lambda)) .
\ee
For instance, for $A = \bar \psi \gamma_\mu \psi$, $d_A=3$.  
Therefore, the terms of higher dimensional operators (with larger $d_i$) 
are expected to be suppressed in the deep Euclidean region, $x \to y$ or $p^2 \to \infty$,  
by powers of $\Lambda_{\rm QCD}^2/p^2$.  
It is known that the OPE may break due to  non-perturbative effects 
at short distances, such as instantons~\cite{Shifman:1978bx}.  
This occurs at a dimension as high as $d=12$.  
In practice, the computation up to such high dimensions are not very easy.  

In the phenomenological side, using the analyticity, the correlation function is expressed  
by a  spectral function which is related to observables of hadrons.  
In momentum space, it takes the form
\be
\Pi (p) 
=
\int^\infty_{0} ds\ \frac{\rho(s)}{s-p^2 - i \epsilon}\, .
\label{Pi_phenomenological}
\ee
In general the integral is contributed from bound states, resonant states and continuum.  
Usually, a parametrization is made with a singly isolated pole for bound or resonant state, 
and a continuum term.  
The latter is defined to be the region $s > s_0$ (threshold parameter) 
where it is assumed that the 
spectral density $\rho(s)$ can be identified with the imaginary part of the correlation function, 
thus calculable by OPE.   
By comparing the two expressions of (\ref{eq_WilsonsOPE}) and (\ref{Pi_phenomenological}), 
one expects to extract the information of resonance state by optimizing the parameters in $\rho(s)$.  
The phenomenological expression involves the integration (sum) over the physical region 
and so is named the sum rule.  

In  comparison of the two correlation functions, the parameters in $\rho(s)$ are 
optimized by requiring several conditions for a reliable sum rule.  
In the widely used Borel analysis, the Borel mass $M$ and the threshold $s_0$ are 
such  parameters.    
The Borel mass $M$ suppresses the continuum contributions 
to extract better low energy information by enhancing   
the contribution from the resonance  as compared to the one from the continuum.  
By varying the parameters $M$ and $s_0$, 
the extracted resonance mass should depend only mildly on them.  
Furthermore, in the spectral sum in the phenomenological side, 
the resonance (so-called pole) contributions  must be sufficiently large 
as compared to the continuum contributions.  

Having the above general remarks, let us turn to the discussion of
exotic hadrons with multiquarks.  
One of unique features is that currents for multiquark states are of higher dimensions, because they couple to multiple number of quarks and antiquarks.
This requires an OPE expansion of (\ref{eq_WilsonsOPE}) with more terms, 
with various vacuum condensates of higher dimensional operators.  
In principle, they are independent and we need more inputs for them.  
One way to reduce the number of input parameters is the vacuum saturation, 
where the vacuum expectation values of higher dimensional operators are approximated 
by the products  of those of lower dimensional  operators.

Another point to be considered is the fact that there are several 
independent currents available for a set of given quantum numbers.  
For instance, for scalar and isoscalar quantum numbers corresponding to 
$\sigma (f_0)$ meson, 
there are five independent currents~\cite{Chen:2007xr}, 
\be
\nonumber\label{define_udud_current} S^\sigma &=& 
(u_a^T C \gamma_5 d_b)(\bar{u}_a \gamma_5 C \bar{d}_b^T - \bar{u}_b \gamma_5
C \bar{d}_a^T)\, ,
\\ \nonumber
V^\sigma &=& (u_a^T C \gamma_{\mu} \gamma_5 d_b)(\bar{u}_a
\gamma^{\mu}\gamma_5 C \bar{d}_b^T - \bar{u}_b \gamma^{\mu}\gamma_5
C \bar{d}_a^T)\, ,
\label{eq_diquark_diquark_currents}
\\
T^\sigma &=& (u_a^T C \sigma_{\mu\nu} d_b)(\bar{u}_a
\sigma^{\mu\nu} C \bar{d}_b^T + \bar{u}_b \sigma^{\mu\nu} C
\bar{d}_a^T)\, ,
\\ \nonumber
A^\sigma &=& (u_a^T C \gamma_{\mu} d_b)(\bar{u}_a \gamma^{\mu} C
\bar{d}_b^T + \bar{u}_b \gamma^{\mu} C \bar{d}_a^T)\, ,
\\ \nonumber
P^\sigma &=& (u_a^T C d_b)(\bar{u}_a C \bar{d}_b^T - \bar{u}_b C
\bar{d}_a^T)\, , 
\ee
where $C$ is the charge conjugation matrix in the Dirac space, $C = i \gamma^0 \gamma^2$.  
Any of the above currents, and hence any linear combinations of them can couple to scalar mesons.  
If the QCD sum rule would work in a reasonable manner, the results for the physical states should not 
depend on the choice of the currents.  
In practice, this is not the case, because  some of them
couple stronger to the physical states than the others.  
Another remark is that the above currents  formed by the products of  diquarks
(for instance $u_a^T C \gamma_5 d_b$ in the first equation of (\ref{eq_diquark_diquark_currents}))
can be equivalently written in terms of $\bar qq$-binaries.  
The two sets of the five currents are  related each other by linear transformations~\cite{Chen:2007xr}.  

By now, there have been many theoretical works 
for the study of exotic hadrons using the QCD sum rule~\cite{Nielsen:2009uh}.  
Recently in Ref.~\cite{Narison:2010pd}, 
$X(3872)$ was studied in the sum rule by using several different types of four quark currents.  
They considered four types of currents made of 
(1) a color $\bar{\mathbf 3}_{\mathrm c}$ diquarks and antidiquark, 
(2)  a color ${\mathbf 6}_{\mathrm c}$ diquarks and antidiquark, 
(3) a $D\bar D^*$ molecular type, and 
(4) a $(\bar cc) (\bar qq)$ type.  
Although their analysis was made up to six dimensions of OPE, 
they employed a double ratio sum rule to reduce  ambiguities 
coming from higher order $\alpha_{\mathrm s}$ corrections.  
They have found a reasonable Borel window to reproduce the experimental data
for the mass.
They have also studied the coupling $X(3872) \to J/\psi \omega$ using 
the above four currents. 
The results seem to be very sensitive to the choice of currents, 
and therefore, a definitive conclusion for the nature of $X(3872)$ is not easily drawn.   
In Ref.~\cite{Chen:2013pya}, analysis was made for $X(3872)$ 
by using a $D\bar D^*$ molecular type currents 
and a hybrid currents of $\bar c G c$ where $G$ is the gluon field, with their mixing.  
They investigated the mixing of the two terms,  and found 
a reasonable Borel window with a finite amount of mixing.  

QCD sum rules have been also applied to  other exotic states.  
$Y(4260)$ was analyzed by employing a molecular type current
of $D_0 \bar D^*$~\cite{Albuquerque:2008up} and $J/\psi f_0$~\cite{Albuquerque:2011ix}.  
Their results are however inconsistent with each other.  
A consistent result was then obtained when a mixed currents of a charmonium and 
diquark-antidiquark current was employed~\cite{Dias:2012ek}.  
For $Z(4430)$, prior to the confirmation of $J^P = 1^+$ by LHCb, 
QCD sum rule analyses were performed by assuming $J^P = 0^-$
and by employing $D_1 \bar D^*$ molecular and diquark-antidiquark 
currents~\cite{Lee:2007gs}.  
An attempt was also made by using a diquark-antidiquark current of $J^P=1^-$, 
but the resulting mass was overestimated by about 400 MeV.  
Recently, analysis was made using a diquark-antidiquark current of $J^P=1^+$,
where two states were simultaneously studied for both  $Z(4430)$ and 
$Z(3900)$~\cite{Wang:2014vha}.  
Thus $Z(4430)$ was regarded as a radial excitation of $Z(3900)$.  

At present, 
it does not seem easy to make conclusive statement for exotic multiquark states
in the QCD sum rule.  
Nevertheless, the QCD sum rule is useful to test at least a consistency of 
hadronic states with QCD, and further in some cases some dynamical 
contents of them.  
Keeping these advantages, there are several recent attempts to 
improve the sum rule.  
One is to employ the maximum entropy method for the physical spectral function 
to be optimized with the OPE correlation function~\cite{Gubler:2010cf}.  
Another is to extend kinematic variables to complex ones~\cite{Araki:2014qya}.  
Both are to improve the extraction of physical information from the 
sum rule equation.
Application to exotic states would be an interesting future problem.

\subsubsection{Lattice QCD} 
\label{sec:LQCD}
\ \ \

The original motivation of the lattice QCD is to develop a scheme for the study of 
a non-Abelian gauge theory for quarks and gluons, 
where the coupling constant grows indefinitely at a certain energy scale 
as in Eq.~(\ref{eq_alpha_S}). 
It is a field theory which is defined on a discretized Euclidean space-time.  
Due to a finite lattice distance $a$, the ultraviolet momentum is cut 
at $\pi/a$,  and there is no divergence problem.  
By making the space-time volume finite the number of degrees of freedom
becomes finite, and 
the theory renders, in principle, unambiguous framework 
without relying on the perturbation expansion.  
In this regard, it is often said that the lattice QCD is the first principle method.  
The path-integration for quantized fields on a lattice 
is performed as a numerical simulation and 
has close analogy with the method in statistical physics.  

In practice, the lattice simulations requires a huge amount of computational resources.
Therefore, at each level of computer power, various methods have been invented 
to obtain the best results within a limited amount of computer resources.  
For a recent review, we refer, for example, to a textbook of Ref.~\cite{Gattringer:2010}.  

The important problems that the lattice QCD can attack in hadron physics are 
\begin{itemize}
\setlength{\itemsep}{1em}
\item 
Spectroscopy:  reproduce observed hadron masses such as pions and protons,
and describe various form factors probed by electromagnetic, weak and strong interactions.  
In the context of the present article, 
it is  important to predict new hadrons, or establish whether for instance multiquark hadrons 
exist or not.  

\item
Interaction: explain hadron interactions such as nuclear force, and describe 
other various forces which are not easily investigated by actual experiments.  

\item 
Non-perturbative dynamics of vacuum:  confinement and chiral symmetry breaking 
is the most important issues to be explained.  

\item
QCD matter:  The phase structure of hadronic matter and nature of phase transitions.  
Associated to this hadron properties under extreme conditions of finite temperature and 
density are also important. 

\end{itemize}

In fact all of them are the important problems  for hadron physics.  
Here, we  briefly present the recent  status of lattice QCD relevant  to our discussions.   
Perhaps the most successful is in the masses of ground state hadrons and their 
interactions.  
Hadron masses are studied by the correlation function in the Euclidean space-time, 
\be
\bra 0 | T J_h(x) J_h(0) |0\ket \to \exp(-E_0 T)\, .
\ee
Here $J_h$ is a hadronic current which couples to
a hadron $h$, and the separation $x$ can be taken 
$x = (T, 0, 0, 0)$.  
By inserting the complete set of the hamiltonian, 
the correlation function is expressed by a sum over 
all eigenstates with the weight factor $\exp(-E_0 T)$.
For sufficiently large $T$, the correlation function is then 
dominated by the state of the lowest energy $E_0$, which is identified with 
the mass of the ground state $h$.  

The current achievement of state-of-the-art is shown in Fig.~\ref{fig_lattice_ground}.  
The left panel shows the masses of 
light flavor ($u, d, s$) hadrons by quenched (CP-PACS2000)~\cite{Aoki:1999yr} and 
full QCD (BMW2008)~\cite{Durr:2008zz}
calculations as compared with experimental data (see also Ref.~\cite{Fodor:2012gf}).   
For CP-PACS2000, the masses of the $\pi$, $K$ and $\phi$ mesons are used as inputs 
while in the BMW2008 the masses of $\pi$, $K$ mesons and $\Xi$ baryon are.  
These hadrons are ground states in that they are the lowest mass states for a given 
flavor quantum numbers, and in the quark model their wave functions are given by 
the lowest $S$-wave states.  
Remarkably both the quenched and full QCD results agrees reasonably well 
with the experimental data, while the agreement of the full QCD results is generally better.   
Now the lattice simulation has been reaching to the precision science at the level of a few percent or less
for those states classified as orbitally ground state by the quark model; 
even electromagnetic effects can be calculated very precisely~\cite{Portelli:2010yn}.

The heavy hadron spectrum is shown in the right panel of Fig.~\ref{fig_lattice_ground}.  
In this case, several excited states are shown with good agreement 
with experimental data, but all of them are below their decay channel thresholds
of open flavors as shown by the dashed lines.  

Resonances may be assigned as orbitally excited states in the conventional quark model.  
Manipulating the interpolating fields compatible with the SU(6) symmetry on the lattice, 
excited states have been studied. The signals were observed in a manner consistent with what is expected 
in the quark model~\cite{Edwards:2011jj}.  
Furthermore, wave functions are also extracted for the Roper resonance of the nucleon 
where they observe clearly monopole type oscillations~\cite{Roberts:2013oea}.  
Quantitatively, however, as compared to the accuracy of the ground states, excited states can not be 
precisely determined suffering from finite volume effects as the pion mass 
approaches the physical value.  
This  implies that excited states are extended spatially more than the ground states 
especially at the physical point with the light pion. 
To see uncertainties which arise from some technical issues, 
we refer to references~\cite{Engel:2010my,Edwards:2011jj} where resonance masses 
are computed at different pion masses and extrapolated to the physical point.   
Typically there is uncertainties of some  hundreds MeV.  

For $X(3872)$, lattice calculation was performed by using an interpolator 
of  $\bar{c}c$, $D \bar D^*$ and diquark-antidiquark type~\cite{Prelovsek:2013xba,Padmanath:2015era}.
They claim that a signal was observed near the experimental mass region for the $I=0$ channel.  
Also, they found that the signal was seen when they include the $\bar{c}c$ term.  
These observations seem consistent with the admixture picture of molecular and $\bar{c}c$ core 
which we discuss in Section~\ref{sec:theory_X3872}.  
Other applications of the lattice method to other $XYZ$ states are shortly mentioned in Section~\ref{sec:Y4260_theory}.  
\begin{figure}[t]
\begin{center}
\includegraphics[width=0.9 \linewidth]{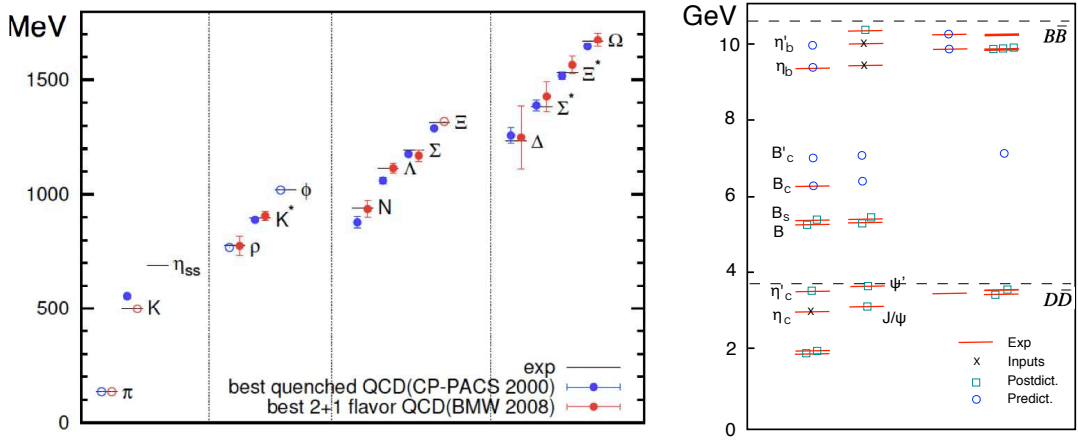}
\end{center}
\vspace{-5mm}
\caption{Masses of low lying hadrons.  
Left: light flavor ($uds$) hadrons from~\cite{Aoki:1999yr,Durr:2008zz}, 
Right: heavy flavor ($cb$) mesons, date taken from~\cite{Agashe:2014kda}}
\label{fig_lattice_ground}
\end{figure}

\subsection{Examples}
\label{sec:Examples}

In this subsection, we discuss $X(3872)$ and $Z_b(10610)^+, Z_b(10650)^+$ as typical examples 
of hadronic molecules to apply the idea explained in Section~\ref{sec:Hadron_model}.  
For $X(3872)$, we also consider an admixture of the $\bar cc$ component.  
After concrete and detailed discussions for those states, we briefly overview various interpretations for 
other states discussed in this article, $Y(4260), Z_c(4430)^+$ and $Z_c(3900)^+$.  
Some remarks on possible dynamical treatment is also addressed.  

\subsubsection{$X(3872)$}
\label{sec:theory_X3872}
\ \ \

As discussed in section~\ref{sec:exp_x3872}, 
observation of the $X(3872)$ triggered
the present activities for exotic hadrons.
Thus this state has been studied most extensively
both experimentally and theoretically.  
Although the spin and parity of  $X(3872)$ have been by now confirmed to be 
$J^{PC} = 1^{++}$ which  can be  accessible by a 
$^3P_1$ state of $\bar cc$, its properties are not  explained by a simple $\bar c c$ configuration. 
  
As compared to the other states of the same $LS$ multiplets 
$^3P_J(n=2)$, the mass of $X(3872)$ is significantly lighter 
than what is predicted as the charmonium.
Its mass is located very close to the threshold of the masses of 
$D$ and $\bar D^*$, 
\be
M(X(3872)) &=& 3871.69 \; {\rm MeV}, 
\nonumber \\
M(D^0 +D^{*0}) &=& 1864.84 + 2006.96 = 3871.80  \; {\rm MeV}, 
\label{massesXDDstar} \\
M(D^{\pm} +D^{* \mp}) &=& 1869.61 + 2010.29 = 3879.87  \; {\rm MeV}.
\nonumber 
\ee
Most notably, the decay rate of~\cite{Abe:2005ix}
\be
\frac{{\mathrm{Br}}(X(3872) \to J/\psi + 3 \pi (\omega))}
{{\mathrm{Br}}(X(3872) \to J/\psi + 2 \pi (\rho))}
\simeq 
1,
\label{Xdecay_2p_3p}
\ee
can not be explained by the isosinglet charmonium 
unless there is a mechanism of large isospin violation.  
If we look at the mass values in Eq.~(\ref{massesXDDstar}), 
we immediately realize the very special situation
where the scale of the isospin violation 
(mass difference in neutral and charged modes)
is significantly larger than the location 
of $X(3872)$ from the $D^0 \bar D^{*0}$ threshold.  

These observations naturally have lead to the idea of 
a hadronic molecule picture of $D^0 \bar D^{*0}$-$D^{\pm}D^{*\mp}$.  
In the threshold region where a new channel opens, 
the new degrees of freedom in the channel will dominate the dynamics 
of the system.  
If there is a suitable interaction they may form 
a bound or resonant state near the threshold 
which may have a very different character from what is expected 
for the ordinary states located  far from the threshold.  
The so-called threshold phenomena have been known 
for long time, for instance, in nuclear physics where alpha-clusters may play a role of 
effective degrees of freedom rather than the nucleons.  
A well known example is the Hoyle state of $^{12}C$ which can be explained by 
three-alpha structure of small binding rather than a single particle 
excitation of nucleons~\cite{Hoyle:1954zz,Tohsaki:2001an}.  

To illustrate the molecular picture for $X(3872)$ with a large isospin violation, 
let us start with a two-channel model of $D^0 \bar D^{*0}$ and $D^\pm \bar D^{*\mp}$.  
The model 
Hamiltonian is given by 
\be
H = 
\left(
\begin{array}{cc}
B_1 & V_{12} \\
V_{21} & B_2
\end{array}
\right) , 
\label{22hamiltonian}
\ee
where $B_{1,2}$ are the unperturbed binding energies
of the neutral and charged states, $D^0D^{*0}$ and $D^{\pm}D^{*\mp}$, respectively, 
$V$ their interaction matrix element, 
\be
V_{12} = V_{21}  = \bra B_1 | V |B_2 \ket .
\ee

Good  isospin symmetry implies that $B_1 \simeq B_2$ and so
\be
|B_1 - B_2| \ll |V_{12}|
\ee
In this case the eigenstates of the hamiltonian (\ref{22hamiltonian}) 
is a superposition of $|B_1\ket$ and  $|B_2\ket$ of almost equal weights,
corresponding to good isospin states.  
Contrary, suppose that the binding energy $B_1$ becomes 
much smaller than the other one $B_2$, as 
$X(3872)$ seems to be in such a case; 
$
0 \simeq B_1 \ll 
B_2 \simeq {\rm a\; few\; MeV}
$.  
If the binding energy $B_1$ is sufficiently small, 
one expects (as the actual wave functions  Eqs.~(\ref{massesXDDstar}) imply), 
that the matrix element $V_{12}$ is getting smaller as proportional to 
$B_1^{1/4}$ (see Eq.~(\ref{WF_psi_B})), and so
\be
|B_1 - B_2| \gg |V_{12}|,
\ee
Thus the eigenstates are (almost) purely $|B_1\ket$ or $|B_2\ket$, 
corresponding to the isospin breaking states.  

So far, the discussions have been made under the assumption that a suitable interaction for $D \bar D^*$
forms a state near the threshold. 
Although near threshold states are sensitive to  input conditions, 
the basic interaction itself is not understood to that accuracy.     
For instance, the $D^*D\pi$ coupling strength obtained by the experimentally observed 
$D^*$ decay is about twice larger than what is expected from SU(4) symmetry.  

The importance of the one pion exchange potential (OPEP) for $X(3872)$ 
was pointed out in Ref.~\cite{Tornqvist:1993ng} and later in Refs.~\cite{Yasui:2009bz} for other exotic states.  
The OPEP is a robust consequence of chiral symmetry and its spontaneous breaking 
for light quarks, and so is the case for the $D$ and $D^*$ mesons.    
In fact, many examples have been discussed for low-lying states of heavy hadron systems 
which are formed by the attractive interaction generated by the OPEP.  
In particular the coupled channel effect induced by the tensor structure of the OPEP
is particularly important as is known for the deuteron~\cite{Ericson:1988gk,Tornqvist:1993ng}.  

On the other hand, in the so-called chiral unitary model only a short range interaction 
is employed without the one pion exchange~\cite{Gamermann:2006nm}.  
This is based on the observation that for the channels which allow 
quark-antiquark annihilation such as $D \bar D^*$ systems, 
vector meson exchange ($\rho$ and $\omega$) provides very strong attraction.  
In practice, the short range nature of the interaction necessitates the introduction 
of a cutoff parameter which controls the strength of quantum corrections.  
By choosing the parameter suitably, the model can generate 
bound and/or resonant states in the threshold region.  
The determination of the interaction between hadrons is therefore very crucial to make 
a realistic description for the new hadrons, and is a good touchstone 
particularly for $X(3872)$.  
Further studies based on QCD with non-perturbative method is indeed needed, 
by for example the lattice simulations.  

The possible existence of $D \bar D^*$ molecular state does not necessarily explain 
the (seeming) absence of $\chi_{c1}(2P)$.  
Because the $\chi_{c1}(2P)$ state can be definitely accessible by 
a $\bar c c$ pair, it would be natural to consider a model 
with the  $\bar c c$ pair and a $D \bar D^*$ molecule.  

Having those observations, Takizawa and Takeuchi proposed to 
describe $X(3872)$ 
as an admixture state of 
charmonium and $DD^*$ molecule~\cite{Takizawa:2012hy}, 
\be
|X\ket 
= c_1 |\bar cc\ket 
+ c_2 |D^0 \bar D^{*0}\ket 
+ c_3 |D^{\pm} \bar D^{*\mp}\ket .
\label{X3872_hybrid}
\ee
The basic assumption of their model is the coupling between 
the charmonium $\bar cc$ and molecule $D\bar D^*$, which is 
parametrized as
\be
\bra D\bar D^*| V | \bar c c\ket 
= \frac{g}{\sqrt{\Lambda}}
\frac{\Lambda^2}{q^2 + \Lambda^2} .
\ee
where $g$ and $\Lambda$ are the coupling constant and the form factor, respectively.  
This interaction is commonly used for both the neutral and charged $D \bar D^*$ molecules.  
The hamiltonian to be diagonalized is then 
\be
H =
\left(
\begin{array}{ccc}
m_{\bar cc} & V & V \\
V & m_{D^0\bar D^{*0}} + K & 0 \\
V  &  0  &  m_{D^\pm\bar D^{*\mp}} + K
\end{array}
\right) ,
\ee
where $K$ is the non-relativistic kinetic energy of the corresponding $D \bar D^*$ state.  
In their model, the interactions between $D \bar D^*$ are not considered
in order to minimize the ambiguity, but with the emphasis on their coupling to 
the $\bar cc$ state.  
Such a coupling effectively plays a role of attractive interaction 
if $\bar cc$ mass is larger than the $D \bar D^*$ masses.  
As shown in Ref.~\cite{Takizawa:2012hy}, 
a small coupling $V$ indeed generates a state corresponding to 
$X(3872)$ at the right position.  

Actual numerical results depend on the choice of parameters.  
Here we present one of their typical results, where the two parameters 
in the transition interaction $V$ in Ref.~\cite{Takizawa:2012hy} 
are chosen as
$g = 0.03$ and $\Lambda= 0.5$ GeV.  
The coefficients $c_i$ are given as 
\be
c_1 = 0.293\, , \; \; c_2 = -0.92\, ,  \; \; c_3 =- 0.259,
\label{mixing_c}
\ee
This in turn gives the probability
\be
P(\bar cc) = 8 \% \, , \; \; 
P(\bar I=0) = 85 \% \, , \; \; 
P(\bar I=1) = 7 \% \, .
\label{eq_Xprobability}
\ee
indicating the non-negligible isospin violation as implied in (\ref{Xdecay_2p_3p}).  
The amount of the $\bar{c}c$ component is not very large, but is 
important to explain the production rate of $X(3872)$ through initial hard processes~\cite{Achasov:2015oia}.  
The result (\ref{eq_Xprobability}) can be well seen in the spectral function as in Fig.~\ref{fig_Xspectrum}, 
showing a sharp narrow peak at the position of $X(3872)$ 
generated by the pole of the neutral $D^0 \bar{D}^{*0}$ origin
only slightly below the threshold.  
The effect of the another pole of  charged $D^\pm \bar{D}^{*\mp}$ origin is not seen
while the evidence of the $\bar{c} c$ core origin with mass 3.95 GeV remains 
as a  wide (almost flat) background.  
This explains the absence of a clear peak of $\bar{c}c$ like state 
at the position expected in the quark model.  

\begin{figure}[t]
\begin{center}
\includegraphics[width=0.5 \linewidth]{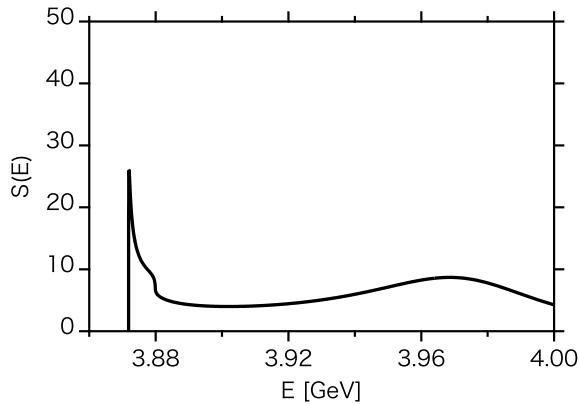}
\end{center}
\vspace{-5mm}
\caption{Spectral function for the $X(3872)$ channel~\cite{Takizawa:2012hy}. 
}
\label{fig_Xspectrum}
\end{figure}

Finally, to show another unique feature of $X(3872)$ let us look at 
its wave function.  
Suppose that the bound states $|B_{1}\ket$ and $|B_{2}\ket$ are loosely bound, meaning that their 
spatial size is sufficiently larger than the interaction range.  
In such a case, (assuming $S$-wave) the normalized wave function can be written 
for the outside of the interaction range as 
\be
\psi_B = \sqrt{\frac{\kappa}{2\pi}} \frac{e^{-\kappa r}}{r}\, , \; \; 
\kappa = \sqrt{2 \mu B},
\label{WF_psi_B}
\ee
where $\mu$ is the reduced mass of the two particles.  
We find the following elementary relation
\be
\bra r^2 \ket^{1/2} = \frac{1}{\sqrt{2} \kappa},
\label{rms_radius}
\ee
from which one can estimate the size of the bound state
using the reduced mass of the $D \bar D^*$ system ($\sim 1$ GeV) 
and the binding energy.  
Thus, from (\ref{massesXDDstar}) 
$B(D^\pm \bar D^{*\mp}) \sim 8$ MeV, 
$B(D^0 \bar D^{*0}) \sim 0.1$ MeV, 
the sizes (diameters) of the $D \bar D^*$ bound states are
\be
r_{\rm charged} = 1.1\;{\rm fm}, \; \; \; 
r_{\rm neutral} = 10\;{\rm fm}.
\label{Xsize}
\ee
The charged and neutral states are significantly different in size;
the neutral one extends further than the charged one 
due to the smaller binding energy.  
The diameter 10 fm corresponds almost to the one of 
a nucleus of mass number around $A \sim 120$, which 
is unexpectedly large as a single hadron.  
These sizes as well as the composition in the $X(3872)$ wave function (\ref{mixing_c}) give an estimation consistent with the large isospin violation.

\subsubsection{$B$ meson molecules for $Z_b(10610)^+$ and $Z_b(10650)^+$} 
\label{sec:Zbtheory}
\ \ \

As discussed in section~\ref{sec:Zb}, the observation of these two resonances 
was made in the anomalously large decay rate 
of the $\Upsilon(5S)$ resonance accompanied by two pions.  
Then it has  lead 
to the idea of hadronic molecules again.  
There are enough reasons for it.
They are:  
\begin{itemize}

\item
Their masses are very close to the $B \bar B^*$ and $B^* \bar B^*$ thresholds, 
indicating loosely bound or resonant states of these mesons.  
\be
M(B) = 5280 \; {\rm MeV}, \; \; \; M(B^*) = 5325 \; {\rm MeV}.
\ee

\item
It does not seem easy to explain the existence of the two states with the same 
quantum numbers, and with a small width in a conventional approach.  

\item
Decays into both $\Upsilon(^3S_1(n)) \pi$ ($n = 1,2,3$) and $h_b(^1P_1(n)) \pi$ ($n = 1,2$)
indicate a large violation of heavy quark spin symmetry, 
because $\Upsilon$ and $h_b$ have opposite heavy spin structure 
of either parallel ($\Upsilon$) or antiparallel  ($h_b$).  

\end{itemize}

These  problems can be resolved by assuming the molecular structure
of $Z_b(10610)$ as $B \bar B^*$ and $Z_b(10650)$ as $B^* \bar B^*$~\cite{Bondar:2011ev}.  
In particular, the last point of the violation of heavy quark spin symmetry 
can be explained by the following decomposition,  
\be
|Z_b(10610)\ket &\simeq& |B \bar B^*\ket =
\frac{1}{\sqrt{2}} | 0_H^- \otimes 1_l^-\ket 
- \frac{1}{\sqrt{2}} | 1_H^- \otimes 0_l^-\ket \, , 
\nonumber \\
|Z_b(10650)\ket &\simeq&  |B^* \bar B^*\ket = 
\frac{1}{\sqrt{2}} | 0_H^- \otimes 1_l^-\ket 
+ \frac{1}{\sqrt{2}} | 1_H^- \otimes 0_l^-\ket \, , 
\label{Zb_HQdecomposition}
\ee
where the subscripts $H$ and $l$ indicate that the spin values (0 or 1) 
are formed by two heavy and light quarks, respectively.  
Using these labels, the spin structures of $B$ and $\bar B^*$ are
\be
| B\ket \sim | [1/2_H, 1/2_l]^0 \ket\, , \; \; \; 
| B^*\ket \sim | [1/2_H, 1/2_l]^1\ket \, .
\ee
where the notation of spin coupling $[s_1, s_2]^s$ is introduced.  
Eqs.~(\ref{Zb_HQdecomposition}) are also regarded as 
the Fiertz transformations of the four quark spins of $B\bar{B}^*$ and $B^*\bar B^*$ into the pair of heavy quark spins and that of light quark spins, namely the spin recoupling.  
An important observation here is that the $B \bar B^*$ and $B^* \bar B^*$ states 
have both heavy quark spin 0 and 1 components with equal weights.  
This explains the decay of $Z_b$'s into both $\Upsilon$ and $h_b$ states.   

There have been many theoretical studies for $Z_b$'s assuming the 
$B^{(*)}$ molecule structure~\cite{Dias:2014pva,Wang:2014gwa,Wang:2014gwa,Wang:2013daa,Wang:2013zra,Ohkoda:2013cea,Dong:2012hc,Ohkoda:2012rj,Mehen:2011yh,Cui:2011fj,Cleven:2011gp,Zhang:2011jja,Bugg:2011jr,Yang:2011rp,Nieves:2011vw,Sun:2011uh,Cleven:2011gp,Mehen:2011yh,Ohkoda:2011vj,Bondar:2011ev,Voloshin:2011qa}.  
Among them we briefly overview the works in Refs.~\cite{Ohkoda:2011vj}.  
The starting point is to set up a suitable 
interaction between the heavy $B^{(*)}$ and $\bar B^*$mesons.  
This has been provided by a one boson exchange potential (OBEP) model 
derived from the Lagrangians
of heavy quark and chiral symmetries as described in Section~\ref{sec:Hadron_model}.  
The coupling strengths are determined by several mesonic processes  which are 
derived by these Lagrangians.  
Thus we have a one-boson-exchange model of $\pi, \rho$ and $\omega$ mesons.  
The form factor can be estimated from a similar OBEP model
for the deuteron; where the cutoff parameter is  scaled in (inversely) proportional to hadron size (nucleon to $B$-meson sizes).  

In Ref.~\cite{Ohkoda:2011vj}, low-lying states of angular momentum up to $L \leq 2$
were investigated.  
For a given set of quantum numbers $J^{PC} (^{2S+1}L_J)$, several coupled channels are  
possible, as pointed out in Section~\ref{sec:Hadron_model}. 
In particular, the coupling of 
different angular momenta $L, L\pm 2$ is important due to the tensor force of the one pion exchange.  
Here we summarize coupled channels for various molecular states in 
Table~\ref{tbl:BB_channels}. 

\begin{table}[t]
\caption{ Possible coupled channels for low lying 
 $B^{(\ast)}\bar{B}^{(\ast)}$ states with 
 $J^{PC}$ ($J \le 2$)~\cite{Ohkoda:2011vj}. 
 Names of some resonances which emerge by solving the coupled channel problem are shown in the last column, $Z$'s and $W$'s in Table~\ref{tbl:W_states}.
 }
\begin{center}
{ \renewcommand\arraystretch{1.5}
\small{
\begin{tabular}{|c|c|c|}
\hline
$J^{PC}$ & Components & \\ 
\hline
$0^{+-}$ & ------ & \\ 
\hline
$0^{++}$ & $B\bar{B}(^{1}S_{0})$, $B^{\ast}\bar{B}^{\ast}(^{1}S_{0})$, $B^{\ast}\bar{B}^{\ast}(^{5}D_{0})$  & \\ 
\hline
$0^{--}$ & $\frac{1}{\sqrt{2}} \left( B\bar{B}^{\ast}+B^{\ast}\bar{B} \right)(^{3}P_{0})$   & 
$W_{b0}$\\ 
\hline
$0^{-+}$ & $\frac{1}{\sqrt{2}} \left( B\bar{B}^{\ast}-B^{\ast}\bar{B} \right)(^{3}P_{0})$, $B^{\ast}\bar{B}^{\ast}(^{3}P_{0})$  & \\ 
\hline
$1^{+-}$ & $\frac{1}{\sqrt{2}} \left( B\bar{B}^{\ast}-B^{\ast}\bar{B} \right) (^{3}S_{1})$, $\frac{1}{\sqrt{2}} \left( B\bar{B}^{\ast}-B^{\ast}\bar{B} \right) (^{3}D_{1})$, $B^{\ast}\bar{B}^{\ast}(^{3}S_{1})$, $B^{\ast}\bar{B}^{\ast}(^{3}D_{1})$  & $Z_b$\\ 
\hline
$1^{++}$ & $\frac{1}{\sqrt{2}} \left( B\bar{B}^{\ast}+B^{\ast}\bar{B} \right) (^{3}S_{1})$, $\frac{1}{\sqrt{2}} \left( B\bar{B}^{\ast}+B^{\ast}\bar{B} \right)(^{3}D_{1})$, $B^{\ast}\bar{B}^{\ast}(^{5}D_{1})$  & \\ 
\hline
$1^{--}$ & $B\bar{B}(^{1}P_{1})$, $\frac{1}{\sqrt{2}} \left( B\bar{B}^{\ast}+B^{\ast}\bar{B} \right)(^{3}P_{1})$, $B^{\ast}\bar{B}^{\ast}(^{1}P_{1})$, $B^{\ast}\bar{B}^{\ast}(^{5}P_{1})$, $B^{\ast}\bar{B}^{\ast}(^{5}F_{1})$  & $W_{b1},W_{b1}^\prime$\\ 
\hline
$1^{-+}$ & $\frac{1}{\sqrt{2}} \left( B\bar{B}^{\ast}-B^{\ast}\bar{B} \right)(^{3}P_{1})$, $B^{\ast}\bar{B}^{\ast}(^{3}P_{1})$ & \\ 
\hline
$2^{+-}$ & $\frac{1}{\sqrt{2}} \left( B\bar{B}^{\ast}-B^{\ast}\bar{B} \right)(^{3}D_{2})$, $B^{\ast}\bar{B}^{\ast}(^{3}D_{2})$ & \\ 
\hline
$2^{++}$ & $B\bar{B}(^{1}D_{2})$, $\frac{1}{\sqrt{2}} \left( B\bar{B}^{\ast}+B^{\ast}\bar{B} \right)(^{3}D_{2})$, $B^{\ast}\bar{B}^{\ast}(^{1}D_{2})$, & \\
  & 
$B^{\ast}\bar{B}^{\ast}(^{5}S_{2})$, $B^{\ast}\bar{B}^{\ast}(^{5}D_{2})$, $B^{\ast}\bar{B}^{\ast}(^{5}G_{2})$  & \\ 
\hline
$2^{-+}$ & $\frac{1}{\sqrt{2}} \left( B\bar{B}^{\ast}-B^{\ast}\bar{B} \right)(^{3}P_{2})$, $\frac{1}{\sqrt{2}} \left( B\bar{B}^{\ast}-B^{\ast}\bar{B} \right)(^{3}F_{2})$, $B^{\ast}\bar{B}^{\ast}(^{3}P_{2})$, $B^{\ast}\bar{B}^{\ast}(^{3}F_{2})$  & \\ 
\hline
$2^{--}$ & $\frac{1}{\sqrt{2}} \left( B\bar{B}^{\ast}+B^{\ast}\bar{B} \right)(^{3}P_{2})$, $\frac{1}{\sqrt{2}} \left( B\bar{B}^{\ast}+B^{\ast}\bar{B} \right)(^{3}F_{2})$, $B^{\ast}\bar{B}^{\ast}(^{5}P_{2})$, $B^{\ast}\bar{B}^{\ast}(^{5}F_{2})$ & $W_{b2},W_{b2}^\prime$\\ 
\hline
\end{tabular}
}
}
\end{center}
\label{tbl:BB_channels}
\end{table}%

The coupled channel problems can be treated  non-relativistically by solving the Schr\"odinger equation,  
which is a good approximation for the heavy $B^{(*)}$ meson molecules with small binding or resonance energies.  
The latter should be checked as a consistency after obtaining the solutions.  
The results are summarized in Fig.~\ref{fig_BBstates}, where theory predictions are shown by 
horizontal bars together with (only) two experimental data corresponding to $Z_b$'s.  

It is interesting to see that the OBEP model for the $B\bar B^*$ interaction 
not only reproduces the observed twin $Z_b$'s (at least qualitatively) but also predicts  
many weakly bound and resonant states near the threshold regions.  
A priori, this is not trivial since the energies of those states depends sensitively on 
the properties of the interaction.  
It is emphasized that the model parameters here are not fixed to reproduce the twin $Z_b$'s
but from the other conditions.  
In addition, two more twin states are predicted for 
${I^G} J^{PC} = 1^+ 1^{--}, 1^+ 2^{--}$.  
Therefore, the molecular structure of $Z_b$ states implies a rich spectrum 
which can be studied experimentally, a challenge in the future.  

\begin{figure}[t]
\begin{center}
\includegraphics[width=0.7 \linewidth]{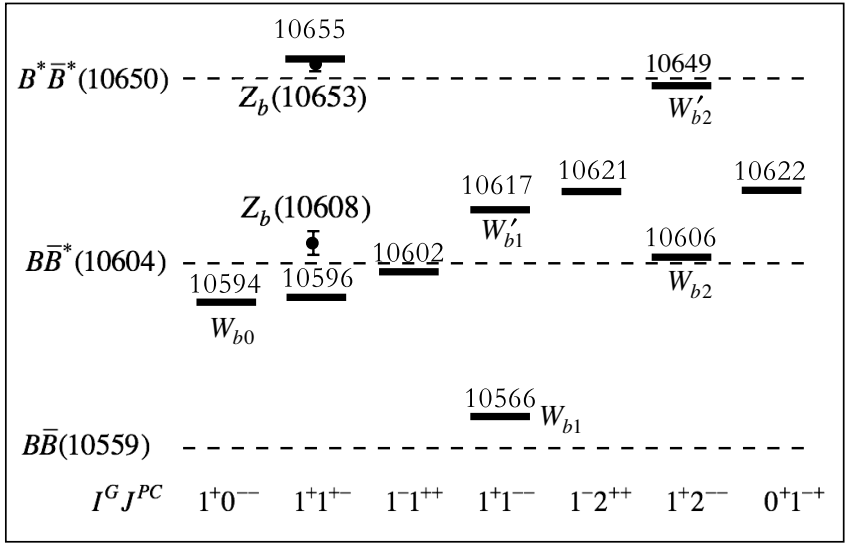}
\end{center}
\vspace{-5mm}
\caption{Low-lying states of $B \bar{B}^*$ molecules~\cite{Ohkoda:2011vj}. 
Some states are labeled as $Z$ and $W$ as shown in Table~\ref{tbl:W_states}.
}
\label{fig_BBstates}
\end{figure}

The molecular structure of the $Z_b$ resonances  shows its evidence 
also in their decay properties.  
As discussed in section~\ref{sec:Zb},  large branching ratios of $Z_b$ decaying into 
$B\bar B^*$ and $B^*\bar B^*$ states have been confirmed, 
as well as  other branching ratios into various channels.  
These are also useful to test the molecular structure of $Z_b$'s.  

\begin{figure}[t]
\begin{center}
\includegraphics[width=0.4 \linewidth]{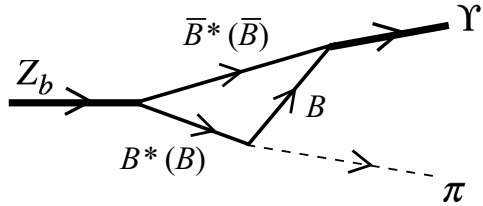}
\end{center}
\vspace{-5mm}
\caption{Decay of $Z_b$ through a hadron triangle diagram. 
}
\label{fig_Zbdecay_diagram}
\end{figure}

In Ref.~\cite{Ohkoda:2013cea}, a theoretical estimate was made by assuming that the 
decays occur through the triangle mesonic diagrams as shown in Fig.~\ref{fig_Zbdecay_diagram}.  
Physically this corresponds to the process of 
(1) $Z_b$'s dissociation into $B\bar B^*$ or $B^* \bar B^*$, 
(2) pion emission by $B^*$, and then
(3) $B\bar B^*$ or $B\bar B$ merging into the final state bottomonium $\Upsilon$.  
Various coupling constants needed to calculate these Feynman diagrams 
are evaluated by other mesonic processes, and then the decay rates 
of $\Upsilon(nS) \pi$ have been studied.  
By introducing a phenomenological form factor 
of the initial vertex for the $Z_b B \bar B^*$ and $Z_b B^* \bar B^*$, 
it was shown that the experimental decay ratios were explained 
qualitatively.
In particular, small branching ratio decaying into $\Upsilon(1S)\pi$, 
which seems opposite to what is naively expected from the largest phase space 
volume for the $1S$ state, was explained~\cite{Ohkoda:2013cea}.  
We may consider the same mechanism for the decay of $\Upsilon(5S)$ to $\pi Z_{b}'$s together with the large decay rate associated with two pion emission. This is an important task in the future to clarify the nature of $\Upsilon(5S)$

The decay properties also provide a good opportunity  to test 
the heavy quark spin symmetry, which 
leads to heavy quark spin selection rules, rigorous relations in the limit of heavy quark mass.  
The prescription to derive them is to decompose the physical states into heavy quark 
basis~\cite{Ohkoda:2013cea,Yamaguchi:2014era}.  
The model-independent relations are then derived for transitions among the states having 
the same orbital structure (dynamical property) of light degrees of freedom.  
For illustration, let us consider radiative decays of $Z_b$'s into $\chi_{b J} \gamma$ ($J=0,1,2$), 
as shown in Fig.~\ref{fig_Zdecays} by solid lines.  
In general, for such decays $M1$ and $M2$ transitions are possible.  
However, it turns out that only $M1$ transitions are available for the present decays.
Too see this point, let us perform a heavy light decomposition for the final state~\cite{Ohkoda:2012rj}
\be
| \chi_{b0} \; \gamma(M1) \ket &=&
(1_H^- \otimes 1_l^-)_{\chi_{b0}} \otimes (0^+_H \otimes 1_l^+)_\gamma 
\nonumber \\
&=& 
\frac{1}{3} (1_H^- \otimes 0_l^-)_{J=1}
- \frac{1}{\sqrt{3}} (1_H^- \otimes 1_l^-)_{J=1}
+ \frac{\sqrt{5}}{3} (1_H^- \otimes 2_l^-)_{J=1},
\label{chi0_gamma_HQdecomp}
\ee
and similarly, 
\be
| \chi_{b1} \; \gamma(M1) \ket 
&=&
-\frac{1}{\sqrt{3}} (1_H^- \otimes 0_l^-)_{J=1}
+ \frac{1}{2} (1_H^- \otimes 1_l^-)_{J=1}
+ \frac{\sqrt{15}}{6} (1_H^- \otimes 2_l^-)_{J=1},
\nonumber \\
| \chi_{b2} \; \gamma(M1) \ket 
&=&
\frac{\sqrt{5}}{3} (1_H^- \otimes 0_l^-)_{J=1}
- \frac{\sqrt{5}}{\sqrt{6}} (1_H^- \otimes 1_l^-)_{J=1}
+ \frac{1}{6} (1_H^- \otimes 2_l^-)_{J=1},
\label{chi12_gamma_HQdecomp}
\ee
The transition occurs through a matrix element of a total rank zero operator ${\cal O}_{M1}$, 
$\bra  \chi_{bL} \; \gamma(M1) | {\cal O}_{M1} | Z_b\ket$, 
and thus the coefficients of  (\ref{Zb_HQdecomposition}),  
(\ref{chi0_gamma_HQdecomp}) and 
(\ref{chi12_gamma_HQdecomp}), 
implies the ratio
\be
\Gamma(Z_b \to \chi_{b0} \gamma) :
\Gamma(Z_b \to \chi_{b1} \gamma) :
\Gamma(Z_b \to \chi_{b2} \gamma) 
=
1:3:5,
\label{Gamma_ratio_Zb_choJ}
\ee
when possible difference of phase space volume and form factor dependence 
is ignored.  

Other molecular partners of $Z_b$, 
one $0^{--}$ state, two $1^{--}$ states and $2^{--}$ states, denoted by $W_{bJ}$, 
decay into $P$-wave of $\Upsilon(1S)\pi$, rendering another example 
subject to heavy quark selection rules, because all of these states have the same 
orbital structure of $P$-wave excitations of $B\bar B^*$ molecules 
(indicated by dashed arrows in Fig.~\ref{fig_Zdecays}).  
The heavy quark decomposition of the above five molecular states 
and the $P$-wave $\Upsilon(1S) \pi$ states were performed in Refs.~\cite{Ohkoda:2012rj,Bondar:2011ev}\footnote{See also Refs.~\cite{Ohkoda:2012rj,Voloshin:2011qa} for gamma decays.}
to find the ratio (excluding again the phase space and form factor effects)
\be
\Gamma(W_{b0}) : \Gamma(W_{b1}^\prime) : \Gamma(W_{b1}) : 
\Gamma(W_{b2}^\prime) : \Gamma(W_{b2}) 
=
4:1:1:3:1\, ,
\ee 
where $W_{b0}$, $W_{b1}^\prime$, $W_{b1}$, $W_{b2}^\prime$ and $W_{b2}$ states are summarized in Table~\ref{tbl:W_states}.
For $S$-wave pion decays into $\Upsilon(1S)$,
the selection rules allow only the transition from the $1^{+-}$ $h_b$ state
while others are forbidden (dotted arrow in Fig.~\ref{fig_Zdecays}). 

\begin{table}[t]
\caption{Predicted bottomonium-like states~\cite{Ohkoda:2012rj,Ohkoda:2011vj}. }
\begin{center}
{ \renewcommand\arraystretch{1.5}
\begin{tabular}{|c|c|c|c|}
\hline
 $W_{bJ}$ & $I^{G}(J^{PC})$ & Main component ($^{2S+1}L_{J}$) & Mass (MeV) \\
 \hline
 $W_{b0}$ & $1^{+}(0^{--})$ & $(B\bar{B}^{\ast} + B^{\ast}\bar{B})$ ($^{3}P_{0}$) & 10594 \\
 $W_{b1}^\prime$ & $1^{+}(1^{--})$ & $(B\bar{B}^{\ast} + B^{\ast}\bar{B})$ ($^{3}P_{1}$) & 10617 \\
 $W_{b1}$ & $1^{+}(1^{--})$ & $B\bar{B}$ ($^{1}P_{1}$) & 10566 \\
 $W_{b2}^\prime$ & $1^{+}(2^{--})$ & $B^{\ast}\bar{B}^{\ast}$ ($^{5}P_{2}$) & 10649 \\
 $W_{b2}$ & $1^{+}(2^{--})$ & $(B\bar{B}^{\ast} + B^{\ast}\bar{B})$ ($^{3}P_{2}$) & 10606 \\
 \hline
\end{tabular}
}
\end{center}
\label{tbl:W_states}
\end{table}%

Finally, let us look at the productions of the above five $W_b$ states in the one pion emission decay
of $\Upsilon(5S)$ state as denoted by the double arrows in Fig.~\ref{fig_Zdecays}.  
These are inversion processes of their decays into $\Upsilon(1S)$ with one pion emission.  
Here the one pion is combined  with the $W_b$'s while in their decays
it is combined with $\Upsilon(1S)$, and so the resulting heavy quark decompositions differ. 
The result of the production rates $P$ of $W_b$'s is
\be
P(W_{b0}) : P(W_{b1}^\prime) : P(W_{b1}) : 
P(W_{b2}^\prime) : P(W_{b2}) 
=
4: 18: 9: 18: 24\, . 
\ee

\begin{figure}[t]
\begin{center}
\includegraphics[width=0.7 \linewidth]{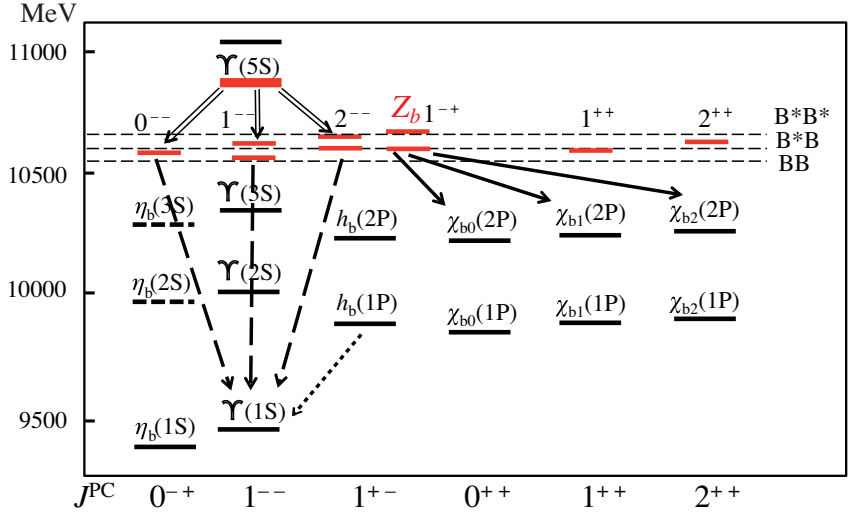}
\end{center}
\vspace{-5mm}
\caption{
Various transitions where heavy quark selection rules are applied \cite{Ohkoda:2012rj,Ohkoda:2011vj}. The black-solid bars are normal bottomonia, the red-solid bars are $B^{(\ast)}\bar{B}^{(\ast)}$ molecules. 
The solid arrows are for the gamma transitions, dashed lines are for one pion emission in P-wave and the dotted line for one pion emission in S-wave.
}
\label{fig_Zdecays}
\end{figure}

\subsubsection{Quick view for $Y(4260)$, $Z_c(4430)^+$, $Z_c(3900)^+$ and dynamical treatments}
\label{sec:quick_view}
\ \ \

In the previous two subsections we have emphasized molecular aspects for $X(3872)$,  
$Z_b(10610)$ and $Z_b(10650)$.  
Here in this subsection, we quickly look at different ideas and 
studies for $Y(4240)$, $Z_c(4430)^+$ and $Z_c(3900)^+$, though our discussions 
can not be a complete list of the previous literatures.  
These idea can be also applied to $X(3872)$ and $Z_b(10610)$ and $Z_b(10650)$.  
As we will see, the fact that there appeared by now many different ideas 
implies that the level of our understanding of these exotic particles are not 
yet satisfactory.  
The difficulty lies in the fact that they couple to various hadronic channels
above the open charm threshold.  
To perform appropriate theoretical studies, their interactions must be known,
which is currently not yet available.  

\paragraph{$Y(4260)$}
\label{sec:Y4260_theory}
\ \ \

The mass of this state lies in between the expected 
$\psi(3S)$ ($\sim 4040$ MeV)
and 
$\psi(4S)$ ($\sim 4420$ MeV) states.  
Near this energy region, there are not open and hidden charm hadronic channels. 
The most prominent feature is that it decays into $J/\psi \pi \pi$, 
while open charm decay rate is suppressed, which is in large contrast 
with ordinary $\psi$ resonances. 
Thus there have been many proposals such as 
orbitally excited tetraquark~\cite{Maiani:2005pe}, 
molecule of a charmonium and a light flavor meson~\cite{Yuan:2005dr, Liu:2005ay}, 
and baryonium of $\Lambda_c \bar \Lambda_c$~\cite{Qiao:2005av}.  
A molecular picture of $D_1(2420) D$ was also tested
 with the decay dynamics to $Z_c(3900)$ and to $J\psi \pi \pi$~\cite{Cleven:2013mka,Li:2013yla}.  
A lattice calculation was performed by using an interpolating field 
of molecular type of open charm, 
claiming that there is a resonance like structure~\cite{Chiu:2005ey}.  
To explain the decay associated with $\pi \pi$, the so-called hadro-charmonium
was proposed~\cite{Dubynskiy:2008mq,Wang:2013kra}.  
Yet, a conventional charmonium explanations were also discussed~\cite{LlanesEstrada:2005hz, Ebert:2008kb}.
In Ref.~\cite{Zhu:2005hp}, Zhu argued that many of these proposals had some difficulties 
to be consistent with observed data.  
He concluded that one plausible candidate was a hybrid with a gluon constituent~\cite{Close:2005iz,Kou:2005gt}.  
In view of these discussions, however,
while experimental observation is conclusive, 
theoretical understanding of this state is not so yet.  

\paragraph{$Z_c(4430)^+$}
\label{sec:z4430_theory}
\ \ \

The observation of this state is the first evidence of genuinely exotic state 
as its charge requires 
a multiquark component, $\bar c c u \bar d$.  
Of course, in principle, it is possible to reach this state without $\bar c c$ which is, however, 
unlikely to the same extent that $J/ \psi$ is not regarded as a light quark and antiquark pair, e.g.,  $\bar u u$.  
Experimentally, as discussed in section~\ref{sec:exp_z4430}, 
it is important that the LHCb establishes it as a resonance with phase rotation in their amplitude analysis 
for the $\psi(2S)\pi^+$ system~\cite{Aaij:2014jqa}.  

From the discovery of this state till 2013 when the spin and parity have been determined 
as $J^P = 1^+$,  there were many speculative discussions on $Z_c(4430)^+$ (with incorrect $J^P$ assignment).  
In Ref.~\cite{Rosner:2007mu}, possible threshold behaviors were discussed because the mass 
4430 MeV is close to the mass of $D^*$ and $D_1(2420)$.  
Naively S-wave states of these mesons provide $J^P = 0^-. 1^-, 2^-$, because 
$J^P$ of $D^*$ and $D_1(2420)$ are $1^-$ and $1^+$, respectively.  
Discussions based on molecular assumptions were made~\cite{Braaten:2007xw,Liu:2007bf,Liu:2008xz}.  
QCD sum rules with different types of interpolating fields~\cite{Lee:2007gs,Bracco:2008jj,Zhang:2009vs}, 
and the lattice QCD calculations were also performed~\cite{Meng:2009qt}.
Many of these studies gave positive results, 
indicating the difficulty in the theoretical study.  

After the spin and parity have been determined to be $J^P = 1^+$, 
together with the discovery of $Z_c^+(3900)$ (see the discussions below), 
yet there are several theoretical studies have appeared.  
Tetraquark calculation was performed with spin-spin interactions included~\cite{Maiani:2014aja}.  
It was also postulated that $Z_c^+(4430)$ could be a radial excitation of $Z_c(3900)$~\cite{Wang:2014vha}.  
Recently, dynamical analyses in scattering processes have been performed based on a coupled channel model~\cite{Guo:2014iya,Baru:2010ww}.  
As we shall discuss shortly below, perhaps, we need more dynamical studies to determine the nature of the observed peak.

\paragraph{$Z_c(3900)^+$}
\label{sec:z3900_theory}
\ \ \

As discussed in section~\ref{sec:z3900}, the peak structure was first 
observed in BESIII~\cite{Ablikim:2013mio} 
and later confirmed by Belle~\cite{Liu:2013dau}, 
in the $J\psi \pi^\pm$ spectrum.  
Because of the similarity of the mass, 
the state, if exist as a resonance, may be regarded as an isospin partner of $X(3872)$.  
Once again, as another candidate of a charged, and hence genuine, exotic state, 
there appeared many theoretical discussions, 
including tetraquark~\cite{Hogaasen:2013nca,Zhao:2014qva,Brodsky:2014xia} and
$DD^*$-molecule~\cite{Xie:2013uha,Wilbring:2013cha,Ke:2013gia,Zhao:2014gqa,Aceti:2014uea}.
For these models to be more established, 
dynamical and symmetry considerations should be useful.  
They include large-$N_C$ aspects in the presence of heavy quarks~\cite{Karliner:2013dqa}, 
the validity of potential tretament~\cite{Braaten:2014qka}, 
four-body dynamics~\cite{Li:2014gra}, 
hadro-charmonium~\cite{Voloshin:2013dpa}, 
and 
stability of diquarks~\cite{Jinno:2015sea}.  
QCD sum rule calculations were also performed using different types of interpolating fields~\cite{Dias:2013xfa,Qiao:2013raa,Kisslinger:2014mwa,Wang:2013vex,Fanomezana:2014qya} indicating the existence of the state.  
Lattice calculations were also performed for masses and scattering 
properties~\cite{Prelovsek:2013xba,Prelovsek:2014swa,Chen:2014afa}.

\paragraph{Dynamical treatments}
\label{sec:dynamical_theory}
\ \ \

In the previous short sections, we have seen many theoretical works, while 
it would be fair to say that we have not yet reached the level of satisfactory understanding.  
Empirically, as discussed in Section~\ref{sec:Introduction_all}, exotic hadrons appear near and above 
threshold where many open hadron channels can enter in the game.  
Dynamical approach is needed where these channels are properly 
included with reliable interactions among them.  
These include two-body (and three and more-body) interactions 
and their coupling to the genuine exotic (and one-body) 
states of tetraquark, pentaquark and so on.  
Eventually, these setups at the phenomenological level should be replaced by 
QCD.  
At the same time, however, we must be able to have reasonable descriptions 
in terms of hadrons as they are directly observed states. 

Once we have various channels coupled in a quantum mechanical manner, 
we must consider carefully the effect of threshold, resonances, interferences
and so on.  
These also require proper treatment of the scattering amplitude 
under suitable conditions such as unitarity.
Different natures of observed peaks can be then properly analyzed and 
interpreted as identified with a resonance peak or a cusp peak.  
Experimentally, this can be done by analyzing the phase dependence of the 
amplitude near the energies of the peak position.  

In the context of $XYZ$, a triangle diagram which couples 
a quarkonium like state (hidden heavy quark) with open heavy quark mesons 
were studied to show explicitly the threshold effects~\cite{Chen:2011pv}.  
The same idea was applied to the peaks of $Z_c^+(3900)$~\cite{Chen:2013coa}. 
A more general discussion was also made in Ref.~\cite{Swanson:2014tra}.
At this moment, however, without enough knowledge of interactions 
between relevant hadron channels, we still need more careful analysis.

\subsection{Future prospects}
\label{sec:future}

So far we have discussed the new experimental observations implying hadronic molecules in several cases.
Hadrons themselves are indeed useful bases as effective degrees of freedom in the sense that they are not only the possible constituent particles inside hadrons but also the asymptotic states which are directly observed.
In this section, we pick up several more exotic states, which have been predicted in theory, but not yet are found in experiments.

First, we discuss the bound/resonant states of $D^{(\ast)}D^{(\ast)}$ molecule states, namely the doubly charmed mesons (section \ref{sec:Tcc_molecule}).
These states have no annihilation channel, and hence should be regarded as the genuinely exotic states.

In addition to the hadronic molecules, however, there are other pictures of exotic hadrons, such as compact multiquark states.
There, colored diquarks are essential ingredients by strongly attractive force.
The diquarks have been discussed for many years, phenomenologically in spectroscopy, productions and decays~\cite{Jaffe:1976ig,Jaffe:1976ih,Jaffe:2005md,Wilczek:2004im,Anselmino:1992vg}, and in lattice simulations~\cite{Anselmino:1992vg,Fleck:1988vm,Close:1980rk,Donoghue:1988ec,Hands:1998kk,Alexandrou:2006cq,Selem:2006nd}.  
Although their clear evidence and nature have not been established,
it is very important to clarify the properties as colorful constituents of hadronic matter for further discussions.
The diquarks can provide the basic degrees of freedom as constituents of exotic hadrons.
We apply the diquark picture to the doubly charmed mesons, and discuss their properties (section \ref{sec:Tcc}).

The properties of the colorful substructure (diquark and/or brown-muck) are important also in the matter states under the extreme conditions, such as high temperature, densities and so on.  
Those are the current issues in relation with physics of relativistic heavy ion collisions and neutron stars.
In fact, the important role of diquarks in the dense matter has been known as the realization of the color superconductivity at high density (see Ref.~\cite{Alford:2007xm} for a review).

\subsubsection{Double charmed molecules} 
\label{sec:Tcc_molecule}

Let us consider a systems of two light and two heavy  anti-quarks  $qq\bar{Q}\bar{Q}$.
When two clusters of $q\bar{Q}$ are developed and are separated sufficiently far in distance, say 1 fm,
it is natural to regard those two clusters as heavy mesons and as the effective degrees of freedom
for the molecular structure; $\bar{P}^{(\ast)}\bar{P}^{(\ast)}$ with $\bar{P} \sim (q\bar{Q})_{\rm spin\, 0}$ and $\bar{P}^{\ast} \sim (q\bar{Q})_{\rm spin \,1}$.

In this case, $\bar{P}^{(\ast)}$ and $\bar{P}^{(\ast)}$ interact through a long distance force of the meson exchanges.
The most important one is the  pion exchange  force as discussed in the previous sections.  
For molecular states, the wave function of $\bar{P}^{(\ast)}\bar{P}^{(\ast)}$  with $I(J^{P})=0(1^{+})$ has four components;
\begin{eqnarray}
\frac{1}{\sqrt{2}} \left( \bar{P} \bar{P}^{\ast} - \bar{P}^{\ast} \bar{P} \right) (^{3}\mathrm{S}_{1}), \,
\frac{1}{\sqrt{2}} \left( \bar{P} \bar{P}^{\ast} - \bar{P}^{\ast} \bar{P} \right) (^{3}\mathrm{D}_{1}), \,
\bar{P}^{\ast}\bar{P}^{\ast} (^{3}\mathrm{S}_{1}), \,
\bar{P}^{\ast}\bar{P}^{\ast} (^{3}\mathrm{D}_{1}),
\end{eqnarray}
with $^{2S+1}L_{J}$ denoting a state of spin $S$, angular momentum $L$ and total spin $J$.
Thus, we have the following $4 \times 4$ matrix for the interaction
\begin{eqnarray}
V_{0(1^{+})}^{\pi}(r) =
\left(
\begin{array}{cccc}
  -C_{\pi}(r) & \sqrt{2} \, T_{\pi}(r) & 2 \, C_{\pi}(r) & \sqrt{2} \, T_{\pi}(r) \\
  \sqrt{2} \, T_{\pi}(r) & -C_{\pi} - T_{\pi}(r)  & \sqrt{2} \, T_{\pi}(r) & 2 \, C_{\pi}(r) - T_{\pi}(r) \\
 2 \, C_{\pi}(r) & \sqrt{2} \,  T_{\pi}(r) & -C_{\pi}(r) & \sqrt{2} \,  T_{\pi}(r) \\
 \sqrt{2} \, T_{\pi}(r) & 2 \, C_{\pi}(r) - T_{\pi}(r) & \sqrt{2} \, T_{\pi}(r) & -C_{\pi}(r) - T_{\pi}(r)
\end{array}
\right) \, ,
\label{eq:pi_pot_1+-}
\end{eqnarray}
with the functions $C_{\pi}(r)$ and $T_{\pi}$(r) in the central and tensor potentials defined in Eqs.~(\ref{eq:C_function}) and (\ref{eq:T_function}) ($r$ is the distance between the two mesons) from the $\pi$ exchange potential \cite{Manohar:1992nd,Ohkoda:2012hv}.
The hamiltonian is given by
\begin{eqnarray}
H_{0(1^{+})}^{\mathrm{HM}}(r) = K(r) + \Delta M + V_{0(1^{+})}^{\pi}(r),
\label{eq:hamiltonian_TQQ_hadron}
\end{eqnarray}
where $K = (K_{0}^{PP^{\ast}}, K_{1}^{P^{\ast}P^{\ast}}, K_{0}^{PP^{\ast}}, K_{1}^{P^{\ast}P^{\ast}})$ is the term of the kinetic energy and $\Delta M = (0,0,\Delta m_{P^{\ast}P},\Delta m_{P^{\ast}P})$ with the $\bar{P}$-$\bar{P}^{\ast}$ mass splitting $\Delta m_{P^{\ast}P}= m_{P^{\ast}}-m_{P}$.
The energy is measured from the $\bar{P}\bar{P}^{\ast}$ threshold.
See Ref.~\cite{Ohkoda:2012hv} for further details including the numerical settings of the parameters.

By diagonalaizing Eq.~(\ref{eq:hamiltonian_TQQ_hadron}),
we find the masses and decay widths for $I(I^{P})=0(1^{+})$ as shown in Fig.~\ref{fig:Tcc_hadronic}.
In our calculation, we obtain a bound state with the mass 3813 MeV (binding energy 60 MeV from $\bar{D}\bar{D}^{*}$ threshold) for $0(1^{+})$.
Therefore, a simple molecular model for double charm can accommodate a stable bound state against the strong decay.  
Similarly we find also other states in several channels (see Ref.~\cite{Ohkoda:2012hv} for details).

\begin{figure}[tb]
  \begin{center}
   \includegraphics[angle=0,width=100mm]{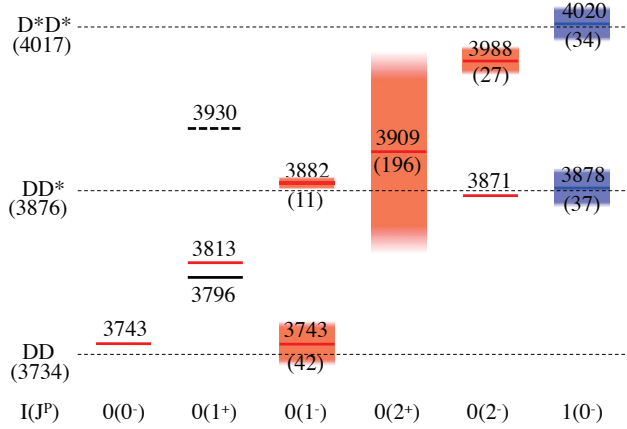}
  \end{center}
 \caption{The predicted mass spectrum of doubly charmed molecules and $T_{cc}$.  The red (blue) bars indicate the mass position and decay widths for $I=0$ ($I=1$) obtained from the hadronic model \cite{Ohkoda:2012hv}. The solid (dashed) black bar indicates the mass position of the $T_{cc}[{\bf 3}_{c}]$ ($T_{cc}[\bar{\bf 6}_{c}]$) with $I(J^{P})=0(1^{+})$ obtained from the diquark model \cite{Lee:2009rt,Lee:2007tn}.The numbers (in parentheses) in the figure indicate the masses (decay widths) in unit of MeV.}
\label{fig:Tcc_hadronic}
\end{figure}

\subsubsection{$T_{cc}$} 
\label{sec:Tcc}

When two $\bar{P}^{(\ast)}$ mesons approach closer,
diquarks $qq$ and $\bar Q \bar Q$ may be more developed to form an exotic tetraquark $qq\bar{Q}\bar{Q}$.
We call this state $T_{QQ}$ to distinguish it from the $\bar{P}^{(\ast)}\bar{P}^{(\ast)}$ hadronic molecules.

$T_{QQ}$ could become stable against the strong decay in a diquark picture, as the argument for the stability goes as follows.  
Apart from the confinement force, 
the dominant interaction energy is supplied by the light diquark $qq$. 
The interaction between a light (anti)quark $q$ ($\bar{q}$) and a heavy (anti)quark $Q$ ($\bar{Q}$) is suppressed 
by ${\cal O}(1/m_{Q})$ with a heavy quark mass $m_{Q}$.
The interaction between two heavy (anti)quarks are further suppressed by ${\cal O}(1/m_{Q}^{2})$.
Therefore, 
the mass of $T_{QQ}$ could be lower than the mass of the possible strong decay channel of two $q\bar{Q}$ mesons, if the $qq$ interaction is attractive;
\begin{eqnarray}
M(qq\bar{Q}\bar{Q}) < M(q\bar{Q}) + M(q\bar{Q}). 
\end{eqnarray}
In this case, the only possible decay is the weak decay of heavy quarks.
We should note that the situation differs for $q\bar{q}Q\bar{Q}$ states.
If $q\bar{q}$ behaves as the Nambu-Goldstone bosons ($\pi$, $K$ and $\eta$ mesons) with small mass, $M(q\bar{q}) \sim 0$,
we expect the following mass relations;
\begin{eqnarray}
M(q\bar{q}) + M(\bar{Q}Q) < M(q\bar{q}Q\bar{Q}) < M(q\bar{Q}) + M(\bar{q}Q),
\end{eqnarray}
hence observe that the $q\bar{q}Q\bar{Q}$ state is not likely to be 
the ground state of the system.

To estimate the mass of $T_{QQ}$, we assume that $ij$ quark pair interact though the color-magnetic interaction (cf. Eq.~(\ref{eq:color-spin_interaction})), 
\begin{eqnarray}
V_{ij}(S)
= 
C_{H}
\vec{s}_{i} \cdot \vec{s}_{j} \frac{1}{m_{i}m_{j}},
\end{eqnarray}
with dynamical (constituent) quark mass $m_{i}$, spin operator $\vec{s}_{i}$~\cite{Lee:2009rt,Lee:2007tn}.
$S$ is the total spin of quark $i$ and $j$.
Obviously, such an interaction is suppressed for heavy quarks of mass 
$m_{c} \simeq 1500$ MeV and $m_{b} \simeq 4700$ MeV ($=m_{Q}$)
as compared to light quarks of mass
$m_{u} \simeq m_{d} \simeq 300$ MeV and $m_{s} \simeq 500$ MeV.  
For quark-quark interaction, from the mass splittings among light hadrons 
with different spins such as $N$-$\Delta$, $\Lambda$-$\Sigma$, $\Lambda_{c}$-$\Sigma_{c}$ and $\Lambda_{b}$-$\Sigma_{b}$, we find $C_{B}/m_{u}^{2}=193$ MeV $(H=B)$ reproduces the observed mass splittings\footnote{For quark-antiquark interaction, we can similarly obtain $C_{M}/m_{u}^{2}=635$ MeV ($H=M$) to reproduce the meson mass splitting $\pi$-$\rho$, $K$-$K^{\ast}$, $D$-$D^{\ast}$ and $B$-$B^{\ast}$.}.
The most attractive energy is in the $ud$ diquark with spin-singlet ($S=0$) and isospin-singlet ($I=0$) (color $\bar{\bf 3}_{c}$).
Thus the lowest mass of $T_{QQ}$ is given as
\begin{eqnarray}
M(qq\bar{Q}\bar{Q}) = 2m_{Q} + 2m_{q} + V_{qq}(S=0),
\end{eqnarray}
which gives lighter mass than $M(q\bar{Q})+M(q\bar{Q}) \simeq 2m_{Q} + 2m_{q}$.
The quantum numbers of the lowest $T_{QQ}$ should be $I(J^{P})=0(1^{+})$.
The total spin $J=1$ is due to the $S$-wave $\bar{Q}\bar{Q}$ pair of color ${\bf 3}_{c}$, satisfying the Pauli principle.
More realistically, we need to consider the contributions of order $1/m_{q}m_{Q}$ for $q\bar{Q}$ and of order $1/m_{Q}^{2}$  for $\bar{Q}\bar{Q}$. 
Finally, we obtain the binding energy of $T_{cc}$ to be about $80$ MeV when measured from the $DD^{\ast}$ threshold, and  about $124$ MeV for $T_{bb}$ from $BB^{\ast}$~\cite{Lee:2009rt,Lee:2007tn}.

The binding energy estimated in the diquark model is eventually the same order of that obtained in the hadronic model (Fig.~\ref{fig:Tcc_hadronic}).
However, the mechanisms are quite different each other; the pion exchange interaction for the former, and the color-spin interaction in diquarks for the latter.
Each model simplifies the real QCD, and the truth must be in between two of them.
Recently the numerical simulation in the lattice QCD has reported the attraction for $DD^{*}$ scattering in the $I(J^{P})=0(1^{+})$ channel corresponding to the one in the $T_{cc}$ channel \cite{Ikeda:2013vwa}.
The final conclusion is, however, not yet available due to the large pion mass in the lattice QCD simulations.

In addition to the color triplet ${\bf 3}_{c}$ configuration for $\bar{Q}\bar{Q}$ (this $T_{cc}$ is denoted by $T_{QQ}[{\bf 3}_{c}]$),
we may consider the exotic color ``antisextet" configuration for $\bar{Q}\bar{Q}$; $T_{QQ}[\bar{\bf 6}_{c}]$~\cite{Hyodo:2012pm}.
The $qq$ ($\bar{Q}\bar{Q}$) pair of ${\bf 6}_{c}$ ($\bar{\bf 6}_{c}$) receives  
attraction in $S=1$ (isospin-singlet) channel with the strength reduced to $1/6$ of the one of $\bar{\bf 3}_{c}$ $qq$ pair (see~Eq.~(\ref{eq:color-spin_interaction}), and also Eqs.~(\ref{eq:lambda_lambda}) and (\ref{eq:sigma_sigma})).
Being of the same total quantum numbers, those two $T_{QQ}$'s may mix, but
the mixing should be suppressed due to the small spin-flip between $\bar{Q}\bar{Q}$ ($S=0$) in $T_{QQ}[{\bf 3}_{c}]$ and $\bar{Q}\bar{Q}$ ($S=1$) in $T_{QQ}[\bar{\bf 6}_{c}]$.
The estimate of the mass of $T_{QQ}[\bar{\bf 6}_{c}]$, however, indicates that $T_{QQ}[\bar{\bf 6}_{c}]$ may not be bound.  
For example, the mass of $T_{cc}[\bar{\bf 6}_{c}]$ ($Q=c$) is 54 MeV above the $DD^{\ast}$ threshold, and it can decays to $D$ and $D^{\ast}$ in $S$-wave.
Thus, some special mechanisms may be required for the $T_{cc}[\bar{\bf 6}_{c}]$ to be a quasi-stable state as a resonance.
Nevertheless, the different color structures are reflected in the differences in the production processes, and hence they may give a hint to understand the color structure inside the hadrons, as mentioned below.

The production of $T_{QQ}$, as well as the $P^{(\ast)}P^{(\ast)}$ states discussed in the previous section, in experimental facilities is an important issue.
One possible method to investigate the double charm production ($cc\bar{c}\bar{c}$) is collider experiments. 
In fact, there are several reports on observations of double charmonia, where $cc\bar{c}\bar{c}$ is combined to two $c\bar{c}$ pairs in the final states.
Then it will be naturally expected that the $cc\bar{c}\bar{c}$ is combined also to produce double charm hadrons containing $cc$ and others with two $\bar{c}q$.
Along this picture, in theoretical studies, the production rates to produce $T_{cc}$ in electron-positron collisions was numerically estimated in the framework based on the NRQCD~\cite{Bodwin:1994jh}.
The interesting result is that the properties of the productions of $T_{QQ}[{\bf 3}_{c}]$
and $T_{QQ}[\bar{\bf 6}_{c}]$ are different in the momentum and angle dependence in the differential cross sections \cite{Hyodo:2012pm}.
It may give us an experimental way to study the internal ``color-exotic" structure in exotic hadrons.
Another possible method to observe $T_{QQ}$ is to investigate the productions of hadrons in relativistic heavy ion collisions.
There, the abundant of quarks are produced, and some of them could be combined to exotic hadrons through the hadronization processes.
The numerical estimates of yields of exotic hadrons, including $T_{cc}$ and other exotic states, was given based on the coalescence model and on the statistical model for LHC and RHIC \cite{Cho:2010db,Cho:2011ew}.
Here, the interesting conclusion is that the yields of exotic hadrons are sensitive to the formation processes; the compact objects as multi-quarks will be produced in the quark-coalescence, and the extended objects as hadronic molecules will be in the hadron-coalescence.
The experimental observations of the exotic hadrons in the relativistic heavy ion collisions may unveil the structures of exotic hadrons in the coming future.

\subsubsection{Baryon-rich exotics} 
\label{sec:baryon}

So far we have discussed the $XYZ$ hadrons with baryon number zero.
The heavy quark symmetry plays the essential role, not only the $XYZ$ hadrons, but also in heavy flavor baryon and heavy flavor nuclei with finite baryon number.
The important thing is the decomposition of the spin structure in the heavy quark limit, as we have discussed in the cases of the $XYZ$ hadrons.
In the heavy flavor baryons with light quarks, $Qqq$, the light degrees of freedom can be regarded as diquarks.
Diquarks are the composite states of two quarks with color non-singlet combinations.
Although the existence of diquarks have been discussed since the time when Gell-Mann invented quarks, 
there are still no significant establishment of their existence nor their properties.
According to the heavy quark symmetry, we can separate the dynamics of two light quarks, $qq$, as a diquark from the heavy quark $Q$.
Therefore, the study of heavy flavor baryons will provide us with a unique opportunity to investigate the diquark dynamics~\cite{Kim:2014qha,Yoshida:2015tia}.
Along this line, the heavy flavor nuclei, as an extension from one baryon number state to multiple baryon number states, are also interesting topics~\cite{Yasui:2013vca,Yamaguchi:2014era}.
In this case, the heavy quark symmetry will affect the spectrum of the heavy flavor nuclei, and the light degrees of freedom will provide us with new dynamical degrees of freedom in describing the several low modes in nuclei.
It is important to note especially the essential degrees of freedom are given by the heavy quark spin and the brown muck spin in the heavy quark limit, where the latter includes the nucleon-hole pairs around the Fermi surface in nuclei.
Thus, the heavy quark symmetry is related also with the breaking of the rotation  symmetry in deformed nuclei, where nucleon-hole pairs are created collectively in the ground state.
The separation of the effective degrees of freedom by the heavy quark spin is a new phenomena, which cannot be seen in light flavor nuclei.
The applications to finite baryon number systems with heavy flavor should be explored both in experimental and theoretical studies in future.

As subjects towards heavier systems, double charm hadrons ($|C|=2$) are important.
They would provide us with information about the heavy {\it quark}-{\it quark} ($QQ$) interaction with color {\it non-singlet} (${\bf 3}_{c}$) channels.
Furthermore, spectroscopy of triply-heavy baryon $\Omega_{QQQ}$ will be important for understanding the color-confinement force in three quarks $QQQ$.
Those topics will open a new horizon for study of heavy hadron systems.

\subsection{Theory summary}
\label{sec:summary_theory}

The question about how hadrons are composed from quarks has been a long-standing problem, since Gell-Mann and Zweig proposed quarks as the most fundamental particles in the strong interaction.
It is remarkable that, in the heavy flavor sector, the recent discoveries of $XYZ$ exotic hadrons, especially the charged $Z$ states, have given the milestone to study the existence of multi-quark states.

In quantum systems, it is generally important to identify the effective degrees of freedom at each and different energy scale, and construct the effective Lagrangian for them appropriately.
The successful example is the chiral perturbation theory at low energies, where the effective degrees of freedom are given by the Nambu-Goldstone bosons generated by the dynamical breaking of chiral symmetry in vacuum~\cite{Weinberg:1978kz,Gasser:1983yg,Gasser:1984gg}. 
In heavy quarkonia, on the other hand, the heavy quark and antiquark are well defined as the effective degrees of freedom due to the suppression of the quark-antiquark annihilation processes, and the interaction is provided by the quark potentials like the color-Coulomb and confinement potentials.

In the $XYZ$ hadrons, however, the questions about the effective degree of freedom and the effective interaction working for them are still under debates. 
One of the keys to approach those questions is the heavy-quark symmetry, where each heavy quark spin and brown muck spin are conserved in the heavy quark limit, $m_{\rm Q} \rightarrow \infty$.
The heavy quark symmetry holds for any hadron systems, including the multi-quark states, the hadronic molecule states, the nuclear systems, and 
hence it will give us a systematic way for understanding.
We need also to investigate the $1/m_{\rm Q}$ corrections beyond the leading order.
The symmetry breaking effects will be important particularly when the mass is close to thresholds.
In particular, if there are open hadronic channels nearby, 
they couple each other and influence the scattering amplitudes showing non-trivial behaviors 
associated with the bound states, resonant states, cusps and so on.  
For the analysis of the $XYZ$ hadrons, these dynamical  effects are very important, and will be an  issue in the future developments.

\section{Summary}
\label{sec:summary_all}

Discovery of $X(3872)$ followed by series of ``$XYZ$" particles have opened a 
new era of hadron spectroscopy. 
A lot of efforts from both experimental 
and theoretical sides have been invested to give proper 
understanding 
of those states. 
Revealing their structures directly corresponds 
to the answer for the fundamental question, 
``What are the most suitable degrees of freedom to form hadrons?", 
when hadrons are composite systems 
of quarks and gluons which are strongly renormalized by QCD interactions.  
The so-called constituent quarks, gluons, and even their composites such as diquarks 
may be such  degrees of freedom.  
Although this is a long-standing question before the current activities started, 
it is amusing to see that the question is revisited by the findings of the $XYZ$ states 
containing heavy quarks.  

Experimentally, not only the identification of peak positions 
but also dynamical properties such as 
decays and branching ratios have been accumulated.  
The latter properties are particularly sensitive to internal structures and 
so are useful to answer the equation raised above.  
As a consequence of the activities,
the role of the molecular states has been recognized to interpret 
$X(3872)$ as the admixture of $J^{PC}=1^{++}$ charmonium and 
$D\bar{D}^{*}$ molecular states and regard $Z_b(10610)^+$ and 
$Z_b(10650)^+$ as $B\bar{B}^*$ and $B^* \bar{B}^*$ molecular states, 
respectively. 
Yet there are several states which are well established by experiments, while little is understood theoretically.  
They are, for instance, $Y(4260)$ and $\Upsilon(5S)$.  
Concerning them the mechanism of large two pion emission rate does not seem to be explained easily, while this property has triggered the discoveries of many exotic states.  
The structure of the charged states $Z_c(4430)^+$ and $Z_c(3900)^+$ is not yet clear neither.  
This is partly the reason that we did not discuss these important states in this article, while we simply mention that further understanding of them brings yet important steps in hadron physics.  

Inspired by the progress, strategies to perform various partner searches 
have been figured out for the quarkonium(-like) states. For that purpose,
higher statistics data is strongly desired because of anticipated dominant 
decay modes and consequent signal yield expectations. 
Yet further activities are expected for double charmed tetraquarks, 
double charmed molecules, heavy flavored baryons, and even heavy flavored 
nuclear matter.  
Theoretically, further mutual developments of the first principle methods 
and effective approaches will become more important.  
All of them bring challenges, 
not only by their own sake, but also to answer the fundamental question.  
After all, composite nature leads to that their properties depends on 
the environments which they interact with.  
This is partly the reason that different approaches are indeed required 
for the study of hadron physics.  

Much more information is hoped for next generation intensity frontier 
projects such as Belle II, upgraded LHCb, and J-PARC experiments.
At the same time, effective theory as well as the lattice QCD approaches
are exhibiting remarkable progress. 
Using the ``$XYZ$" states as  interesting objects to be discussed, 
the non-perturbative QCD mechanism continue to be revealed.

\section*{Aknowledgments}
The authors would like to thank support provided by Grant-in-Aid for Scientific Research
on Priority Areas "Elucidation of New Hadrons with a Variety of Flavors (X00: 21105001,
A01: 21105002, E01: 21105006) from Japan Society for the Promotion of Science (JSPS).
This work is also supported in part by Grant-in-Aid for Science Research (C) 26400273 (A.~H.), 25247036 (S.~Y.) and by Grant-in-Aid for Science Research (B) 15K17641 (S.~Y.).

\bibliographystyle{ptephy}
\bibliography{references_151101}

\end{document}